\documentclass[11pt]{article} 

\usepackage{amsmath,amssymb,amsfonts,amsthm,makeidx,graphicx}
\usepackage{txfonts}
\usepackage{helvet}
\usepackage[dvipsnames]{xcolor}
\usepackage{bbold}
\usepackage{feynmp-auto}
\usepackage{slashed}
\usepackage{subcaption}
\usepackage{hyperref}

\usepackage[capitalise]{cleveref}
\usepackage{tikz}
\usetikzlibrary{decorations.pathmorphing}
\usetikzlibrary{decorations.markings}
\usetikzlibrary{calc,tikzmark,fit,shapes.geometric,matrix,decorations.markings,arrows.meta,decorations.pathmorphing,patterns,positioning,snakes}
\tikzset
  {midarrow/.style={decoration={markings,mark=at position 0.5 with
     {\arrow[thin,xshift=2pt]{Triangle[length=4pt,#1]}}},postaction={decorate}}
  }

\tikzset{
proton/.style = {circle, draw=black, thin, fill=black!20!white, minimum size=#1,
              inner sep=0pt, outer sep=0pt},
proton/.default = 6pt % size of the circle diameter 
}

\tikzset{
blob/.style = {circle, draw=black, thin, preaction={fill, black!20!white}, pattern=north east lines, minimum size=#1,
              inner sep=0pt, outer sep=0pt},
blob/.default = 6pt % size of the circle diameter 
}

\tikzset{
wc/.style = {circle, fill, minimum size=#1,
              inner sep=0pt, outer sep=0pt},
wc/.default = 4pt % size of the circle diameter 
}

\tikzset{vector/.style={decorate, decoration=snake}}

\def\shifta{-4.5} % eq 14
\def\shiftb{-2.5} % eq 20
\def\shiftc{-2.0} % eq 36

\graphicspath{{./figures/}}
\usepackage{geometry}
\usepackage{authblk}
\usepackage{cite}
\usepackage{textgreek}

\title{\vspace*{-2.5cm}\hfill
{\normalsize KA-TP-24-2026, P3H-26-082, MITP-26-048}\\[3cm] 
Effective Quantum Field Theory and SMEFT}

\author[1,2]{Anisha}%

\author[2]{Anke Biek\"otter}%

\author[3]{Matthias K\"onig}%

\affil[1]{\footnotesize Karlsruhe Institute of Technology, Institute for Astroparticle Physics, Hermann-von-Helmholtz-Platz 1, 76344~Eggenstein-Leopoldshafen, Germany}
\affil[2]{Karlsruhe Institute of Technology, {Institute for Theoretical Particle Physics}, Wolfgang-Gaede-Straße 1, 76131~Karlsruhe, Germany}
\affil[3]{Johannes Gutenberg University Mainz, PRISMA$^{++}$ Cluster of Excellence \& Mainz Institute for Theoretical Physics, Staudingerweg~9, 55128~Mainz, Germany}

\begin{document}

\maketitle

\begin{abstract}
We introduce the basics of Effective Field Theories, using the Standard Model Effective Field Theory as our central example.
The first part of this chapter gives a pedagogical introduction to EFTs -- including matching, resummation and the construction of an operator basis -- with explicit examples and provides a starting point for newcomers. 
The second part reviews the current status of SMEFT and collects suitable references for those who want to go deeper. 
\end{abstract}

%--------------------------------
\section{Introduction}
\label{sec:intro}

In physics, we often rely on the idea that, to describe a given phenomenon, we need not account for every detail of the underlying system, but only for those effects that are most relevant for the process we study. For example, we do not need quantum field theory to describe (most of) the phenomena of the macroscopic world around us. The reason for this is that the energy and length scales of the quantum world are nowhere near the energy and length scale of our surrounding objects. As a result, we can describe their physics using an approximate and much simpler theory than the Standard Model of particle physics~(SM).

Effective Field Theories~(EFTs) apply this principle of distilling the essence of a theory relevant at a certain scale to field theory, and are most commonly used in the context of quantum field theory~(QFT). 
Physical processes described in QFTs typically receive contributions from a multitude of effects, 
which can make exact treatments challenging or even intractable.
These contributions occur at different energy scales and those associated with scales close to the scale of the process under consideration tend to dominate, while contributions from other scales are suppressed.
Therefore, it can be useful to define an effective description which identifies the leading contributions and neglects subdominant ones. 
This can be done systematically by identifying an appropriate parametric limit and expanding around it. 
The resulting effective description is an approximation, in which subleading effects can be included systematically through higher-order terms in the expansion. 
Every QFT that is not a ``Theory of everything'' should be considered an EFT. In that sense, we should consider the SM as an EFT of a more complex QFT which can solve the phenomena unexplained by the SM such as why there is a matter-antimatter--asymmetry or how neutrino masses are generated.

The most common application of effective QFTs in particle physics is the calculation of scattering processes mediated by particles with mass $M$ much heavier than the scattering energy $E$, $M \gg E$. This scale hierarchy then allows for an expansion in the ratio $E/M$. This is the example we will focus on here, but we stress that there are different EFT constructions in which the expansion takes on a different form. 
In the calculation of physical processes based on Feynman diagrams, energy scales enter through masses of particles or kinematic invariants. Propagators with large virtualities (i.e. large denominators) act as suppression compared to propagators close to their mass shell, and thus the guiding principle to construct an EFT is to identify field modes giving rise to highly virtual propagators and removing them from the theory. 
We will see in this chapter that this does not only simplify our model, but in the context of radiative corrections it also enables us to perturbatively control large logarithms of the scale ratios.

An important use case of EFTs in particle physics is for physics beyond the Standard Model~(BSM): 
In the absence of a direct discovery of a new particle, for instance at the Large Hadron Collider (LHC) which probes energy scales up to the TeV scale, 
BSM physics is typically assumed to originate from a scale far above the EW scale.  
The assumed hierarchy between the masses of the new heavy particles and those of the SM thus represents a natural application of the EFT formalism: 
If current experiments do not run at energies high enough to observe a new particle directly, the first imprints of new physics~(NP) might be limited to modifications of interactions between the SM particles. 
An EFT is an elegant parameterization of these low-energy effects of the heavy NP, since their construction is model-agnostic. \textit{Any} given NP model (falling under the core assumptions) can be mapped to a universal, agreed-upon set of parameters. 
Crucially, the effective description does not need to be re-derived for different NP models due to its universality. 
This also implies that the phenomenology of the EFT is fully agnostic to the complete model. Predictions for observables can be expressed in terms of the parameters of the EFT, and differences between concrete NP models manifest themselves in the values of the latter.
The price to pay for this universality is the vast number of parameters of an EFT. In practice, determining the EFT parameters corresponding to a specific NP model can be extremely cumbersome, especially beyond leading order. 
Analogously, low-energy observables depend on numerous EFT parameters and in the presence of a deviation of the data from the SM prediction, it can be hard to pinpoint which parameter of the EFT is non-zero. 
Often, simplifying assumptions on the EFT have to be made to render analyses feasible, which reduces the universality of the EFT to studying subsets of the full parameter set. These reductions  of the parameter space are typically motivated by concrete NP scenarios.

In this chapter, we aim to give an overview of the EFT framework and its applications in the context of physics beyond the SM. We begin by introducing core EFT principles in~\cref{sec:principles} and discussing the matching between ultraviolet~(UV)-complete theories and their EFT in~\cref{sec:matching}.
In~\cref{sec:resummation}, we discuss how EFT helps to resum large logarithms through the renormalization group evolution~(RGE). 
In~\cref{sec:basis}, we explain how to form an operator basis of an EFT. 
SMEFT, including its operator basis, flavor structures, field redefinition, input schemes, higher-order contributions, global fits, and UV model connections, is discussed in~\cref{sec:SMEFT}.
We conclude in~\cref{sec:conclusions}.

An extensive list of excellent introductions to 
EFTs and SMEFT is already available, and any short selection is necessarily incomplete. We therefore restrict ourselves here to a few representative examples:
For excellent lecture notes on EFTs, we refer the reader to Refs.~\cite{Burgess:2007pt,Manohar:2018aog,Cohen:2019wxr,Neubert:2019mrz}. 
A comprehensive SMEFT review can be found in~\cite{Brivio:2017vri}.
Visual learners may benefit from the summer school lecture recordings on general EFT topics and SMEFT in Refs.~\cite{ICTS:2024satpp,Brivio:2023ggi}.

%--------------------------------
\section{Core Principles}
\label{sec:principles}
%--------------------------------

\begin{figure}
    \centering
\includegraphics[width=0.4\linewidth]{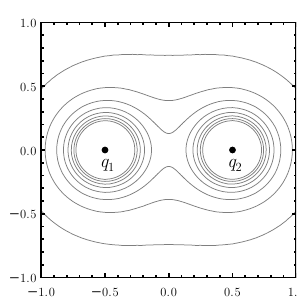}
    \qquad
    \includegraphics[width=0.4\linewidth]{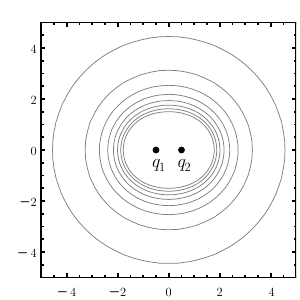}
    \caption{The magnitude of the near (left) and (far) electric field of two point charges $q_1$ and $q_2$. The axes are in units of the distance $d$ between the charges. As we increase the distance from the charge distribution, the field looks more and more like that of a point charge.}
    \label{fig:electric_field}
\end{figure}

EFTs approximate multi-scale theories in a given regime. 
Before moving to QFT, we consider a simpler example from classical field theory: the electric field of two point charges. In~\cref{fig:electric_field}, we show the electric field of two point charges $q_1$ and $q_2$ at two different length scales: at a length scale of the distance $d$ between the two point charges as well as at a length scale significantly larger than the distance between the charges. As one can see, at scales close to $d$, the features of having two separate point charges are clearly visible and need to be taken into account in a valid description of the near field. However, at large distances the far field could be described by a simpler theory: the electric field of one point charge $q_1 + q_2$ in the center. A multipole expansion of a charge distribution is a visual example of an EFT in a context other than QFT. 
%-----------------------------
%-----------------------------
This is a common approach to solving complicated problems in not just physics, but science as a whole. Effective descriptions simplify an analysis by only including the information that is relevant at a scale characteristic to the phenomenon we are studying. For instance, we do not consider electroweak~(EW) bosons explicitly in nuclear physics and we do not need to consider the SM in chemistry.

Now let us move to effective quantum field theories, where the effective theory takes three ingredients: (i) the \textbf{low-energy degrees of freedom} (light fields or soft modes) relevant at the given scale, (ii) the \textbf{symmetries} that limit their possible interactions (see~\cref{sec:basis} for a discussion of how symmetries enter an EFT basis)
and (iii) an \textbf{expansion parameter}~$\lambda$, which is often written as a ratio of a given scale $E$ relative to some cutoff $\Lambda$,  $\lambda=E/\Lambda$. An EFT can be seen as a Taylor series in the parameter $\lambda$.

The parameter $\lambda$ lets us distinguish between \textbf{hard and soft modes}, i.e.\ momenta above and below the cutoff $\Lambda$, respectively. 
In many applications of EFTs, separating these modes is equivalent to 
separating \textbf{heavy and light fields}, i.e.\ fields with masses heavier or lighter than the cutoff $\Lambda$.
While in principle the latter is a statement about the number of fields in the Lagrangian and modes refer to the momentum regions of a single field, both of these terminologies refer to the same approximation. 
In tree-level diagrams with soft external momenta, the momenta of virtual particles are fixed by momentum conservation to be soft as well. Therefore, hard modes are represented in the matrix elements by heavy fields with masses $M \geq \Lambda$ only. 
Once we move beyond tree level, the effective theory is determined by the momenta and not the masses of the fields. 
Therefore, we will use the more general terminology of hard and soft modes in this chapter.

The division of a field $\phi$ into its modes above the validity cutoff (hard), $\phi_h$, or below (soft), $\phi_s$ can be schematically defined in momentum space as
\begin{align}
    \phi(x) = \int \frac{d^dk}{(2\pi)^d}e^{-ik\cdot x}\tilde\phi(k) \left[  \theta_\Lambda(k) + \bar\theta_\Lambda (k)  \right] \equiv \phi_s(x) + \phi_h(x)\,.
\end{align}
Here $\theta_\Lambda(k)$ and $\bar\theta_\Lambda(k) = 1 - \theta_\Lambda(k)$ represent the distinction between soft and hard regions, and their precise form depends on the type of EFT we are trying to construct. \footnote{In order to separate modes by their virtuality, one could take $\theta_\Lambda(k) = \theta(\Lambda^2 - k^2)$ with the Heaviside function $\theta$, but one typically avoids hard cutoffs because they lead to problems with gauge invariance. In a modern context, the separation is achieved differently and for now the $\theta_\Lambda$ are understood to be abstract.} 
At energies below the cutoff $\Lambda$ only the soft modes $\phi_s$ are the dynamical degrees of freedom. 
Therefore, all physics at energies below $\Lambda$ is described by $n$-point correlators of the soft modes which, as for any QFT, we can write in terms of the generating functional $Z$ as
\begin{align}
    \left\langle 0|T\left\{  \phi_s(x_1), \ldots,\phi_s(x_n) \right\}|0\right\rangle
    =
    \frac{1}{Z[0]} \prod_{i=1}^n \left. \left(-\frac{i\delta}{\delta J_s(x_i)}\right) Z[J_s] \right|_{J_s=0}\,,
\end{align}
where the generating functional includes only sources for the soft modes $J_s$
\begin{align}
    Z[J_s] = \int \mathcal D\phi_s \mathcal D\phi_h\,\exp\left\{ iS(\phi_s,\phi_h) + i \int d^4x J_s(x)\phi_s(x) \right\}
\end{align}
and $S(\phi_s,\phi_h)$ is the action of the theory, which can be written in terms of the Lagrangian $\mathcal{L}$, as
\begin{align}
    S(\phi_s,\phi_h) = \int d^4x \, \mathcal L(\phi_s,\phi_h,\partial_\mu \phi_s, \partial_\mu\phi_h)\,.
\end{align}
Carrying out the path integral over the hard modes,
\begin{align}
    \int \mathcal D\phi_h\,\exp \left\{ i S(\phi_s,\phi_h) \right\} \equiv \exp\left\{ iS_\Lambda(\phi_s)  \right\}\,, \label{eq:path_int_over_hard_modes}
\end{align}
defines the Wilsonian effective action $S_\Lambda$, which depends only on soft modes - the hard modes are said to have been \textit{integrated out}. A priori, this object  contains an  infinite number of non-local interaction operators~$O$ between the soft modes (and their derivatives),
\begin{align}
    O \sim \phi_s(x) \ldots \phi_s(x + \Delta x_n)\,.
\end{align}
Restricting ourselves to effective theories in which the momenta of $\phi_s$ are isotropically small, $\Delta x_i$ are distances shorter than those associated with the chosen cutoff,
\begin{align}
    \Delta x_i^\mu \lesssim \frac{1}{\Lambda}\,.
\end{align}
The wavelengths of the soft modes on the other hand are larger than the $\Delta x_i$ by construction. As such, they cannot resolve the non-localities of the scale $\Delta x_i$, enabling an expansion around $\Delta x_i = 0$. In this way, we define the power-counting of the effective theory, with the expansion parameter being the ratio of the low scattering energies and the cutoff, schematically $\lambda \sim E/\Lambda$. Employing this power-counting turns $S_\Lambda$ into a series of local operator products, which can be written as 
\begin{align}
    S_\Lambda = \int d^4x\sum _i C_i O_i(x) \equiv \int d^4x\, \mathcal L_\mathrm{eff}\,,
    \label{eq:eff_Lag}
\end{align}
where $O_i$ are the local operator products of soft modes and $C_i$ are their associated coupling constants, the so-called Wilson coefficients. 
While \cref{eq:eff_Lag} may not look impressive, it encapsulates the core principles of EFTs: It tells us, that below a certain energy threshold (here denoted as $\Lambda$), we are able to formulate an approximate theory with a Lagrangian $\mathcal L_\mathrm{eff}$, built of operators that contain only the soft modes. The effects of hard modes on low-energy processes are then encoded in the Wilson coefficients, which multiply the operators. This \emph{factorization} is an important property of EFTs, which can be exploited to resum problematic terms in the perturbative expansion, as we will explore in 
\cref{sec:resummation}.

The construction of effective field theories closely follows the logic of Appelquist-Carazzone decoupling theorem~\cite{Appelquist:1974tg}. In the presence of a clear hierarchy of the scales between heavy modes and soft modes, the effects of heavy modes on the low-energy processes can be absorbed into the parameters related with the soft modes as well as the coupling coefficients of the local effective operators, $C_i$. However, there are caveats to the decoupling theorem which can lead to the non-decoupling effects\footnote{Such effects can arise, for instance, when the heavy modes acquire the masses through the symmetry-breaking mechanism of the low energy theory~\cite{Brivio:2017vri}.}.

While the number $i$ of effective operators in~\cref{eq:eff_Lag} is formally infinite, the operator product expansion can be truncated by power-counting. Operators are products of fields and derivatives, both of which carry mass dimensions, that can only be saturated by dimensionful quantities at scales below $\Lambda$. For derivatives, this is trivial as they correspond directly to the momenta of the fields they act on. For field operators, the scaling can be deduced from the fact that they enter scattering amplitudes through their two-point functions. For the example of a scalar we can then see:
\begin{align}
    \phi_s^2 \sim \left\langle 0|T\left\{ \phi_s(x) , \phi_s(y) \right\} \right\rangle =\int \frac{d^4k}{(2\pi)^4}e^{-ik\cdot(x-y)}\frac{i}{k^2-m^2}\quad
    \Rightarrow \phi \sim k,m\,,
\end{align}
meaning that the field scales like its mass dimension. The argument proceeds analogously for fields of higher spin.

From the above scaling argument, an operator of mass dimension $j$ scales with $j$ powers of the small momenta (or masses). The action being given by $\int d^4x \, \mathcal L_\mathrm{eff}$ means that the couplings of operators with mass dimensions higher than four have inverse dimension of energy. By virtue of the expansion in $\Delta x_i$, they have to scale as $\Lambda^{j-4}$, meaning that the corresponding terms in the Lagrangian are formally power-suppressed by $\lambda^{j-4}$ with respect to the dimension-four terms. 

Having established that higher number of fields and derivatives correspond to operators being more strongly suppressed, it makes sense to group the operator product expansion by mass dimension:
\begin{align}
    \mathcal L_\mathrm{eff} = \sum_{j}^\infty \frac{1}{\Lambda^{j-4}}\sum_{i=1}^{n_j}C_{i}^{(j)}O_{i}^{(j)}\,,
\end{align}
where $n_j$ is the number of operators at mass-dimension $j$. In writing the above expression, the Wilson coefficients are dimensionless\footnote{Note that operators of mass dimensions $j < 4$ are generically \emph{superleading} in this notation, but their coupling coefficients can carry suppressions to compensate.}. The sum over the mass dimensions $j$ is formally infinite, but in practice is cut off by specifying an order in power-counting to determine a precision goal, leaving us with a finite number of operators. Note also that $n_j$ is finite for each value of $j$.

EFTs can be built in two ways: either \textbf{top down} or \textbf{bottom up}. 
In the top-down approach, the full theory is known and we use the EFT to calculate its limit in a certain regime. For instance, we might want to calculate the low-energy limit of a theory. 

The \textbf{bottom-up} approach is often used for BSM searches. 
We can construct an EFT as a universal description of the low-energy effects of a yet unknown theory containing heavy new particles. The EFT approach allows us to construct all interactions possible at a given order in our expansion parameter $\lambda$. 
Comparing the EFT to data, we can then get an estimate of the scale at which we expect NP.

\begin{figure}
    \centering
    \centering
% colors used in tables and graphs
\definecolor{AccentColor}{rgb}{0.2,0.4,.5}
\newcommand{\shadecolor}[1]{AccentColor!#1!white}

\def\zl{-.75}
 % placement for the ticks and labels
\def\ylo{0}
\def\yd{3.5}
\def\yc{6.2}
\def\yb{9.2}
\def\ya{12.5}
\def\y0{13.2}
\begin{tikzpicture}[x={\linewidth/16},y=1cm]
 % axis and boxes
 \fill[\shadecolor{10}](\ya,0)rectangle(\yb,1);  % SMEFT
 \fill[\shadecolor{15}](\yb,0)rectangle(\yc,1);  % LEFT
 \fill[\shadecolor{20}](\yc,0)rectangle(\yd,1);  % HQET
 \fill[\shadecolor{25}](\yd,0)rectangle(\ylo,1); % XPT
 \draw[thick,-Latex]  (\ylo,0) -- (\y0,0);
 % ticks
 \node at (\y0,-.4){$E$};

 \draw[thick] (\ya,-.1)--(\ya,.1); % TeV
 \draw[thick, dotted] (\ya, 0)--(\ya,1); 
 \node at (\ya, -.4){TeV};
 
 \draw[thick] (\yb,-.1)--(\yb,.1); % EW 
 \draw[thick, dotted] (\yb, 0)--(\yb,1);
 \node at (\yb, -.4){$\mu_\mathrm{EW}$};

 \draw[thick] (\yc,-.1)--(\yc,.1); % B-mass
 \draw[thick, dotted] (\yc, 0)--(\yc,1);
 \node at (\yc, -.4){$m_b$};

 \draw[thick] (\yd,-.1)--(\yd,.1); % QCD scale
 \draw[thick, dotted] (\yd, 0)--(\yd,1); 
 \node at (\yd, -.35){$\Lambda_\mathrm{QCD}$};
 
 % labels in boxes
 \node at (\ya*0.5+\yb*0.5,.5){{SMEFT/HEFT}};
 \node at (\ya*0.5+\yb*0.5,\zl){TeV particles};
 \node at (\yc*0.5+\yb*0.5,.5){{WET/LEFT}};
 \node at (\yc*0.5+\yb*0.5,\zl){$W, Z, h, t$};
 \node at (\yc*0.5+\yd*0.5,.5){{HQET}}; 
 \node at (\yc*0.5+\yd*0.5,\zl){{$b$-quarks}}; 
 \node at (\yd*0.5+\ylo*0.5,.5){{$\chi$PT}};
 \node at (\yd*0.5+\ylo*0.5,\zl){{quarks, gluons}};
 
 \node at (-1.1,\zl){Integrated out:};
 \end{tikzpicture} 
    \caption{Summary of particle physics EFTs relevant at different energy scales. } 
    \label{fig:EFT_flowchart}
\end{figure}
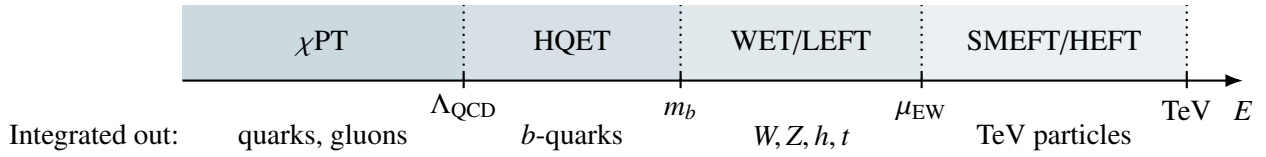

We will now list some EFTs which are widely used in particle physics. We schematically show at which energy scales these theories are relevant in~\cref{fig:EFT_flowchart}.
\begin{itemize}
    \item Standard Model Effective Field Theory (\textbf{SMEFT})~\cite{Buchmuller:1985jz,Brivio:2017vri}: Approximates the effects of potential NP at the TeV scale above/around the EW scale. 
    SMEFT assumes a doublet structure for the SM Higgs boson and hence a linear realization of electroweak symmetry breaking~(EWSB). 
    SMEFT is extensively used for model-agnostic NP searches at the LHC. We will discuss this EFT in~\cref{sec:SMEFT}.
    \item Higgs Effective Field Theory (\textbf{HEFT})~\cite{Feruglio:1992wf,Grinstein:2007iv,Alonso:2023upf}: Just like SMEFT, HEFT approximates the effects of potential NP at the TeV scale above/around the EW scale. Compared to SMEFT, HEFT is more general as it incorporates also non-linear realizations of EWSB.
    \item Weak effective theory (\textbf{WET}) or Low-energy effective field theory (\textbf{LEFT})~\cite{Jenkins:2017jig,Jenkins:2017dyc,Buchalla:1995vs}: Approximates the SM below the EW scale. The fields of the $W$, $Z$ and Higgs bosons as well as the top quark are integrated out. 
    LEFT and WET describe the same physics and differ by conventions. LEFT is the name of the complete, non-redundant basis, while WET is the historical, process-organized version of the same EFT.
    \item Heavy-Quark Effective Theory (\textbf{HQET})~\cite{Isgur:1989vq,Georgi:1990um,Neubert:1993mb}: This theory is most-commonly used to describe the decay of heavy mesons, like the $B$-meson, where the scale hierarchy is given between  the large mass of the decaying particle and the light masses of the decay products. In its construction, the heavy $b$ quark cannot fully be integrated out since it appears in the external state, and instead only the small quantum fluctuations around its mass shell are kept.
    \item Chiral perturbation theory (\textbf{$\mathbf{\chi}$PT})~\cite{Weinberg:1978kz,Gasser:1983yg,Gasser:1984gg}: Approximates the SM below the confinement of the strong interactions. Note that the low-energy degrees of freedom (mesons) are different from the high-energy ones (quarks and gluons).
    \item Soft collinear Effective Theory (\textbf{SCET})~\cite{Bauer:2000ew,Bauer:2000yr,Bauer:2001ct,Becher:2014oda}: Approximates the behavior of light particles at high energies, separating the large scattering energies from the small masses of the scattered particles. It is typically used to resum large double logarithms occurring in QCD corrections to collider physics processes. Similarly to HQET, it is constructed by not integrating out full fields from a UV theory, but only certain Fourier modes.
\end{itemize}

%------------------------------------
\section{Matching}
\label{sec:matching}
%------------------------------------
After deciding on an upper bound for the mass dimension of operators we want to consider, the effective Lagrangian contains a finite set of Wilson coefficients $C_i^{(k)}$. 
If the full theory is known and we take the top-down approach, the Wilson coefficients can be fixed by comparing the full-theory predictions to those found from the effective theory, and tweaking the couplings of the latter such that both theories agree at low energies. This procedure is called \emph{matching}. 

There are multiple ways of performing the matching. The most well-known one is to diagrammatically compute $S$-matrix elements for low-energy processes from both theories and equate them to determine the effective couplings. This is the approach we will outline in detail here. However, it should be noted, that one can also directly evaluate the path integral in~\cref{eq:path_int_over_hard_modes}, which is often referred to as functional matching. Its details are beyond the scope of this manuscript and we refer the reader to~\cite{Gaillard:1985uh,Henning:2014wua,Drozd:2015rsp,Henning:2016lyp,Fuentes-Martin:2016uol,Zhang:2016pja,Kramer:2019fwz,Dittmaier:2021fls,Cohen:2020fcu}.

Consider the full and effective theories, represented by their Lagrangians $\mathcal L_\mathrm{UV}$ and $\mathcal L_\mathrm{eff}$, respectively. The two Lagrangians are structurally different for two reasons. First, the full theory contains both hard and soft modes $\mathcal L_\mathrm{UV} = \mathcal L_\mathrm{UV}(\phi_s,\phi_h)$, while the effective Lagrangian only contains the latter, $\mathcal L_\mathrm{eff} = \mathcal L_\mathrm{eff}(\phi_s)$. 
Second, the two Lagrangians contain different interactions between the soft modes. By construction, the two theories are expected to disagree in their predictions for processes at high energies (above the cutoff $\Lambda$) but agree at low energies in an expansion in $E/\Lambda$. 
In order to determine the coupling constants of the effective theory, we can therefore compare predictions for low-energy matrix elements, 
\begin{align}\label{eq:basic_matching}
    \big\langle 0 \big| \mathcal L_\mathrm{UV} \big| \phi_s^1\ldots \phi_s^n\big\rangle
    =
    \big\langle 0 \big| \mathcal L_\mathrm{eff} \big| \phi_s^1\ldots \phi_s^n\big\rangle \,.
\end{align}
For such an equation to exist, it is clear that the matrix elements can be only those with soft modes $\phi_s$ in the external states (as the hard modes do not exist in the effective theory). 
By the fact that the two theories agree at low energies, the matrix-elements on the left-hand side need to involve contractions of the hard modes $\phi_h$. Matrix elements involving \emph{only} soft modes are identical by construction and lead to trivially fulfilled equations. 
It is therefore sufficient to only evaluate the matrix elements involving contractions of hard modes for the matching procedure. This procedure takes a different shape at tree- and loop-level and it is instructive to discuss them separately. Before doing so, it is important to recall that the hard modes $\phi_h$ are not restricted to modes with masses $M  \gtrsim \Lambda$, but include light or massless fields of large \emph{virtualities} $k^2\gtrsim\Lambda^2$, as well. In the case of matching a BSM theory to an EFT, the hard modes are not just the heavy new fields, but also the large-virtuality modes of the SM particles.

\subsection{Tree-Level Matching}
We first discuss the matching of the EFT Wilson coefficients at tree level. 
Because the external momenta can only be soft, the momenta of virtual particles in tree-level diagrams are fixed to be soft by momentum conservation. 
In this case, the hard modes in the matrix elements are represented exclusively by heavy fields, with masses $M \gtrsim\Lambda$. 
Note however, that not \emph{all} virtual particles in the graphs have to be the heavy fields, unless the external states are kept off-shell, see \cref{sec:greensandOS} for details.

As a classic example, let us consider the SM after EWSB. 
The hierarchy of scales between the masses of the Higgs, $W$ and $Z$ bosons as well as the top quark, which are collectively of the order of what we call the EW scale, $\mathcal{O}(100~\mathrm{GeV})$, and all other particle masses, justifies the use of an EFT description for the scattering of those lighter particles. The appropriate EFT is the LEFT, where the EW-scale particles have been integrated out.

In the following section we will perform the matching of the (LEFT) Wilson coefficients for the example of the weak decay $b\to c\bar u d$, which is the underlying partonic transition governing the hadronic $B\to D$ decay. The example is particularly instructive because its flavor structure allows for only a single diagram at tree-level. In the SM, it is governed by the exchange of a virtual $W$ boson. 
Since the $W$ boson is much heavier than the external quarks, $p_q \ll m_W$, it can be integrated out. Our task is now to find the low-energy Lagrangian and its effective couplings. First, we anticipate it to be of the form
\begin{align}\label{eq:firstEffLag}
    \mathcal L_\mathrm{eff} = -\frac{c_A}{4}F_{\mu\nu}F^{\mu\nu} - \frac{c_G}{4}G^a_{\mu\nu}G^{a\mu\nu} + \sum_{f \neq t} \bar f (c_f i\slashed D - c_{m_f} m_f) f + \sum_{k=5}^\infty \frac{1}{m_W^{k-4}}\sum_ {i=1}^{n_k}C_{i}^{(k)}O_{i}^{(k)}\,,
\end{align}
where the sum in the third term runs over all fermions in the theory except the top quark. Here and throughout the rest of the chapter, we use the positive sign convention for the covariant derivatives, 
\begin{align}
D_\mu \psi = \partial_\mu\psi + i \sum_i g_i T_i^{(\psi)} A^i_\mu \psi\,,
\end{align}
where the sum runs over all the gauge groups $G_i$, with $g_i$ and $A_\mu^i$ being the corresponding gauge couplings and fields, and $T_i^{(\psi)}$ is the generator of the representation under which the field $\psi$ transforms in $G_i$.
In this expression we have accounted for the possibility that the kinetic and mass terms of the light degrees of freedom receive matching corrections. Additionally, we have identified the hard scale with the mass of the $W$ boson, $\Lambda = m_W$. Restricting ourselves to tree-level for now, we see that the only diagrams that can give rise to matching equations of the type in~\cref{eq:basic_matching} are those involving the weak coupling $g_L$: This is because the relevant matrix elements need to have the light particles as external states and involve the heavy $W$ as a virtual state. The only term in the Lagrangian coupling these two is the weak gauge interaction, and we need to insert it at least twice for the heavy $W$ to not appear as external field. At the lowest order in $g_L$ coupling, this corresponds to the following amplitude:
\begin{align} \label{eq:amp_4fermi_treelevel}
   i\mathcal A^\mathrm{(tree)}_\mathrm{UV}(b \to c \bar u d) =\qquad
   \raisebox{\shifta ex}{\includegraphics[scale=0.30]{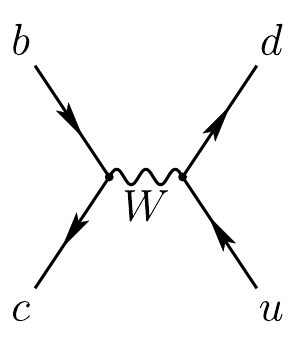}}\qquad
   =
      \bar u_c \frac{ig_L}{\sqrt{2}}\gamma_\mu P_L u_b \frac{i}{q^2-m_W^2} \bar u_d \frac{ig_L}{\sqrt{2}}\gamma^\mu P_L v_u \,,
\end{align}
where $q = p_b - p_c = p_u+p_d$ is the momentum transfer. For the on-shell decay of a $b$-quark, $q^2 \ll m_W^2$. For the EFT to reproduce this amplitude in the low-energy limit, it needs to contain an interaction between four fermions, $O \sim (\bar c  \Gamma b)(\bar d \Gamma u)$, where $\Gamma$ denotes any insertions of Dirac structures and derivatives. From the amplitude it is easy to see that a single operator suffices, so that the dimension-six part of $\mathcal L _\mathrm{eff}$ becomes
\begin{align}
    \mathcal L_\mathrm{eff}^{k=6} = \frac{1}{m_W^2}C_1 (\bar c\gamma_\mu P_L b)(\bar d \gamma^\mu P_L u) \ + \ \ldots \,,
\end{align}
where we have dropped the superscript from the Wilson coefficient as all considered operators and corresponding Wilson coefficients in this section are of mass dimension six.
The dots indicate the fact that the Lagrangian in principle contains many more operators at dimension six, but only the interaction term shown is required for this example. For the matching procedure we now equate the EFT matrix element to that of the full theory, expanding in $E/M$. The matching equation becomes:
\begin{align} \label{eq:tree_match_result}
    \frac{-ig_L^2}{2m_W^2}~(\bar u_c \gamma_\mu P_L u_b)(\bar u_d \gamma^\mu P_L v_u) \ + \  \mathcal{O}(q^4/m_W^{4})
    &=
    \frac{iC_{1}}{m_W^2}~(\bar u_c \gamma_\mu P_L u_b)(\bar u_d \gamma^\mu P_L v_u) \,, &
    \Rightarrow \quad  C_{1} &= -\frac{g_L^2}{2}\,.
\end{align}
Note that there are no tree-level amplitudes that generate non-trivial contributions to the other coefficients in the Lagrangian in~\cref{eq:firstEffLag}, so they are given by
\begin{align}
    c_A = c_f = c_{m_f} = 1\,.
\end{align}
This concludes the tree-level matching of the relevant operators.

\subsection{Loop-Level Matching}
At the loop level, it is less obvious how the amplitudes of the full theory map onto the matching coefficients of the effective theory and its matrix elements. While external momenta are small, loop graphs have momenta unfixed by momentum conservation. The integrals over these additional momenta sweep over regions in which they are of the order of the scattering energy, of the order of the heavy mass, or anything in between and beyond. 
Recalling that the effective theory encapsulates only the short-distance dynamics in matching coefficients, we see that a loop amplitude of the full theory splits into two pieces, depending on the scaling of the loop momentum: a contribution that corresponds to a loop correction to the matching coefficients ($l^2 \sim \Lambda^2$) and a loop-level matrix element of the effective theory ($l^2 \ll \Lambda^2$).

\begin{figure}
    \centering
    \begin{subfigure}[t]{0.65\textwidth}
    \centering
      \includegraphics[scale=0.35]{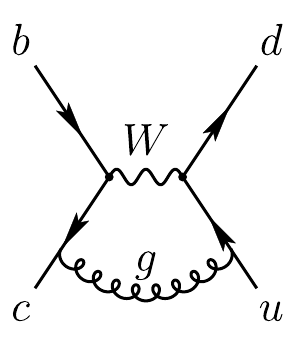}\quad 
    \includegraphics[scale=0.35]{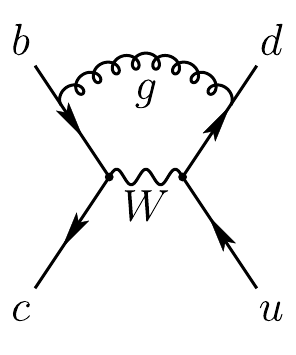}\quad 
    \includegraphics[scale=0.35]{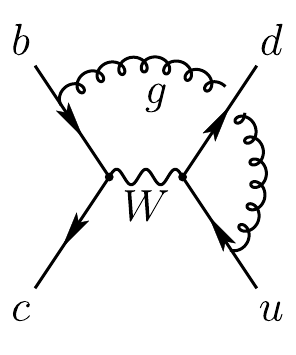}\quad 
    \includegraphics[scale=0.35]{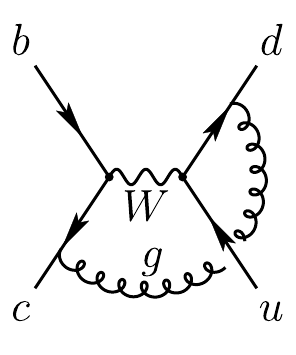}\quad 
      \caption{Box diagrams.}
      \label{fig:oneloopbcud_boxes}
    \end{subfigure}%
\begin{subfigure}[t]{0.32\textwidth}
    \centering
    \includegraphics[scale=0.35]{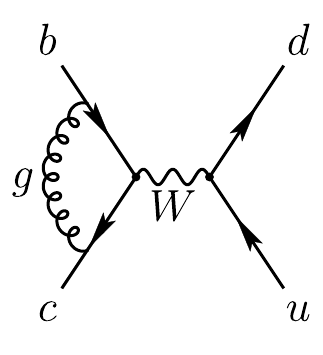}\quad
    \includegraphics[scale=0.35]{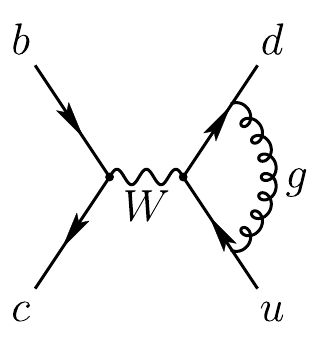}
      \caption{Vertex corrections.}
      \label{fig:oneloopbcud_vertex}
    \end{subfigure}%
    \caption{One-loop QCD diagrams contributing to the $b\to c\bar u d$ transition in the SM. Loop corrections to the quark wave functions are not shown. }
    \label{fig:oneloopbcud_full}
\end{figure}
To understand this systematically, let us investigate the example from above, but now at the next order in the strong coupling $g_s$. The diagrams contributing to the transition at one-loop order in the full EW SM are shown in \cref{fig:oneloopbcud_full}. Consider the first box-type diagram. In order to simplify the following argument, we are going to implement some assumptions on the kinematics. Notice that in QCD, quarks do not exist on-shell but rather as part of hadronic bound states. To model this, we will here set all external quark momenta to zero, but introduce a small off-shellness to each quark propagator in the form of a fictitious mass parameter $m_q$. The amplitude is then
\begin{align} \label{eq:boxgraph}
   \raisebox{\shiftb ex}{\includegraphics[scale=0.25]{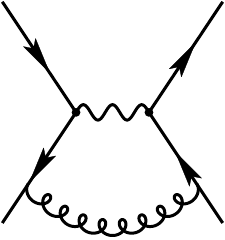}} = \big(\bar u_c \gamma_\mu \gamma_\nu \gamma_\rho P_L u_b\big)\big(\bar u_d \gamma^\rho \gamma^\nu \gamma^\mu P_L v_u\big)\frac{g_L^2 g_s^2}{2d}T^a_{ij}T^a_{mn}\int\frac{d^dl}{(2\pi)^d} \frac{1}{l^2-m_W^2}\left(\frac{1}{l^2-m_q^2}\right)^2\,,
\end{align}
where the color indices $i,j,m,n$ correspond to the quarks $c$, $b$, $d$ and $u$, respectively. The loop integral is straightforward to evaluate, yielding
\begin{align}\label{eq:fullboxintegral}
     \int\frac{d^dl}{(2\pi)^d} \frac{1}{l^2-m_W^2}\left(\frac{1}{l^2-m_q^2}\right)^2 
     = \frac{i}{16\pi^2 m_W^2}\left( \frac{1}{1-r} + \frac{\log r}{(1-r)^2} \right)
     = \frac{i}{16\pi^2 m_W^2}\left( 1 + \log r \right) + \mathcal{O}(r) \,,
\end{align}
where we have introduced the scale ratio $r=m_q^2/m_W^2$, in which we have expanded in the second step. Importantly, this expression contains a logarithm of the ratio $r$. Since $r\ll 1$ by assumption, this is a \textit{large logarithm} appearing at one-loop order in perturbation theory. We will learn in the course of this chapter how EFTs can elegantly address this issue.

In the previous section we have argued that the effective theory is one in which short-distance effects are encapsulated in contact interactions. In momentum space, this means that propagators which scale as inverse powers of the hard scale are shrunk into points. For the loop integral in our example above, this means that in regions where the loop momentum scales like the hard scale $l^2\sim m_W^2$, every propagator is of $\mathcal{O}(1/m_W^2)$. In this region, the whole graph is shrunk into a point. If $l^2\sim m_q^2$ on the other hand, only the $W$-boson propagator is hard. Diagrammatically, we can write
\begin{align}
    \raisebox{\shiftb ex}{\includegraphics[scale=0.25]{match_ll_uv3_nolabel.pdf}} \quad \to \quad 
    \raisebox{\shiftb ex}{\includegraphics[scale=0.25]{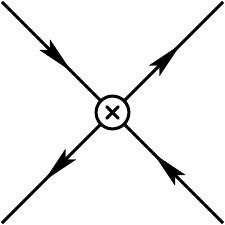}} \ + \ 
    \raisebox{\shiftb ex}{\includegraphics[scale=0.25]{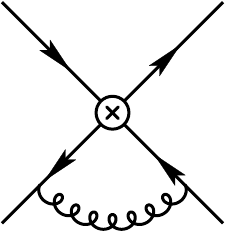}}\,,
\end{align}
where the first graph on the right-hand side is understood as only including the one-loop part in the effective coupling denoted by the crossed circle. The second graph on the right-hand side involves only the tree-level piece of the effective coupling, which we have determined already in \cref{eq:tree_match_result}. To now determine the matching coefficients at one-loop order, we can simply compute the graphs on the right-hand side and equate them to all diagrams appearing on the left-hand side. However, the loop graph in the EFT contains no new information, so a more economical way of performing the matching would be to directly calculate only that part of the loop amplitude corresponding to the first graph on the right-hand side.

The tool that achieves exactly that is called the \emph{method of regions}~\cite{Beneke:1997zp,Smirnov:2002pj}. What we want to achieve is to compute the loop integral in \cref{eq:fullboxintegral} for the region where $l^2\sim m_W^2$. To do so, we \emph{assume} this scaling and expand around the scaling $l^2\sim m_W^2 \gg m_q^2$. At leading order in this expansion we then find
\begin{align} 
    \left.\int\frac{d^dl}{(2\pi)^d} \frac{1}{l^2-m_W^2}\left(\frac{1}{l^2-m_q^2}\right)^2 \right|_\mathrm{hard}
    =\int\frac{d^dl}{(2\pi)^d} \frac{1}{l^2-m_W^2}\left(\frac{1}{l^2}\right)^2 + \ldots \, ,
\end{align}
where the dots correspond to terms with integrands to higher order in $m_q^2/m_W^2$, $m_q^2/l^2$. We employ dimensional regularization using $d=4-2\epsilon$. Due to the loop momentum scaling like the hard scale, we call this the \emph{hard region} of the loop integral. Evaluating the hard region integral, we obtain
\begin{align}
    \int\frac{d^dl}{(2\pi)^d} \frac{1}{l^2-m_W^2}\left(\frac{1}{l^2}\right)^2
    = \frac{i}{16\pi^2m_W^2}\left(\frac{1}{\epsilon} + \log \frac{\mu^2}{m_W^2} + 1 \right) \,.
    \label{eq:MOR_hard}
\end{align}
The two things we should take notice of, are that firstly this integral no longer contains the extreme scale ratio $r = m_q^2/m_W^2$, since the expansion has removed $m_q^2$ from the propagator denominators. Secondly, this integral is divergent. To reproduce the full one-loop integral in~\cref{eq:fullboxintegral}, we have to also evaluate the \emph{soft region}, in which the scaling is $m_W^2 \gg l^2\sim m_q^2$:
\begin{align}
        \left.\int\frac{d^dl}{(2\pi)^d} \frac{1}{l^2-m_W^2}\left(\frac{1}{l^2-m_q^2}\right)^2 \right|_\mathrm{soft}
    =-\frac{1}{m_W^2}\int\frac{d^dl}{(2\pi)^d} \left(\frac{1}{l^2-m_q^2}\right)^2 + \ldots
    =-\frac{i}{16\pi^2m_W^2}\left(\frac{1}{\epsilon} - \log \frac{m_q^2}{\mu^2} \right) \,.
    \label{eq:MOR_soft}
\end{align}
Again, this result depends only on the scale $m_q^2$ and does not involve the ratio $r$, and is divergent as well. One checks quickly that the hard and soft regions sum to the full expression in~\cref{eq:fullboxintegral} after expanding in $q^2/m_W^2$. In particular, the divergences cancel and the logarithms combine to the $\log r$ from the full expression. The hard region is exactly what matches into the one-loop matching coefficients of the effective theory then, whereas the soft regions just correspond to the one-loop matrix elements of the EFT.

Before working out the full one-loop matching for our example, we should address the question whether there might be contributions from other regions, meaning in which $l^2\gg m_W^2$ or $l^2 \ll m_q^2$. In both cases, we arrive at completely scaleless integrals, which in dimensional regularization can be set to zero.

In order to work out the full one-loop matching we therefore now evaluate all diagrams in \cref{fig:oneloopbcud_full} in the hard region. From the outset, we can see that the graphs in \cref{fig:oneloopbcud_vertex} are scaleless in the hard region, since none of the propagators contain a hard scale. This means they cannot contribute to the matching, and we only need to evaluate the four box-type diagrams in~\cref{fig:oneloopbcud_boxes}.

For this calculation, we need to take care of one additional subtlety: The Dirac structure in the hard region expression in~\cref{eq:boxgraph} does not match that of our operator $O_1 = (\bar c\gamma_\mu P_L b)(\bar d \gamma^\mu P_L u)$, but it can be related to it, using the completeness of the Dirac basis. However, this completeness is incompatible with dimensional regularization, and so whatever identity we employ is incorrect by a term of $\mathcal{O}(\epsilon)$. For this term, one has to make a scheme choice, which needs to be kept consistently throughout the calculation. For the structures appearing in the box graphs, we have:
\begin{align}\label{eq:reductions}
    \begin{aligned}
        \Big[\gamma_\mu \gamma_\nu \gamma_\rho P_L \otimes \gamma^\mu \gamma^\nu \gamma^\rho P_L\Big]
            &\to (16+\kappa\epsilon)\, \Big[\gamma_\mu P_L \otimes \gamma^\mu P_L \Big] \,, \\
        \Big[\gamma_\mu \gamma_\nu \gamma_\rho P_L \otimes \gamma^\rho \gamma^\nu \gamma^\mu P_L \Big]
            &\to (4-[12+\kappa]\epsilon)\, \Big[\gamma_\mu P_L \otimes \gamma^\mu P_L \Big]\,,
    \end{aligned}
\end{align}
where the parameter $\kappa$ encodes the Dirac reduction scheme choice. 
Notice that the two reduction identities involve the same parameter since they can be rewritten into each other by anticommuting Dirac matrices, which is an operation that is compatible with dimensional regularization. The parameter $\kappa$ will drop out of any physical quantity computed from the EFT.

Armed with these reductions we just need to carefully evaluate the box diagrams, including the color algebra, which always involves the combination
\begin{align}
    T^a_{ij}T^a_{mn} = \frac{1}{2} \left(\delta_{in}\delta_{mj} - \frac{1}{N_c} \delta_{ij} \delta_{mn} \right)\,.
\end{align}
Note that only the second term corresponds to the operator $O_1$, in which the color indices of the $b$ and $c$ quark are contracted among each other, as are those of the $u$ and $d$ quarks. To capture the other term, a new operator is needed, in which the color index of the $b$ quark is contracted with that of the $d$ quark. Using Fierz identities~\cite{Nieves:2003in}, the new operator can be written such that the quarks with contracted color indices are in the same current, yielding:
\begin{align}
    O_2 = (\bar c^j\gamma_\mu P_L b^i)(\bar d^i \gamma^\mu P_L u^j) = (\bar c\gamma_\mu P_L u)(\bar d \gamma^\mu P_L b)\,.
\end{align}
We can now finally give the result for the hard one-loop amplitude, found from the four box diagrams in \cref{fig:oneloopbcud_full} and the tree-level graph, which yields:
\begin{align}\label{eq:boxeshard}
\begin{aligned}
        i\mathcal A_\mathrm{hard} = -\frac{ig_L^2}{2m_W^2}&\left\{ 
        (\bar u_c \gamma_\mu P_L u_b)(\bar u_d \gamma^\mu P_L v_u)
            \left[ 1 - \frac{\alpha_s}{4\pi}\left( \frac{1}{\epsilon}+\log\frac{\mu^2}{m_W^2}+\frac{\kappa}{6}+\frac{5}{2} \right) \right] \right. \\
        &\qquad +\left.
        (\bar u_c \gamma_\mu P_L v_u)(\bar u_d \gamma^\mu P_L u_b) 
            \left[\frac{3\alpha_s}{4\pi}\left( \frac{1}{\epsilon}+\log\frac{\mu^2}{m_W^2}+\frac{\kappa}{6}+\frac{5}{2} \right) \right]
        \right\} \,.
\end{aligned}
\end{align}
This result maps onto the matching coefficients of the effective theory directly once the divergences are absorbed into counterterms of the effective theory. In practice, one simply drops the $1/\epsilon$ poles. And so our effective Lagrangian at dimension six becomes
\begin{align}
    \mathcal L_\mathrm{eff}^{k=6} = \frac{1}{m_W^2}C_1 (\bar c\gamma_\mu P_L b)(\bar d \gamma^\mu P_L u) 
    +
    \frac{1}{m_W^2}C_2 (\bar c\gamma_\mu P_L u)(\bar d \gamma^\mu P_L b) 
    \ + \ \ldots \,,
\end{align}
with the one-loop matching coefficients
\begin{align} \label{eq:loop_match_result}
\begin{aligned}
    C_1 &= -\frac{g_L^2}{2} \left[ 1 - \frac{\alpha_s}{4\pi}\left( \log\frac{\mu^2}{m_W^2}+\frac{\kappa}{6}+\frac{5}{2} \right) \right] \,, \\
    C_2 &= -\frac{g_L^2}{2} \left[ \frac{3\alpha_s}{4\pi}\left(\log\frac{\mu^2}{m_W^2}+\frac{\kappa}{6}+\frac{5}{2} \right)\right]\,.
\end{aligned}
\end{align}
These coefficients encapsulate the short-distance dynamics, at energy scales around the mass of the heavy $W$ boson. This is reflected by the fact that the only (physical) scale appearing in them is $m_W$. The dependence on the low scale $m_q^2$ is carried by the matrix elements of the EFT.

%------------------------------------
\section{Resummation of Large Logarithms}
\label{sec:resummation}
%------------------------------------

In the previous section, we have seen an explicit example of a process containing a large logarithm appearing in a loop-level quantity. This is a relatively generic feature of any multiscale problem, in which the different scales are strongly separated. In our example, the two scales were given by the mass of the $W$ boson, $m_W\sim 80~\mathrm{GeV}$ and the off-shellness of the quarks bound in the hadrons, $m_q^2\sim 250~\mathrm{MeV}$. Since the large logarithmic correction arises from QCD, it is only suppressed by $\alpha_s$, rendering it potentially problematic: a quick back-of-the-envelope estimate shows us that, relative to tree-level, this correction is very roughly of the size
\begin{align}
    \frac{\alpha_s}{\pi}\log\frac{m_W^2}{\Lambda_\mathrm{QCD}^2} \sim 0.5\,,
\end{align}
leading to an extremely poor convergence of the perturbative expansion: At any order in the perturbative expansion, we would expect a term of the form $\alpha_s^n \log^n r$, adding a sizable correction comparable to the previous order.

Schematically, a quantity expanded in $\alpha_s$ will be of the form
\begin{align}\label{eq:M_fixed_order}
    \mathcal M \sim 1\ +\ \frac{\alpha_s}{\pi}\left( b_{1,1} \log r + c_1 \right)
    \ +\ \frac{\alpha_s^2}{\pi^2}\left( b_{2,2} \log^2 r + b_{2,1}\log r + c_2 \right) \ + \ \ldots
\end{align}
where the $b_i$ are the coefficients of the logarithmic terms and $c_i$ constants at any loop order. Each term in this sum is of higher order in perturbation theory. To account for the size of the large logarithms, it would be desirable to reorganize the expansion such that the logarithmic terms count as the leading term, such that
\begin{align}\label{eq:M_resummed}
    \mathcal M \sim \left( 1+ \frac{\alpha_s b_{1,1}}{\pi}\log r + \frac{\alpha_s^2 b_{2,2}}{\pi^2}\log^2 r + \ldots \right) 
    \ +\ \left( \frac{\alpha_s}{\pi} c_1 + \frac{\alpha_s^2 b_{2,1}}{\pi^2} \log r\right)
    \ +\ \frac{\alpha_s^2}{\pi^2} c_2  \ + \ \ldots \,,
\end{align}
implying a power-counting scheme in which $\alpha_s \sim \eta$, $\log r \sim \eta^{-1}$, such that $\alpha_s\log r = \mathcal{O}(1)$. Immediately one notices that this form is only achievable if we can predict the coefficients $b_{i,i}$ to all orders to even obtain the leading term in the expansion. Applying the renormalization group to the framework of effective theories allows us to do just that, deriving expressions in the form of~\cref{eq:M_fixed_order} and from that deriving the coefficients $b_{i,i}$ to reshuffle the series and \emph{resumming} the large logarithms.

In the following, we will show that in an EFT multi-scale objects factorize into coefficients depending only on the hard scale and operators depending only on the soft scale and that large logs can be absorbed into the renormalization group connecting the two objects. This allows us to evaluate the coefficients and operators at their respective natural scale. 

\subsection{Renormalization of the Effective Theory}

\begin{figure}
    \centering
    \includegraphics[scale=0.35]{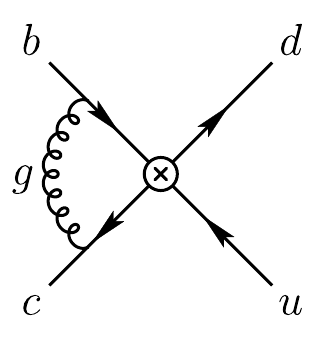} \quad 
    \includegraphics[scale=0.35]{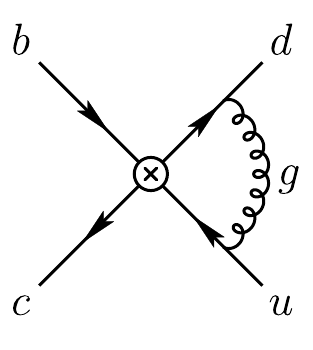} \quad 
    \includegraphics[scale=0.35]{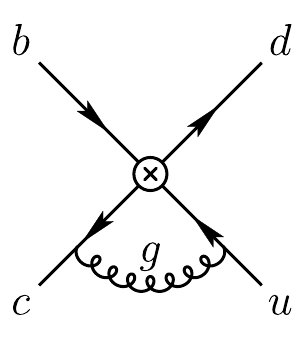} \quad 
    \includegraphics[scale=0.35]{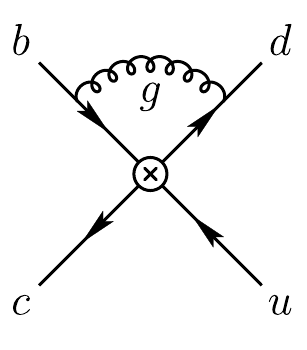} \quad 
    \includegraphics[scale=0.35]{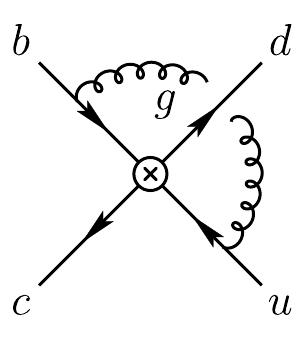} \quad 
    \includegraphics[scale=0.35]{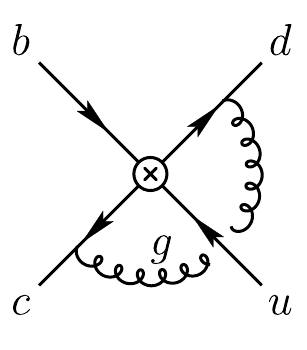}
    \caption{One-Loop diagrams in the effective theory contributing at $\mathcal{O}(\alpha_s)$ to the counterterms $Z_1$ and $Z_2$. Wave-function diagrams are not shown.}
    \label{fig:oneloopbcud_eft}
\end{figure}
Having performed the matching at one-loop order in the strong coupling constant in the previous section, we can now proceed to renormalize the effective theory. The relevant effective Lagrangian is given by
\begin{align}
    \mathcal L_\mathrm{eff} 
    =  -\frac{1}{4}F_{\mu\nu}F^{\mu\nu} - \frac{1}{4}G^a_{\mu\nu}G^{a\mu\nu} + \sum_f \bar f (i\slashed D - m_f) f
    + \frac{1}{m_W^2}C_1 (\bar c\gamma_\mu P_L b)(\bar d \gamma^\mu P_L u) 
    +
    \frac{1}{m_W^2}C_2 (\bar c\gamma_\mu P_L u)(\bar d \gamma^\mu P_L b)\,,
\end{align}
with the matching conditions in~\cref{eq:loop_match_result}. While we did not dwell on it, the matching coefficients $c_A$, $c_G$, $c_f$ and $c_{m_f}$ that we wrote in the Lagrangian in~\cref{eq:firstEffLag} stay trivial at order $\alpha_s$ since there are no one-loop graphs containing the $W$~boson and gluons. Indeed, even if they are non-trivial, they can be removed by redefining the fields, gauge couplings and masses of the theory, such that the Lagrangian can always be written in the form above. The next step is now to introduce renormalization constants. Writing only the relevant terms, we need
\begin{align}
    \mathcal L_\mathrm{eff}^\mathrm{ren} = \sum_q Z_q \bar qi \slashed D q + \frac{\sqrt{Z_bZ_cZ_uZ_d}}{m_W^2}
    \left[
        Z_1 C_1 (\bar c \gamma_\mu P_L b) (\bar d \gamma^\mu P_L u)
    +   Z_2 C_2 (\bar c \gamma_\mu P_L u) (\bar d \gamma^\mu P_L b)
    \right]\,,
\end{align}
with each renormalization constant of the form $Z_i = 1 + \delta_i$ at one-loop order. Computing the wave-function counterterms for the quarks is a straightforward exercise, yielding the well-known result
\begin{align}
   Z_q \equiv Z_b = Z_c = Z_u = Z_d = 1 - \frac{\alpha_s C_F}{4\pi\epsilon}\,.
\end{align}
The computation of $Z_1$ and $Z_2$ is slightly more involved. First, the Feynman rule for the one-loop counterterms is
\begin{align}
    \raisebox{\shiftc ex}{\includegraphics[scale=0.25]{match_tl_eft_nolabel.pdf}}
    = i(\delta_1 + 2\delta_q)~C_1\, (\bar u_c \gamma_\mu P_L u_b)(\bar u_d \gamma^\mu P_L v_u) 
   \ +\ i(\delta_2 + 2 \delta_q)~C_2\, (\bar u_c \gamma_\mu P_L v_u)(\bar u_d \gamma^\mu P_L u_b) \,.
\end{align}
To determine the values for $\delta_1$ and $\delta_2$ we now evaluate the one-loop diagrams shown in \cref{fig:oneloopbcud_eft}, add to them the counterterm and require the $1/\epsilon$ terms to cancel. In doing so, special care needs to be taken to ensure that the divergences found from the loop graphs are of UV nature. To do so, we introduce IR regulators to any massless or on-shell propagators in the graph and send them to zero after the loop integration is performed. The loop integrals we encounter in the calculation of the vertex diagrams are all of the form
\begin{align}
    \int \frac{d^dl}{(2\pi)^d} \frac{l^2}{l^2-\Lambda_\mathrm{IR}^2} \left(\frac{1}{l^2 - m_q^2} \right)^2 = \frac{i}{16\pi^2\epsilon} \ + \ \mathrm{(finite)}\,,
\end{align}
where the first term corresponds to the gluon propagator, regularized with a fictitious mass $\Lambda_\mathrm{IR}$ and the numerator follows from the quark propagators. We see that in this case, the regulator $\Lambda_\mathrm{IR}$ is not needed since the off-shellness $m_q^2$ sufficiently regularizes the integral in the IR, but for generic cases it is prudent to keep it.

Carefully evaluating all one-loop graphs in \cref{fig:oneloopbcud_eft}, performing the same reductions and color algebra as in \cref{sec:matching}, we obtain for the divergent part of the one-loop amplitude
\begin{align}
    i\mathcal A_\mathrm{eff}^\mathrm{div} = 
    i (\bar u_c \gamma_\mu P_L u_b)(\bar u_d \gamma^\mu P_L v_u) \,
        \frac{\alpha_s}{4\pi\epsilon}(C_1 - 4 C_2)
    \ + \
    i (\bar u_c \gamma_\mu P_L v_u)(\bar u_d \gamma^\mu P_L u_b) \,
        \frac{\alpha_s}{4\pi} (3C_1 - 4 C_2)\,.
\end{align}
Adding to this the counterterm amplitude, inserting $\delta_q$ and solving for $\delta_1$ and $\delta_2$ we obtain for the counterterms:
\begin{align}
    Z_1 &= 1 - \frac{\alpha_s}{4\pi\epsilon} \left( 1 - 3 \frac{C_2}{C_1} \right) \,,&
    Z_2 &= 1 - \frac{\alpha_s}{4\pi\epsilon} \left( 1 - 3 \frac{C_1}{C_2} \right) \,.
\end{align}
A feature quite generic to renormalization problems in EFT is the fact that the renormalization is not just multiplicative: The counterterm of $C_1$ depends on $C_2$ and vice versa. When solving the renormalization group~(RG) equations, this means that $C_1$ will be generated by the running even if it vanishes at the boundary condition, as long as $C_2$ is non-zero - we say it is \emph{radiatively generated}.

We can now derive the evolution equations for the coefficients $C_i$. They are found by imposing the bare coefficients to be independent of the renormalization scale $\mu$,
\begin{align}
    \mu \frac{d}{d\mu}C_i^\mathrm{bare} = \mu \frac{d}{d\mu} Z_i C_i \stackrel{!}{=} 0 
    \qquad  \Rightarrow  \qquad 
    \mu \frac{d}{d\mu} C_i = - \frac{C_i}{Z_i} \mu \frac{d\alpha_s}{d\mu}\, \frac{dZ_i}{d\alpha_s}
    = 2 \alpha_s \epsilon \frac{C_i}{Z_i} \, \frac{dZ_i}{d\alpha_s}\,,
\end{align}
where we have used the fact that the counterterms $Z_i$ depend on $\mu$ only through the running strong coupling constant in the second step, and have subsequently inserted the leading-order relation for the scale-dependence of $\alpha_s$. It is convenient to organize the effective couplings $C_i$ in a vector, $\vec C = (C_1, C_2)$ which translates the system of RG equations into
\begin{align}
    \mu \frac{d}{d\mu}\vec C = \frac{\alpha_s}{2\pi}
    \begin{pmatrix}
        -1 &  3 \\
         3 & -1
    \end{pmatrix} \cdot \vec C\,.
\end{align}
Using the running of the strong coupling,
\begin{align}\label{eq:strongcouplingRGE}
    \mu \frac{d\alpha_s}{d\mu} = - \frac{\alpha_s^2}{2\pi}\beta_0\,,
\end{align}
we find
\begin{align}
    \vec C(\mu^2) &= V^{-1}\cdot
    \begin{pmatrix}
        R^{-4/\beta_0} & 0 \\
        0 & R^{2/\beta_0}
    \end{pmatrix}
    \cdot V \cdot \vec C(\mu_0^2)\,, &
    \text{with}\quad    V &= \begin{pmatrix}
        -1 & 1 \\ 1 & 1
    \end{pmatrix}\,, &
    \text{and}\quad
    R &= \frac{\alpha_s(\mu_0)} {\alpha_s(\mu)}\,.
\end{align}
Solving the RG \cref{eq:strongcouplingRGE} allows us to rewrite $R$ into
\begin{align}
    R &= \frac{1}{1-\beta_0 L(\mu_0^2,\mu^2)}\,, &
    L(\mu_0^2, \mu^2) = \frac{\alpha_s(\mu^2)}{4\pi}\log \frac{\mu^2}{\mu_0^2}\,.
\end{align}
We can now combine these results to evolve the coefficients $C_i$ from the matching scale $\mu_0^2 = m_W^2$ to the low scale $\mu^2 = m_q^2$. This way, the solution to the evolution equation can be written as
\begin{align}\label{eq:resummed_wilson}
    \vec C(m_q^2) = V^{-1} \cdot \left\{ 
        \sum_{k=0}^\infty \frac{(-\beta_0\, L(m_W^2, m_q^2))^k}{\Gamma(1+k)}\,
        \mathrm{diag}\left(
            \frac{\Gamma(1+ 4/\beta_0)}{\Gamma(1-k+4/\beta_0)},\, 
            \frac{\Gamma(1-2/\beta_0)}{\Gamma(1-k-2/\beta_0)}
        \right)
    \right\} \cdot V\cdot \vec C(m_W^2)\,.
\end{align}
The above expression has two important features. First, it contains an infinite sum of the large logarithms $L(m_W^2, m_q^2)=\alpha_s/(4\pi)\,\log (m_q^2/m_W^2)$,  and gives a closed-form expression for the coefficients $b_{k,k}$ that we wanted to find in \cref{eq:M_resummed}. Second, the boundary condition, $C_i(m_W^2)$ is free of large logarithms since the logarithms in~\cref{eq:loop_match_result} vanish for the scale choice $\mu^2 = m_W^2$, we say that $m_W$ is the \emph{natural scale} for $C_i$.

\subsection{Matrix Element}

To obtain the complete resummed prediction, we still need to determine the one-loop matrix element of the effective theory. This amounts again to evaluating the graphs in \cref{fig:oneloopbcud_eft} along with the tree-level and counterterm graphs, yielding:
\begin{align}\begin{aligned}
    i\mathcal{A}_\mathrm{eff} =& \frac{i}{m_W^2}(\bar u_c \gamma_\mu P_L u_b)(\bar u_d \gamma^\mu P_L v_u) \left\{ 
        C_1(\mu^2)\left[
            1 + \frac{\alpha_s}{4\pi}\left(\log \frac{\mu^2}{m_q^2} + \frac{\kappa-15}{6}\right)\right] - C_2(\mu^2)\frac{\alpha_s}{4\pi}\left(3\log \frac{\mu^2}{m_q^2}+\frac{14+\kappa}{4} \right)
    \right\} \\
        &+\frac{i}{m_W^2}(\bar u_c \gamma_\mu P_L v_u)(\bar u_d \gamma^\mu P_L u_b)
    \left\{
        C_2(\mu^2)\left[
            1 + \frac{\alpha_s}{4\pi} \left( \log \frac{\mu^2}{m_q^2}-\frac{66+7\kappa}{12} \right)
        \right]
        -C_1(\mu^2)\frac{\alpha_s}{4\pi}\left( 3\log \frac{\mu^2}{m_q^2}+\frac{9+\kappa}{2}\right)
    \right\}\,.
\end{aligned}\end{align}
Notice that the amplitude depends on the parameter $\kappa$, since the loop graphs require the reductions of Dirac structures in~\cref{eq:reductions} already entering the matching calculation. If we insert the matching results in~\cref{eq:loop_match_result} into the above expression and expand to $\mathcal{O}(\alpha_s)$, we see that $\kappa$ cancels as it should.

The EFT amplitude explicitly contains logarithms with a natural scale $\mu^2 = m_q^2$. By evaluating the expression at its natural scale, and using the RG-evolved matching coefficients in~\cref{eq:resummed_wilson}, we obtain the final result for the resummed amplitude, which contains the large logarithms to all orders. After defining
\begin{align}
    \Delta = \left(1 - \frac{\alpha_s \beta_0}{4\pi}\log \frac{m_q^2}{m_W^2} \right)^{{1}/{\beta_0}}\,,
\end{align}
the amplitude can be written as:
\begin{align} \label{eq:resummed_full}
\begin{aligned}
    i\mathcal{A}_\mathrm{res} = - \frac{ig_L^2}{2m_W^2}&\left\{
        (\bar u_c \gamma_\mu P_L u_b)(\bar u_d \gamma^\mu P_L v_u) 
        \left[
            \frac{1+\Delta^6}{2\Delta^2} - \frac{\alpha_s}{4\pi}
            \left(
                \frac{1+9\Delta^6}{2\Delta^2}
                  + \frac{1-\Delta^6}{8\Delta^2}\kappa
            \right)
        \right]
    \right. \\
        &+
    \left.
        (\bar u_c \gamma_\mu P_L v_u)(\bar u_d \gamma^\mu P_L u_b)
        \left[
            \frac{1-\Delta^6}{2\Delta^2} + \frac{\alpha_s}{4\pi}
            \left(
                \frac{11\Delta^6-5}{2\Delta^2} 
                - \frac{3(1-\Delta^6)}{8\Delta^2}\kappa
            \right)            
        \right]
    \right\}\,.
\end{aligned}
\end{align}
This is our final answer for the resummed amplitude, and it deserves a number of comments.

First, we observe that this expression has achieved precisely the goal we stated in the beginning: The logarithmic terms no longer appear as part of the perturbation series and instead are contained in the expression to all orders in $\alpha_s$. In our counting scheme $\alpha_s \sim \eta$, $\log r\sim \eta^{-1}$, and as a result the expression $\Delta$ is of leading order. 
We refer to the orders in $\eta$ as \emph{leading logarithm} (LL) for $\mathcal{O}(\eta^0)$, \emph{next-to-leading logarithm}  (NLL) for $\mathcal{O}(\eta^1)$, shown in round brackets in~\cref{eq:resummed_full}, and so on. 
When expanding in the coupling instead, we refer to the counting scheme as \emph{fixed-order} perturbation theory instead, where the orders are typically denoted by \emph{leading order} (LO), \emph{next-to-leading order} (NLO), and so on.
The fixed-order result can be obtained from~\cref{eq:resummed_full} by inserting $\Delta$ and expanding in $\alpha_s$.

The resummation of large logarithms was enabled by the separation of scales. In the full theory, we were computing loop matrix elements depending on different scales, which naturally contained logarithms of this scale ratio. Constructing an effective theory separates the scales, meaning that multi-scale objects turn into products of coupling coefficients depending only on the hard scale, and operator matrix elements depending only on the soft scale. Schematically, the effective theory produces amplitudes in the form
\begin{align}
    \mathcal A(m_q^2,m_W^2) = \sum_i C_i(\mu^2, m_W^2) \cdot \langle O_i \rangle (\mu^2, m_q^2)\,.
\end{align}
This expression is an example of a \emph{factorization theorem}. 
In each of the components, the logarithms have been turned into divergences: 
\begin{align}
\begin{aligned}
  \text{in }C_i:\qquad
  \log \frac{m_q^2}{m_W^2} &\to \frac{1}{\epsilon} + \log \frac{\mu^2}{m_W^2}\, , \\
  \text{in }\langle O_i\rangle:\qquad
    \log \frac{m_q^2}{m_W^2} &\to -\frac{1}{\epsilon} + \log \frac{m_q^2}{\mu^2}\,.
\end{aligned}
\end{align}
In the matching coefficients $C_i$, the soft scale $m_q^2$ is absent and they feature a divergence instead.
The same is true for the operator matrix elements, which lack any information about the UV scale $m_W^2$ and see it replaced by a divergence, as well.

The new divergences appearing in the factorization theorem are a direct consequence of removing scales acting as  regulators in the integrals of the component functions, for example see~\cref{eq:fullboxintegral} and compare to~\cref{eq:MOR_hard,eq:MOR_soft}. 
Without RG evolution, this expression simply reproduces the amplitude of the full theory after it is expanded in the scale ratio $r=m_q^2/m_W^2$, including the large logarithms. Very importantly, this stays true for \emph{any} choice of $\mu^2$, since the product is independent of this scale, meaning that there is no choice of $\mu^2$ for which both $C_i$ and $\langle O_i \rangle$ are simultaneously free of large logarithms. The key improvement then comes from the fact that we can use the renormalization group to connect the two objects, allowing us to evaluate \emph{each} component function at their respective natural scale,
\begin{align}
    \mathcal A^\mathrm{res}(m_q^2,m_W^2) = \sum_{i,j} C_i(m_W^2, m_W^2)\cdot U_{ij}^\mathrm{RG}(m_W^2,m_q^2) \cdot \langle O_j \rangle (m_q^2, m_q^2) \,,
\end{align}
with $U_{ij}^\mathrm{RG}(m_W^2,m_q^2)$ determined by the evolution equations. This shows that separating scales renders the individual objects more divergent and enables us to use the renormalization group to sum the logarithms associated to that splitting. Any large logarithm can be treated in this way, by identifying and expanding around the parametric limit in which it turns into a divergence, and by constructing and renormalizing an effective theory around this limit.

Finally, we note that the result in~\cref{eq:resummed_full} is only complete for the leading term in each of the square brackets. The LL term contains the tree-level contribution along with all the logarithms obtained from the one-loop RG evolution, but the NLL piece (the second term in each square bracket) is lacking the terms from two-loop running, of the form
$\alpha_s^2 \log r$. This is a generic feature - we need the RG equations at one loop-order higher than the fixed-order terms. 
This means that, for a resummed prediction at LL accuracy, we need tree-level matching and one-loop running, for NLL we need one-loop matching and two-loop running. The dependence of the resummed expression on the scheme parameter $\kappa$ is a symptom of the fact that we are lacking the full NLL expressions. Indeed, by expanding the term proportional to $\kappa$, we see that it is of the form $\alpha_s^{n+1}\log^n r$ to any order in $\alpha_s$.

%------------------------------------
\section{Operator Basis}
\label{sec:basis}
%------------------------------------
For an EFT construction to be complete, it is important that the effective Lagrangian contains all possible operators, meaning that the set of operators forms a basis. In the previous section, we have seen an example in which integrating out a particle yielded an effective Lagrangian for which the number of operators differed depending on the order in perturbation theory to which the matching was computed. We have also seen how the renormalization group generated non-zero matching coefficients for operators that are not generated in the matching at tree-level.

In order to construct a complete basis, it is not sufficient to look at operators generated by a specific matching calculation, since we can never guarantee that we find a complete basis. Instead, we should always write down all operators compatible with the symmetries and power-counting of the EFT. The most naive approach to do so would be to construct all products of fields and covariant derivatives, in which all occurring indices are contracted. However, in doing so we might generate operators that are not linearly independent of each other. Our set of operators then contains redundant operators and does not form a basis. 
Furthermore, not all operators are physical, meaning that they generate vanishing matrix elements.
Hilbert-series techniques allow finding the number of independent operators at arbitrary mass dimension~\cite{Henning:2015alf}, without actually determining the basis itself.

In this section we will learn how to identify and eliminate redundant operators, and how to deal with subtleties that follow from that.

\subsection{Derivative Operators and Integration-by-Parts Identities}

Different operators containing identical fields and number of derivatives can often be related to one another by taking note of the fact that the action is an integral of the Lagrangian,
\begin{align}
    S = \int d^4x \mathcal L \,.
\end{align}
This means that terms of the form $X_\mu \partial^\mu Y$ can always be rewritten using integration by parts (IBP),
\begin{align}
    \int d^4x \ X_\mu \partial^\mu Y  =  \int d^4x \left( \partial^\mu \left[ X_\mu Y \right]  - Y \partial^\mu X_\mu  \right)\,.
\end{align}
The first term on the right-hand side is a total derivative, and corresponds to a surface term at infinity. It can be dropped since we assume fields to vanish at infinity. In momentum space, we can see this term generates an interaction proportional to the sum of all momenta in a vertex, and therefore vanishes by momentum conservation. Therefore, in a Lagrangian, we can rewrite operators using
\begin{align}
    \Delta \mathcal L = X_\mu \partial^\mu Y = -Y \partial^\mu X_\mu \,.
\end{align}
As a simple example, in a real-scalar theory, the two possible dimension-six operators containing four scalars and two derivatives,
\begin{align}\label{eq:Lag_scalars_dim6_two_ops}
    \mathcal L_6 = C_1 \varphi^3 \partial^2 \varphi + C_2 \varphi^2 (\partial_\mu\varphi)(\partial^\mu \varphi)\,,
\end{align}
can be reduced to just one by considering the total-derivative term and requiring it to vanish,
\begin{align}
    \partial^\mu (\varphi^3 \partial_\mu \varphi) = \varphi^3 \partial^2\varphi + 3\varphi^2(\partial_\mu\varphi)(\partial^\mu\varphi) \stackrel{!}{=} 0\,,
\end{align}
leading to
\begin{align} 
    \mathcal L_6 = \left(C_1 - \frac{C_2}{3}\right) \varphi^3 \partial^2\varphi \equiv C\, \varphi^3\partial^2\varphi \,.
\end{align}
These identities are exact, and in particular hold in dimensional regularization. The different Lagrangians obtained by them are equivalent, in that they predict the same physical observables. 

\subsection{Field Redefinitions}
When using IBP identities to relate derivative operators to each other, we are left with a choice of a target operator basis. In the example above, we could have written the final Lagrangian in terms of the second operator in~\cref{eq:Lag_scalars_dim6_two_ops} instead of the first one. 
In this subsection, we will show that it is beneficial to apply IBP to generate as many operators as possible of the form $\chi\partial^2\varphi$, as we can subsequently remove them by an appropriate field-redefinition.

Since the fields are integration variables in the path integral, we are allowed to redefine them. Consider again the real scalar theory from above, given after IBP relations by the Lagrangian
\begin{align}\label{eq:LagafterIBP}
    \mathcal L = \frac{1}{2}(\partial_\mu\varphi)(\partial^\mu\varphi)  - \frac{\lambda}{4!}\varphi^4 + C \varphi^3\partial^2\varphi\,,
\end{align}
where $\lambda$ is not to be confused with the power-counting parameter introduced previously. Consider a field redefinition 
\begin{align}
    \varphi \to \varphi + \chi\,,
\end{align}
where $\chi$ is an unspecified scalar object, which we assume to be of subleading order in the EFT power-counting. Inserting this redefinition into the kinetic Lagrangian of the scalar, and expanding in the power-counting, yields
\begin{align}
    \frac{1}{2}(\partial_\mu\varphi)(\partial^\mu\varphi) 
    \quad \to \quad
    \frac{1}{2}(\partial_\mu\varphi)(\partial^\mu\varphi) 
    -
    \chi (\partial^2\varphi) + \mathcal{O}(\chi^2)\,,
\end{align}
where we have used IBP reductions to rewrite the second 
term on the right-hand side. Notice how this specific term is now of the form of the dimension-six operator in the Lagrangian in~\cref{eq:LagafterIBP} for a particular choice of $\chi$: By choosing 
\begin{align}
    \chi = C \varphi^3\,,
\end{align}
the field redefinition exactly cancels this operator. The insertion into the quartic interaction then produces a new operator, leaving us with the final Lagrangian after field redefinitions:
\begin{align} \label{eq:LagAfterFieldRedef}
     \mathcal L = \frac{1}{2}(\partial_\mu\varphi)(\partial^\mu\varphi) -\frac{\lambda}{4!}\varphi^4  - \frac{\lambda C}{3!}\varphi^6 \,. 
\end{align}
This result motivates the preferred target operators for the IBP reductions: If we aim to produce as many operators as possible of the form $\chi\partial^2\varphi$, we can subsequently remove them by an appropriate field-redefinition. Observant readers may have noticed that the derivative structure in the operator we have removed bears a striking resemblance to the equation of motion (EOM) of the scalar field,
\begin{align}\label{eq:realscalareom}
    \partial^2\varphi = - \frac{\lambda}{3!}\varphi^3 +
    6C\left( \varphi (\partial_\mu\varphi)(\partial^\mu\varphi)+\varphi^2\partial^2\varphi \right)
    \,.
\end{align}
This pattern extends also to fields of other spins: For Dirac fermions, operators of the form $\bar\chi\slashed\partial\psi$ can be removed by a field redefinition of $\psi$. Similar rules hold for vector fields, we give a full list in \cref{tab:fieldRedefs}.
\begin{table}[t]
\renewcommand{\arraystretch}{1.2}
\centering
\begin{tabular}{ccc}
\hline
Field Type & Redundant operators & Field redefinition \\ \hline
Real scalar $\varphi$ & $\chi D^2\varphi$ & $\varphi \to \varphi + \chi$ \\ 
Complex scalar $\phi$ & $\chi D^2{\phi} + (D^2{\phi^\dagger})\Delta$ & $\phi \to \phi + \frac{1}{2}(\chi^\dagger + \Delta)$ \\ \hline
Majorana fermion $\eta$ & $\chi \slashed D{\eta} + {\eta^T}\overleftarrow{\slashed{D}}\Delta$ & $\eta\to\eta + i C(\chi^T-\Delta)$ \\
Dirac fermion $\psi$ & $\chi \slashed D{\psi} +{\bar\psi}\overleftarrow{\slashed{D}}\Delta$ & $\psi\to\psi - \frac{i}{2}(\bar\chi + \Delta)$ \\ \hline
Real vector field $A$ & $\chi_\nu D_\mu{F^{\mu\nu}}$ & $A_\mu \to A_\mu - \chi_\mu$ \\
Complex vector field $A$ & $\chi_\nu D_\mu{F^{\mu\nu}} + (D_\mu{F^{\dagger\mu\nu}})\Delta_\nu$ & $A_\mu \to A_\mu -\frac{1}{2}( \chi_\mu^\dagger + \Delta_\mu) $ \\ \hline
\end{tabular}
\caption{Field redefinitions needed to remove redundant operators involving a given field type, taken from~\cite{Fuentes-Martin:2022jrf}. For a hermitian Lagrangian, the coefficients $\chi$ and $\Delta$ for each field should be conjugates of each other, but the operators are written in a way such that the Lagrangian does not have to be in a manifestly hermitian form.}
\label{tab:fieldRedefs}
\end{table}

Because of their similarity to the fields' equations of motion, redundant operators of the form discussed here are often referred to as "EOM operators". It is commonly stated that these operators can also be removed by simply deriving the EOMs and inserting them in the operator. We stress here that this produces the correct results \emph{only} for operators at the highest order in power-counting. In our example, where the redundant operator appears at dimension six and we do not consider higher-dimensional operators, the EOM method indeed produces the correct answer: One quickly checks that inserting~\cref{eq:realscalareom} into the redundant dimension-six operator in the Lagrangian in~\cref{eq:LagafterIBP} produces the last operator in the Lagrangian in~\cref{eq:LagAfterFieldRedef}. Importantly, the term in the EOM proportional to $C$ produces a higher-dimensional operator $\propto C^2$
\begin{align}
    \left. \delta\mathcal L \right|_\mathrm{EOM} = \frac{24C^2}{5}\varphi^5\partial^2\varphi = - \frac{4\lambda C^2}{5}\varphi^8 \,.
\end{align}
which in the second step we have further reduced by using the EOM once more.
However, if we work with field redefinitions, we find the dimension-eight Lagrangian to be
\begin{align}
    \left. \delta\mathcal L \right|_\mathrm{redef} = - \frac{9\lambda C^2}{10}\varphi^8 \,.    
\end{align}
In order to verify which expression is the correct one, we can compute six- and eight-scalar scattering from the unreduced Lagrangian in~\cref{eq:LagafterIBP} and see which of the interaction terms reproduce the amplitude. We leave this exercise to the reader, but if done carefully, one finds the Lagrangian $\left.\delta\mathcal L\right|_\mathrm{redef}$ to be the correct one. The reason for the discrepancy is that the celebrated equivalence between EOMs and field redefinitions holds only for the terms \emph{linear} in the field-redefinition, such that in the redefinition
\begin{align}
    \mathcal L(\varphi) \quad \to \quad \mathcal L(\varphi+\chi)\,,
\end{align}
only the linear term in an expansion in $\chi$ is correctly reproduced by the EOMs.

\subsubsection{Green's Basis and On-Shell Basis}\label{sec:greensandOS}
Besides producing operators with fewer derivatives, removing EOM-type operators with field-redefinitions is desirable for another reason. 
Consider again the example of the Lagrangian in~\cref{eq:LagafterIBP}. 
The redundant operator contains four scalar fields, and so it generates a four-point amplitude. 
The corresponding Feynman rule for it is proportional to the squared momenta of the four scalars $\sim p_1^2+p_2^2+p_3^2+p_4^2$. For on-shell massless fields this operator generates no on-shell matrix elements as all $p_i^2 = 0$.

The operator basis we find when writing every possible operator, including derivatives, and reducing it only using IBP relations, is called the Green's basis. It contains redundant operators generating vanishing on-shell matrix elements. This means, when matching to a Green's-basis Lagrangian, we have to keep all external fields to be off-shell. If we instead remove all redundant operators, we find what is usually called an on-shell basis. In matching to the on-shell basis we can of course assume external fields to be on-shell, but it comes with the added complication of having to evaluate more diagram topologies, as we will see in the following.

To understand the issue, let us understand diagrammatically how the Green's-basis operators and on-shell operators generate equivalent matrix elements. As an example, let us compute the six-scalar amplitude first from the Lagrangian in~\cref{eq:LagafterIBP}. In this basis, it is generated by two insertions of quartic interactions. Limiting ourselves to the contribution involving exactly one insertion of the dimension-six operator, it can be written as the time-ordered product,
\begin{align} \label{eq:topology_redundant}
    \raisebox{\shiftb ex}{\includegraphics[scale=.4]{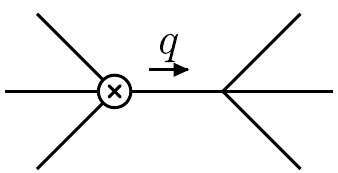}}\quad = \quad  \frac{\lambda C}{4!} \left\langle \varphi_1 \varphi_2 \varphi_3 \varphi_4 \varphi_5 | T\{\varphi^4, \varphi^3\partial^2\varphi\} | \varphi_0\right\rangle\,.
\end{align}
Assuming external particles to be on-shell, the field with the derivative acting on it can only be the one propagating between the two quartic vertices, as for all others the Feynman rule produces the vanishing squared momentum of an on-shell field. In the $T$-product, this means that the operator $\partial^2\varphi$ can only be contracted with one of the $\varphi$ from the second quartic, producing a propagator $i/q^2$, and thus canceling the $-q^2$ factor from the Feynman rules. Afterwards, we just contract the six remaining scalars with all external states. Explicitly, the calculation plays out as:
\begin{align}
        \frac{\lambda C}{4!} \left\langle \varphi_1 \varphi_2 \varphi_3 \varphi_4 \varphi_5 | T\{\varphi^4, \varphi^3\partial^2\varphi\} | \varphi_0\right\rangle
        = \frac{-i\lambda C}{4!}\cdot 4\cdot \left\langle \varphi_1 \varphi_2 \varphi_3 \varphi_4 \varphi_5 | T\{\varphi^3, \varphi^3\} | \varphi_0\right\rangle
        = -i\lambda C \frac{4\cdot 6!}{4!}= -120 i\lambda C \,.
\end{align}
It is straightforward to see that the final expression is precisely what we find if we simply compute this amplitude from the on-shell Lagrangian in~\cref{eq:LagAfterFieldRedef}.

The important takeaway from this example is that the derivative in the dimension-six operator cancels exactly the pole of the intermediate propagator. The fact that there is no $1/q^2$ dependence in the amplitude explains how the contribution can be described by an operator of the EFT after field-redefinitions, even though the virtual propagator does not belong to a heavy particle. This is true for EOM-operators in general: Their on-shell matrix elements vanish, and matrix elements with off-shell fields can be described by "non-EOM" operators. For this reason, EOM operators are redundant and can be dropped.

When performing matching calculations directly to the on-shell basis, we can therefore not restrict our attention to graphs in which all internal lines are hard (i.e.\ heavy or highly virtual), but instead need to allow for further interactions of the light particles, as in \cref{eq:topology_redundant}. This is inconvenient not just because we have to compute more diagrams, but also because said extra diagrams do not necessarily contribute to the matching in their entirety. Recall the cancellation between a $q^2$ from the interaction and a $1/q^2$ from the light propagator. When computing these graphs, we have to carefully isolate the terms in which the cancellation occurs and which no longer contain $1/q^2$. To see what we mean by this, consider an example where the above diagram is the result of a heavy scalar integrated out at one-loop level, for instance through the following graph:
\begin{align} 
    \raisebox{\shiftb ex}{\includegraphics[scale=.4]{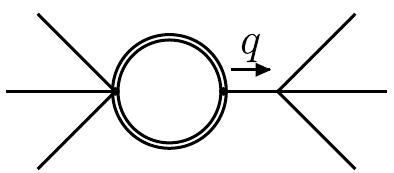}}\quad \sim \quad  
    \left[ \int\frac{d^dl}{(2\pi)^d}\frac{1}{l^2-M^2}\frac{1}{(l+q)^2-M^2} \right]\cdot \frac{1}{q^2}\,,
\end{align}
where the double line refers to the heavy scalar.
For the matching calculation, we expand the integrand around the hard region $l^2\sim M^2\gg q^2$ and rewrite the loop integral to make the $q^2$ dependence explicit, finding 
\begin{align}
    \left. \left[ \int\frac{d^dl}{(2\pi)^d}\frac{1}{l^2-M^2}\frac{1}{(l+q)^2-M^2} \right]\cdot \frac{1}{q^2} 
    \right|_\mathrm{hard}
    =
    \left[\frac{1-\epsilon}{M^2}\left(1 + \frac{q^2}{6M^2}\epsilon \right) \int\frac{d^dl}{(2\pi)^d}\frac{1}{l^2-M^2}\right]\cdot \frac{1}{q^2}\,.
\end{align}
This expression shows that only one of the two terms in the square bracket cancels the neighboring propagator, and thus the on-shell matching contribution to the $\varphi^6$ operator is given just by this term. Importantly, the first term in the square bracket corresponds to the one-loop matching of the $\varphi^4$ interaction and cannot be included in the matching of the $\varphi^6$ interaction - you can tell by the fact that it is a non-local contribution, meaning it has a non-polynomial dependence on the external momenta through  the factor $1/q^2$. Very roughly speaking, the correspondence of full-theory graph and the EFT graphs can be visualized as
\begin{align} 
    \raisebox{\shiftb ex}{\includegraphics[scale=.4]{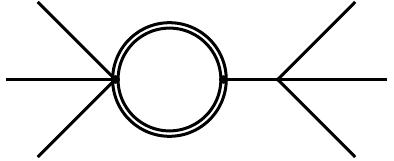}}\quad \to \quad  \raisebox{\shiftb ex}{\includegraphics[scale=.4]{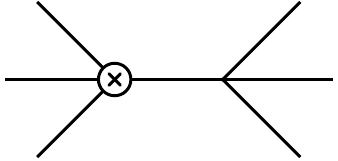}}\quad + \quad  \raisebox{\shiftb ex}{\includegraphics[scale=.4]{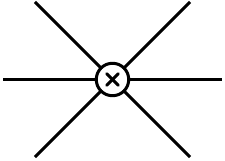}}\,,
\end{align}
where the crossed circle denotes the matching contribution arising from the loop graph. In realistic scenarios,
the differentiation is often subtle, relying on advantageous choices of the kinematic variables to make the cancellation explicit. For this reason, it can be easier to match to an off-shell basis where redundant topologies like~\cref{eq:topology_redundant} do not need to be considered. Of course this comes at the expense of having to write down (and often later reduce) redundant operators.

\subsection{Evanescent Operators}
When performing calculations beyond tree-level, divergences appearing in loop integrals are most-commonly treated in dimensional regularization (dim-reg), by extending the number of spacetime dimensions to $d=4-2\epsilon$ and expanding around the $\epsilon\to 0$ limit. The manipulations to the operators we have employed up to this point are exact, meaning that they are not affected by any subtleties that dim-reg introduces.

However, this changes once we use identities that rely on the completeness of the spacetime algebra. 
When extending the number of spacetime dimensions to a non-integer value, there is no longer a way to meaningfully define a Levi-Civita symbol, which effectively renders the basis of Dirac and Lorentz structures infinite. 
 In this subsection, we will show that identities relying on the completeness of the spacetime algebra lead to an $\mathcal{O}(\epsilon)$ difference between two operators in $d$ dimensions. While this difference is harmless at tree-level when $\epsilon \to 0$, at loop level the $\mathcal{O}(\epsilon)$ piece can multiply a $1/\epsilon$-pole from a loop integral, leaving a finite leftover in the $\epsilon \to 0$ limit.
 These contributions can be absorbed as an evanescent counterterm into the coupling of a physical operator of the theory~\cite{Buras:1989xd,Dugan:1990df,Herrlich:1994kh}. This means that (at loop level) the scheme used to relate the Dirac and Lorentz structures is part of the definition of an operator basis.

We will show this explicitly using the commonly-used example of Fierz relations,  following~\cite{Fuentes-Martin:2022vvu}: 
Consider two operators, written in the unbroken phase of the SM:
\begin{align}
    \begin{aligned}
    O_1 &= (\bar\ell \gamma_\mu P_L \ell)(\bar e \gamma^\mu P_R e)\,, \\
    O_2 &= (\bar\ell P_R e)(\bar e P_L \ell)\,.
    \end{aligned}
\end{align}
Using Fierz transformations, the two operators can be related to each other, 
\begin{align} \label{eq:o1o2_fierz_id}
    O_2 = -\frac{1}{2}O_1\,.
\end{align}
By this equation, the two Lagrangians
\begin{align}
\begin{aligned}
    \mathcal L^{(1)}_\mathrm{\ell e} &= C_1 (\bar\ell \gamma_\mu P_L \ell)(\bar e \gamma^\mu P_R e) + C_2 (\bar\ell P_R e)(\bar e P_L \ell) \,,  \\
    \label{eq:twoLagrangians_eva}
    \mathcal L^{(2)}_\mathrm{\ell e} &= \left(C_1-\frac{1}{2}C_2 \right) (\bar\ell \gamma_\mu P_L \ell)(\bar e \gamma^\mu P_R e)  \,,
\end{aligned}
\end{align}
should be equivalent and should give rise to identical matrix elements. Let us now test that at the example of the matrix element $\langle \ell H B| \mathcal L_{\ell e}^{(i)} | \ell \rangle$, where $B$ denotes the gauge boson of hypercharge $U(1)_Y$, not to be confused with the $B$ meson. 

\begin{figure}
    \centering
    \begin{subfigure}[t]{0.49\textwidth}
    \centering
      \includegraphics[scale=0.4]{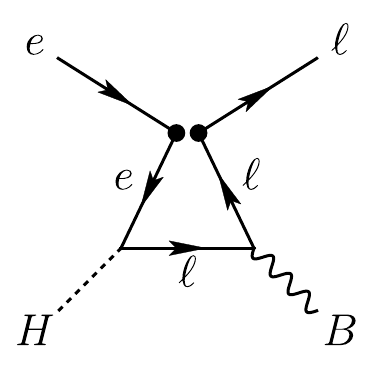} \qquad 
    \includegraphics[scale=0.4]{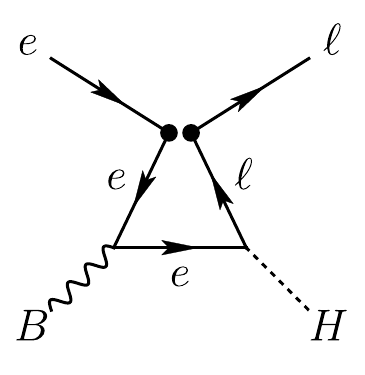} \qquad     
      \caption{$O_1$ contributions.}
    \end{subfigure}%
    ~
    \begin{subfigure}[t]{0.49\textwidth}
        \centering
            \includegraphics[scale=0.4]{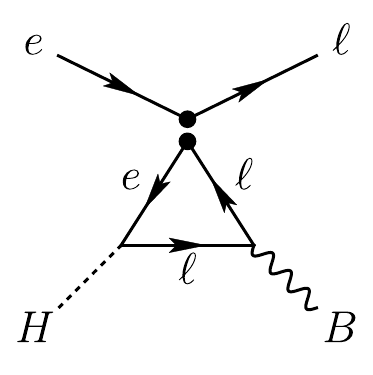} \qquad 
    \includegraphics[scale=0.4]{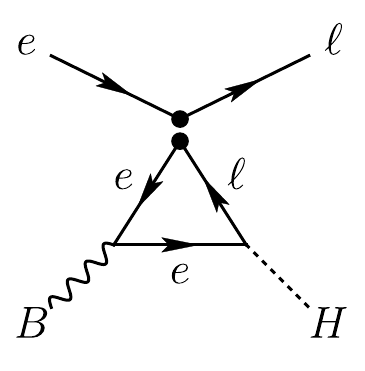}
    \caption{$O_2$ contributions.}
    \end{subfigure}%
    \caption{One-Loop diagrams in the effective theory generated by insertions of the four-lepton operators $O_1$ and $O_2$. The double dots indicate the contraction of spinor indices in the four-fermion operators.}
    \label{fig:evanescent_loops}
\end{figure}

The computation of the matrix element amounts to evaluating the one-loop diagrams shown in \cref{fig:evanescent_loops}, with the crossed circle denoting insertions of the operators $O_1$ and $O_2$. The details of the calculation are left as an exercise to the reader, but the important realization is that the ordering of the spinors leads to the fact that the matrix elements of $O_1$ are given by one continuous chain of Dirac matrices, whereas the operator $O_2$ turns the loop into a closed trace of Dirac matrices. This trace turns out to be proportional to $q\cdot\varepsilon^\ast$, where $q_\mu$ and $\varepsilon_\mu$ are the momentum and polarization vectors of the gauge boson, and hence the trace - and with it the matrix element of $O_2$ - vanishes. The sum of the two graphs can then be written as a function of the Wilson coefficients $C_{1,2}$:
\begin{align}
    i\mathcal A(C_1,C_2) =  C_1 y_e g_Y (Y_\ell+Y_e)(\bar u_\ell \slashed q \slashed \varepsilon^\ast P_R v_e)\ \epsilon \int\frac{d^dl}{(2\pi)^d}\frac{1}{l^2}\frac{1}{(l+q)^2}\,,
\end{align}
where $Y_\ell$ and $Y_e$ are the $U(1)_Y$ hypercharges of the left- and right-handed leptons, respectively, and the momentum of the external Higgs has been set to zero. Comparing the Lagrangians $\mathcal L_{\ell e}^{(i)}$ in \cref{eq:twoLagrangians_eva}, it is clear that the matrix element should obey
\begin{align}
    \mathcal A(C_1,C_2) = \mathcal A \Big(C_1 - \frac{C_2}{2},0 \Big)\,,
\end{align}
for the two Lagrangians to give identical predictions. The only way this relation can still hold is if $\mathcal A(C_1,C_2) = 0$. Note that the matrix element is proportional to the dimensional regulator, which is expected - the Fierz identity holds exactly in the limit of $\epsilon\to 0$. However, the loop integral is UV-divergent,
\begin{align}
    \int\frac{d^dl}{(2\pi)^d}\frac{1}{l^2}\frac{1}{(l+q)^2} = \frac{i}{16\pi^2}\left(\frac{1}{\epsilon} + \mathcal{O}(\epsilon^0)\right) \,,
\end{align}
and so we find\footnote{Note that the sign of this amplitude and all expressions derived from it depends on the sign convention for the covariant derivative. We remind the reader that we use the positive convention, $D_\mu f = \partial_\mu + i g_Y Y_f B_{Y\mu} f$ for $f=e,\ell$.}
\begin{align}
    i\mathcal A(C_1,C_2) =  \frac{iC_1 y_e g_Y}{16\pi^2} (Y_\ell+Y_e)(\bar u_\ell \slashed q \slashed \varepsilon^\ast P_R u_e)\,,
\end{align}
which is non-zero even in the limit of $\epsilon\to 0$. 
Note that this does not mean that we are not allowed to use Fierz and similar reductions when working beyond tree-level.
The fact that the difference between the two Lagrangians is given by an $\mathcal{O}(\epsilon)$ term hitting the UV-divergence is the first hint towards the solution of the problem: It means that the non-zero contribution is a \emph{local} one, and just like regular UV-divergences, can be absorbed by a redefinition of some other coupling in the theory.

To systematically treat this problem, we begin once again with the Lagrangian $\mathcal L_{\ell e}^{(1)}$ and use the Fierz relation in \cref{eq:o1o2_fierz_id}, but afterwards immediately add back the difference,
\begin{align}
    O_2 = - \frac{1}{2}O_1 + \left( \frac{1}{2}O_1 + O_2 \right) \equiv -\frac{1}{2}O_1 + E_{12}\,,
\end{align}
yielding the Lagrangian
\begin{align}
    \mathcal L_{\ell e}^{(2+E)} = \left(C_1-\frac{1}{2}C_2 \right) (\bar\ell \gamma_\mu P_L \ell)(\bar e \gamma^\mu P_R e)  + C_2 E_{12}\,.
\end{align}
This Lagrangian is obviously identical to $\mathcal L_{\ell e}^{(1)}$ even beyond the $\epsilon\to 0$ limit, but the difference between $\mathcal L_{\ell e}^{(1)}$ and $\mathcal L_{\ell e}^{(2)}$ is explicitly captured by the operator $E_{12}$. We call such operators \emph{evanescent operators}. They are called this way because they only appear at an intermediate step: They do not produce tree-level matrix elements, since they are formally of $\mathcal{O}(\epsilon)$. Once a theory is renormalized and the $\epsilon\to 0$ limit can be taken, they will no longer play a role and can be dropped. Their non-vanishing contributions are generated from UV-divergences of loop integrals, which can be absorbed by counterterms. Contrary to the usual UV-counterterms, the ones found here are rendered finite by the $\mathcal{O}(\epsilon)$ prefactor of the evanescent matrix elements.

In our particular example, the matrix element of the evanescent operator is simply given by the difference 
\begin{align}
    i\mathcal A_E = i\mathcal A(C_1,C_2) - i\mathcal A\Big(C_1-\frac{1}{2}C_2,0\Big) 
    = -\frac{3iC_2}{64\pi^2}y_e g_Y [\bar u_\ell\slashed q \slashed\varepsilon^\ast P_R u_e] \,,
\end{align}
where we have inserted the hypercharges $Y_e = -1$, $Y_\ell = -1/2$.
We now need to find the effective operator producing a matrix element of this form. It is given by the hypercharge-dipole operator
\begin{align}
    \mathcal L_\mathrm{dip} = C_3 F_{\mu\nu} (\bar\ell H\sigma^{\mu\nu}P_R e) + \mathrm{h.c.}\,,
\end{align}
generating the matrix element
\begin{align}
    i\mathcal A_3 = -i C_3 \bar u_\ell \slashed q\slashed\varepsilon^\ast P_R u_e\,.
\end{align}
Next, we introduce a (finite) counterterm to $C_3$ and match it to capture the contribution from $i\mathcal A_E$. We find:
\begin{align}
    C_3 &\to C_3 + \delta_e^{\mathrm{[ev]}}\,,&
    \text{with}\quad \delta_e^{\mathrm{[ev]}} &= \frac{3C_2}{64\pi^2} y_e g_Y\,.
\end{align}
After having shifted the effective coupling $C_3$, we can safely drop $E_{12}$, since its contribution is now contained in $C_3$.

To summarize, when reducing operators using identities valid only in $d=4$, the differences are captured by evanescent operators. Their matrix elements generate finite shifts to the couplings of physical, i.e. non-evanescent, operators. After these counterterms are computed, evanescent operators can be dropped. A few important comments are in order. 
\begin{itemize}
    \item First, it is important to note that, for a complete treatment, the above analysis needs to be repeated for every matrix element that $E_{12}$ can produce, which then in turn generates shifts to other effective operators as well. An obvious example is the matrix element in which we replace the hypercharge gauge boson by a $W$ boson, which would generate a shift to the weak dipole operator $W^a_{\mu\nu} (\bar\ell \tau^a H\sigma^{\mu\nu}P_R e)$.
    \item Second, it is important to note that the value of $C_3$ changes depending on how we write the four-lepton operators, as we have seen above: The two physically equivalent Lagrangians are (focusing only on the operators in this example):
    \begin{align}
        \begin{aligned}
            \mathcal L^{(1)}_{\ell e+ \mathrm{dip}} &= C_1 O_1 + C_2 O_2  + \left[ C_3 O_3 + \mathrm{h.c.} \right] \,, \\
            \text{and}\quad \mathcal L^{(2)}_{\ell e+ \mathrm{dip}} &= \left(C_1-\frac{1}{2}C_2\right) O_1 + \left[ \left(C_3 + \frac{3C_2}{64\pi^2} y_e g_Y  \right) O_3 + \mathrm{h.c.} \right] \,.
        \end{aligned}
    \end{align}
    This means that the question "What is the coupling of the operator $O_3$?" can only be meaningfully answered in the context of a full operator basis. \footnote{Strictly speaking, $\mathcal L^{(1)}$ does not constitute a basis since $O_1$ and $O_2$ are linearly dependent in $d=4$.}

    \item Reduction identities that are only exact in $d=4$ and thus require the introduction of evanescent operators, are:
    \begin{enumerate}
        \item Fierz-transformations, since they rely on the completeness of the Dirac basis.
        \item Dirac-reductions  relying on $\gamma_5 = -\frac{i}{4!}\epsilon^{\mu\nu\alpha\beta}\gamma_\mu\gamma_\nu\gamma_\alpha\gamma_\beta$ to relate structures to each other. Commonly used examples are
            \begin{align}
            \begin{aligned}
                \sigma^{\mu\nu}\gamma_5 &= \frac{i}{2}\epsilon^{\mu\nu\alpha\beta}\sigma_{\alpha\beta} \,, \\
                \gamma^\mu \gamma^\nu \gamma^\rho &= g^{\mu\nu}\gamma^\rho - g^{\mu\rho}\gamma^\nu + g^{\nu\rho}\gamma^\mu + i \epsilon^{\mu\nu\rho\sigma}\gamma_\rho\gamma_5\,.
            \end{aligned}
            \end{align}
        \item Contractions of multiple Levi-Civita objects 
        \begin{align}
            \epsilon^{\mu\nu\rho\sigma}\epsilon_{\alpha\beta\delta\lambda}
            =\left|
            \begin{array}{cccc}
                 \delta^\mu_\alpha & \delta^\mu_\beta  & \delta^\mu_\delta & \delta^\mu_\lambda \\
                 \delta^\nu_\alpha & \delta^\nu_\beta  & \delta^\nu_\delta & \delta^\nu_\lambda \\
                 \delta^\rho_\alpha & \delta^\rho_\beta  & \delta^\rho_\delta & \delta^\rho_\lambda \\
                 \delta^\sigma_\alpha & \delta^\sigma_\beta  & \delta^\sigma_\delta & \delta^\sigma_\lambda 
            \end{array}
            \right| \, .
        \end{align}
        Note that these can appear even in fully bosonic operators, and so the appearance of evanescent operators is not exclusive to fermionic interactions.
    \end{enumerate}
\end{itemize}

%------------------------------------
\section{SMEFT}
\label{sec:SMEFT}
%------------------------------------
The SM has been extremely successful in describing experimental data across a wide range of phenomena and phase-space regimes.  
Its Lagrangian is given as 
\begin{align}\label{eq:SMlag}
	\mathcal{L}_\text{SM} =& -\frac{1}{4} G_{\mu \nu}^A G^{A\mu \nu}-\frac{1}{4} W_{\mu \nu}^I W^{I \mu \nu} -\frac{1}{4} B_{\mu \nu} B^{\mu \nu}
    +\sum_{\psi=q,u,d,l,e} \overline \psi\, i \slashed{D} \, \psi 
	+ (D_\mu H^\dagger)(D^\mu H)
    -\mu_H^2\, H^\dagger H +\lambda_H (H^\dagger H)^2 \nonumber \\ 
	& - \Big[ y_d \, \overline d\,  H^{\dagger}q +  y_u\, \overline u\, \widetilde H^{\dagger}q + y_e \, \overline e\, H^{\dagger}l + \hbox{h.c.}\Big]\,,
\end{align}
where $G$, $W$, $B$ denote the field-strength tensors of the gauge fields and $q, l$ and $u,d,e$ are the left-handed and right-handed fermion fields, respectively. The covariant derivative is defined as 
\begin{align}\label{eq:covderivative}
D_{\mu}&=\partial_{\mu}+ig_sT^{A}G^{A}_{\mu}+i\frac{g_{W}}{2}\sigma^{I}W^{I}_{\mu}+ig_{Y}Y B_{\mu} \, ,
\end{align} 
where $T^{A}$ and $\sigma^{I}$ are the $SU(3)_{c}$ and $SU(2)_{L}$ generators and $Y$ denotes the hypercharge. $g_{s},g_{W}, g_{Y}$ are the $SU(3)_{c}\times SU(2)_{L}\times U(1)_{Y}$ gauge coupling constants.
$H$ denotes the $SU(2)_{L}$ doublet, which is expanded around the vacuum expectation value $v$ as
\begin{align}
H = \begin{pmatrix}
	-i G^{+} \\
	\frac{1}{\sqrt{2}}(v+h+i G^{0})
	\end{pmatrix}\,,
    \label{eq:Higgs_doublet}
\end{align}
where $h$ is the SM Higgs boson while $G^{0}$, $G^{+}$ denote the neutral and charged Goldstone bosons respectively.

While the SM explains most experimental data very well, there are plenty of observations which lead us to believe that the SM is currently incomplete. 
For instance, we do not yet have a quantum theory of gravity, we have not discovered a candidate for dark matter, and there is no explanation for either the observed matter--anti-matter asymmetry of the Universe or the origin of neutrino mixing. 
Moreover, the SM comes with theoretical hierarchy problems such as leaving us without an explanation for the observed flavor hierarchies or the strong $CP$ problem. 
Due to the fact that current experiments see no evidence for new GeV-scale particles, one intriguing possibility is that at least some new particles are too heavy to be directly produced in ongoing experiments
and that there is a mass gap between the SM and BSM particles. 
If this is the case, we can describe the low-energy effects of the heavy NP with an EFT built from the fields of the SM~\cite{Buchmuller:1985jz}. 
We call the corresponding EFT Standard Model Effective Field Theory (SMEFT). 

SMEFT respects the SM $SU(3)_C \times SU(2)_L \times U(1)_Y$ symmetry as well as Lorentz invariance and, in contrast to the more general theory of Higgs Effective Field Theory~(HEFT), assumes that the Higgs boson is in the doublet structure of \cref{eq:Higgs_doublet} such that EWSB is linearly realized. 
The SMEFT Lagrangian is given by
%------------
\begin{align}\label{eq:SMEFTlag}
    \mathcal{L}_\text{SMEFT} &= \mathcal{L}_\text{SM} 
    +  \frac{C^{(5)}}{\Lambda} {O}^{(5)}
    + \sum_i \frac{C_i^{(6)}}{\Lambda^2} {O}^{(6)}_i
    +  \sum_j \frac{C^{(7)}_j}{\Lambda^3} {O}^{(7)}_j
    + \sum_k \frac{C_k^{(8)}}{\Lambda^4} {O}^{(8)}_k
    + \cdots \, ,
\end{align}
%------------
where, as before, the superscript on the Wilson coefficients and operators corresponds to the operator mass dimension $d$ and all operators are suppressed by appropriate powers $d-4$ of the new-physics scale $\Lambda$, the expansion parameter of SMEFT.
At mass dimension five, there is a single operator, the so-called Weinberg operator which generates Majorana neutrino masses~\cite{Weinberg:1979sa}. 
Operators of odd mass dimensions violate baryon and/or lepton number conservation (they lead to odd $(\Delta B -\Delta L)/2$~\cite{Kobach:2016ami}, where $B$ and $L$ refer to the baryon and lepton number) and are thus often neglected in phenomenological analyses when these effects do not play a dominant role.
Therefore, general SMEFT analyses typically include only even-dimensional operators and start from mass dimensions six and eight as the lowest-order contributions.

%------------------------------------
\subsection{SMEFT Operator Bases }
\label{sec:SMEFT_OperatorBasis}
%------------------------------------

As discussed in \cref{sec:basis}, the construction of EFT operator bases requires the choice of a complete and non-redundant set of operators. The operators that differ by total derivatives or that are related by field redefinitions, EOMs, IBP or Fierz identities must be removed to form a complete basis. Such elimination of redundancies avoids overcounting independent operators while preserving potential high-energy effects on low-energy physical observables.  At dimension six, many interaction structures are possible following the SM gauge symmetry. Several operator bases have been introduced in the literature, each suited for different applications. For instance, the Strongly Interacting Light Higgs (SILH)~\cite{Giudice:2007fh,Elias-Miro:2013mua} and Hagiwara-Ishihara-Szalpski-Zeppenfeld (HISZ)~\cite{Hagiwara:1993ck} bases are commonly used in studies related to Higgs and EW gauge bosons interactions.
However, the most commonly used basis for dimension-six SMEFT is the so-called Warsaw basis~\cite{Grzadkowski:2010es}. 
For a single fermion family, the Warsaw basis has $59$ independent operators, for three fermion families there are $2499$ operators. The Warsaw basis is typically grouped into eight classes according to the field content of the operators:
(i) $X^3$ (ii) $H^6$ (iii) $H^4 D^2$ (iv) $X^2H^2$ (v) $\psi^2 H^3$ (vi) $\psi^2 X H$ (vii) $\psi^2 H^2 D$ (viii) $\psi^4$, 
where $X$, $H$, $D$ and $\psi$ refer to 
field strength, the Higgs doublet, covariant derivative and fermions, respectively. This classification is useful phenomenologically since different classes contribute to different sets of physical observables such as EW precision observables, Higgs production and decay observables and four-fermion contact interactions.

At dimension-eight, several operator bases have been developed for different phenomenological applications. The operator set introduced by Almeida, Éboli, Gonzalez-Garcia and Mizukoshi~\cite{Eboli:2003nq,Eboli:2006wa,Eboli:2016kko,Durieux:2024zrg} is widely used in studies of anomalous quartic gauge couplings and has been extensively adopted in experimental vector-boson-scattering analyses~\cite{ATLAS:2026wew,CMS:2025dbm}. 
A dedicated dimension-eight basis for anomalous neutral triple gauge couplings was constructed in Ref.~\cite{Degrande:2013kka,Ellis:2022zdw} and has been used in experimental diboson analyses~\cite{CMS:2025cxp,ATLAS:2025ply}. 
Complete and non-redundant bases of dimension-eight SMEFT operators have been constructed in Refs.~\cite{Murphy:2020rsh,Li:2020gnx}. 
In addition, the dimension-eight Green’s basis has been developed, first for the bosonic sector~\cite{Chala:2021cgt} and subsequently for the complete SMEFT operators, including fermionic interactions~\cite{Ren:2022tvi}.
For mass dimensions up to dimension twelve, using the techniques of Hilbert Series and Young tableaux, non-redundant operator bases are already known~\cite{Weinberg:1979sa,Grzadkowski:2010es,Lehman:2014jma,Henning:2015alf,Li:2020xlh,Liao:2020jmn,Harlander:2023psl}.

%------------------------------------
\subsection{Flavor Assumptions}
\label{sec:SMEFT_flavorAssumptions}
%------------------------------------
The $2499$ operators at dimension-six are different flavor combinations of common operator structures. For one fermion generation only and assuming real Wilson coefficients, the number of independent operators reduces to $59$. 
As a result, the assumption of flavor symmetries can drastically reduce the number of free parameters in our theory. 
Assuming a certain flavor structure can be motivated from a UV model perspective, for instance if new heavy particles couple predominantly to top quarks or the third generation. 

In this subsection, we use an explicit example to explore how the assumption of flavor symmetries reduces the operator basis. 
Let us consider the operator
\begin{align}
    O_{\substack{Hu\\ij}} &= \left( H^\dagger i \overleftrightarrow{D}_\mu H \right) \left( \bar{u}_i \gamma^\mu u_j \right) \, ,
\end{align}
with $( H^\dagger i \overleftrightarrow{D}_\mu H ) = i (H^\dagger D_\mu H - (D_\mu H)^\dagger H)$. 
The operator $O_{Hu}$ is hermitian
\begin{align}
    O_{\substack{Hu\\ij}}^\dagger = -i (H^\dagger D_\mu H - (D_\mu H)^\dagger H) \left( \bar{u}_j \gamma^\mu u_i \right) = O_{\substack{Hu\\ ji}} \, .
\end{align}
For a single fermion generation, there is thus only one independent Wilson coefficient. 
For three fermion generations, the Wilson coefficients need to fulfill $C_{ij} = C_{ji}^\dagger$. Therefore, the diagonal elements have to be real
\begin{align}
C_{ij} \sim 
\begin{pmatrix}
r & c & c \\
& r & c \\
& & r
\end{pmatrix}
\, ,
\end{align}
where $r$ and $c$ stand for real and complex entries of the Wilson coefficients and the lower left triangle is defined by $C_{ij} = C_{ji}^\dagger$. This leaves us with three real and three complex coefficients and a total number of nine (six real and 3 imaginary) independent Wilson coefficients for this operator structure. 

From experimental data, we know that there can only be a small amount of flavor violation~\cite{Calibbi:2017uvl,Silvestrini:2018dos}. 
Therefore, in phenomenological studies we often reduce  the number of Wilson coefficients by employing flavor symmetries. 
The strictest assumption we can make is a $U(3)^5$ symmetry, i.e.\ a $U(3)$ symmetry for each of the fermion fields
\begin{align}
    U(3)^5 &= U(3)_q \times U(3)_u \times U(3)_d \times U(3)_l \times U(3)_e \, .
\end{align}
Under the $U(3)$ transformation, each of the fermion fields $f$ is transformed according to
\begin{align}
    f \to \Omega_f f \, , \qquad \Omega_f \subset U(3), \qquad \text{where } \qquad \Omega_f^\dagger \Omega_f = \mathbb{1} \, .
\end{align}
Let us now consider how the fermion bilinear $\bar{u}_i \gamma_\mu u_j$ with arbitrary $i,j$ behaves under this transformation
\begin{align}
    \bar{u}_i \gamma_\mu u_j &\to (\bar{u} \Omega_u^\dagger)_i \gamma_\mu (\Omega_u u)_j \nonumber \\
    &= \bar{u}_a \Omega_{u,ai}^\dagger \gamma_\mu \Omega_{u,jb} u_b \nonumber  \\
    &= (\bar{u}_a \gamma_\mu u_b)\, \Omega_{u,ai}^\dagger \Omega_{u,jb} \, .
\end{align}
The fermion bilinear is not generally invariant under the $\Omega_u$ transformation. However, the sum over the diagonal component is, as we can see after using $\Omega_u^\dagger \Omega_u = \mathbb{1}$
\begin{align}
   \delta^{ij} \bar{u}_i \gamma_\mu u_j & \to (\bar{u}_a \gamma_\mu u_b)\, \Omega_{u,ai}^\dagger \Omega_{u,ib} \nonumber \\
   &= (\bar{u}_a \gamma_\mu u_b) \delta^{a b} \, .
\end{align}
As only the combination $\delta^{ij} \bar{u}_i \gamma_\mu u_j$ is invariant under the $U(3)$ transformation, we only have a single independent Wilson coefficient for operators including this fermion bilinear. 

Bilinears with two different fermion fields are not $U(3)^5$ symmetric 
\begin{align}
    \bar{q}_i d_j \to (\bar{q} \Omega_q^\dagger)_i (\Omega_d d)_j \, .
    \label{eq:qd_bilinear}
\end{align}
Therefore, operators containing two different fermion fields within the same bilinear are completely forbidden under this symmetry. 

Another popular choice for a flavor assumption employed in phenomenological studies is Minimal Flavor Violation~(MFV)~\cite{Gerard:1982mm,Chivukula:1987py,Hall:1990ac,DAmbrosio:2002vsn}. Under the MFV assumption the SM Yukawa couplings are promoted to spurions of the $U(3)^5$ breaking. 
We assign the following transformation to the Yukawas
\begin{align}
    Y_f \to \Omega_f Y_f \Omega_f^\dagger \, .
\end{align}
With this transformation, we now find for a bilinear equivalent to \cref{eq:qd_bilinear}
\begin{align}
    \bar{q}_i Y_{d,ij} d_j &\to (\bar{q}_a \Omega_{q,ai}^\dagger) ( \Omega_{q,im} Y_{d,mn} \Omega_{d,nj} ) (\Omega_{d,jl} d_l ) \nonumber \\
    &= \bar{q}_a  (\Omega_{q,ai}^\dagger  \Omega_{q,im} ) Y_{d,mn} (\Omega_{d,nj} \Omega_{d,jl}) d_l \nonumber \\
    & = \bar{q}_a Y_{d,al} d_l \, .
\end{align}
Operators now require $Y_f$ insertions to be invariant under MFV. Since $Y_d, Y_e \ll 1$, the operators requiring insertion of the down-type or lepton Yukawa couplings are naturally suppressed with respect to operators invariant without the insertion of a Yukawa coupling. 

Another popular flavor assumption is $U(2)^3 \times U(3)^2$~\cite{Barbieri:2012uh}, where we assume a different behavior for the third quark generation. 

%------------------------------------
\subsection{Redefinitions of SM Fields and Parameters}
\label{sec:SMEFT_fieldRedef}
%------------------------------------
SMEFT operators induce corrections to the theoretical predictions for physical observables and can therefore affect collider phenomenology. In the EW broken phase, dimension-six operators modify the relations between the fields and parameters of the SM Lagrangian and the corresponding physical observables. Before discussing the phenomenological implications of these corrections, it is therefore useful to examine these field and parameter redefinitions arising from the SMEFT Lagrangian truncated at dimension six. In particular, these lead to shifts in the SM fields, the vacuum expectation value (vev), the gauge and Yukawa couplings at ${\mathcal{O}}(1/\Lambda^{2})$~\cite{Alonso:2013hga,Dedes:2017zog,Brivio:2020onw}. These redefinitions introduce additional modifications alongside the direct SMEFT corrections to SM predictions for physical processes.  Starting with the scalar sector, the SM scalar potential is modified with the operator of class $H^6$ described as
\begin{align}
	V(H) = -\mu_H^2 \, H^\dag H +\lambda_H (H^\dag H)^2 - \frac{C_H}{\Lambda^2}(H^\dag H)^3 \,.
\end{align} 
The presence of $O_{H}$ shifts the minimum of the potential to the true vev $v_{T}$
\begin{align}
	 \frac{v_T^2}{2} &\equiv \frac{v^2}{2}\left[1+\frac{3}{4\lambda_{H}}\bar{C}_{H} + \mathcal{O}\Big(\frac{1}{\Lambda^{4}}\Big)\right]\,
\end{align}
where $v=\sqrt{\mu_{H}^{2}/\lambda_{H}}$ is the SM vev and we define $\bar{C}_i=C_i v_T^2/\Lambda^2$.
With the presence of the operators of class $H^4D^2$, the scalar kinetic term receives modifications from 
\begin{align}\label{eq:SMEFThiggskin}
	{\cal L}_{\text{SMEFT}} &\supset (D_\mu H^\dagger)(D^\mu H) + \frac{C_{H \Box}}{\Lambda^2} \left( H^\dagger H \right) \Box \left( H^\dagger H \right) + \frac{C_{HD}}{\Lambda^2} \left( H^\dagger D^\mu H \right)^* \left( H^\dagger D_\mu H \right)\,.
\end{align}
In the broken phase, these modify the kinetic term of the physical Higgs boson as
\begin{align}
 \frac{1}{2} \partial_\mu h \partial^\mu h \to \frac{1}{2} \partial_\mu h \partial^\mu h\left(
1-2\, \bar{C}_{H,\textrm{kin}}\right)
\quad\textrm{with} \quad \bar{C}_{H,\textrm{kin}} = \Big(\bar{C}_{H \Box}-\frac{\bar{C}_{HD}}{4}\Big)\,,
\end{align}
To restore the canonical normalization of the kinetic term, we can redefine the physical Higgs field as $h \to h \big(1+\bar{C}_{H,\textrm{kin}}\big)$. This redefinition consequently induces modifications to all Higgs-related couplings. By taking these modifications into account and working in the unitary gauge for simplicity, the scalar doublet $H$ takes the form 
\begin{align}
H =\frac{1}{\sqrt{2}} \begin{pmatrix}
	0 \\
	v_{T}+(1+\bar{C}_{H,\textrm{kin}})h
\end{pmatrix}\,,
\end{align}
and the modified potential yields the corrected Higgs boson mass as
\begin{align}
m_H^2 &= 2\lambda_H v_T^2 \left(1-\frac{3v^2}{2 \lambda_H}\bar{C}_H + 2 \bar{C}_{H,\textrm{kin}} \right)\,.
\end{align}
In the gauge sector, the gauge fields and couplings are modified by operators belonging to the class $H^2X^2$. Before identifying the normalized physical gauge bosons, these modifications must be removed through appropriate field redefinitions. The relevant dimension-six operators are
\begin{align}
	{\cal L}_{\text{SMEFT}} &\supset \frac{C_{HG}}{\Lambda^2} H^\dagger H G_{\mu \nu}^A G^{A\mu \nu} + \frac{C_{HW}}{\Lambda^2} H^\dagger H W_{\mu \nu}^I W^{I \mu \nu}  + \frac{C_{HB}}{\Lambda^2} H^\dagger H B_{\mu \nu} B^{\mu \nu} + \frac{C_{HWB}}{\Lambda^2} H^\dagger \tau^I H W^I_{\mu \nu} B^{\mu \nu} \,.
\end{align}
These operators modify the gauge boson kinetic terms and consequently require redefinitions of the corresponding gauge fields. In the broken phase, the corrections to the SM gauge kinetic terms are
\begin{align}
\label{eq:SMEFTgaugekin}
{\cal L}_{\text{SMEFT}} &\supset 
-\frac{1}{4}W^I_{\mu\nu} W^{I\mu\nu} \left(1-2\bar{C}_{HW}\right) 
-\frac{1}{4}B_{\mu\nu} B^{\mu\nu} \left(1-2\bar{C}_{HB}\right)
-\frac{1}{4}G^a_{\mu\nu} G^{a\mu\nu} \left(1-2\bar{C}_{HG}\right)
-\frac{\bar{C}_{HWB}}{2} W_{\mu \nu}^3 B^{\mu \nu} 
\, .
\end{align}
The diagonal corrections are absorbed through the following field redefinitions
\begin{align}\label{eq:SMEFTbosonredef}
W^I_\mu  &\to  W^I_\mu \left(1 + \bar{C}_{HW} \right), &
B_\mu  &\to  B_\mu \left(1 + \bar{C}_{HB} \right), &
G_\mu^A &\to G_\mu^A \left(1 + \bar{C}_{HG}\right)\,.
\end{align}
The gauge fields enter the covariant derivative in combination with their corresponding gauge couplings, see~\cref{eq:covderivative}. To preserve the form of the covariant derivative under the field rescalings, the gauge couplings are redefined inversely as
\begin{align}\label{eq:SMEFTgauge}
g_{W} &\to  g_{W} \left(1 - \bar{C}_{HW} \right), &
g_{Y}  &\to  g_{Y} \left(1 - \bar{C}_{HB} \right), &
g_s &\to g_s \left(1 -\bar{C}_{HG} \right)\,.
\end{align}
These shifts induced by $O_{HW},O_{HB},O_{HG}$ are therefore absorbed into the gauge fields and couplings and have no further impact on the gauge sector. They only contribute to physical processes through the  modifications of Higgs-gauge interactions.
In the neutral EW sector, $O_{HWB}$ induces off-diagonal kinetic mixing between $W^{3}_{\mu}$ and $B_{\mu}$ as shown in~\cref{eq:SMEFTgaugekin}. This additional off-diagonal mixing is removed by the non-orthogonal transformation
\begin{align} \label{eq:SMEFTneutralkin}
\begin{pmatrix}W^3_\mu \\ B_\mu \end{pmatrix} 
\longrightarrow
\begin{pmatrix}1& - \bar{C}_{HWB}/2 \\  - \bar{C}_{HWB}/2 & 1\end{pmatrix}
\begin{pmatrix}W^3_\mu \\ B_\mu \end{pmatrix}\,.
\end{align}
Together,~\cref{eq:SMEFTbosonredef,eq:SMEFTgauge,eq:SMEFTneutralkin} bring the gauge kinetic terms into canonical and diagonal form up to corrections of $\mathcal{O}(1/\Lambda^4)$. The electric-charge eigenstates $W^\pm$ are defined in the usual way 
\begin{align}
W^{\pm}_{\mu}= \frac{1}{\sqrt{2}}(W^1_{\mu}\mp i W^{2}_{\mu}) \,.
\end{align}
After applying the field and gauge coupling redefinitions, the gauge boson mass terms arising from the SM Higgs kinetic term and $H^4D^2$ operators, see~\cref{eq:SMEFThiggskin}, take the form
\begin{align}
{\cal L}_{\text{SMEFT}} &\supset  \frac14 g_W^2 v_T^2  W_\mu^+ W^{-\,\mu} + \frac{1}{8} v_T^2  (g_W W^3_\mu -g_Y B_\mu)^2 \Big(1+\frac{1}{2} \bar{C}_{H D}\Big)+ \frac{1}{8} v_T^2 \bar{C}_{HWB}(g_W W^3_\mu -g_Y B_\mu)(g_Y W^3_\mu -g_W B_\mu)\,,
\end{align}
where the terms appearing at $\mathcal{O}(1/\Lambda^4)$ are neglected. For the neutral gauge bosons, the mass matrix is diagonalized using the $2\times2 $ rotation matrix by introducing the physical $Z$ and photon fields as
\begin{align}
\begin{pmatrix}W_\mu^3 \\ B_\mu\end{pmatrix} 
=
\begin{pmatrix}c_w& s_w\\ - s_w& c_w\end{pmatrix}
\begin{pmatrix}Z_\mu \\ A_\mu\end{pmatrix}\,,
\end{align}
where $s_w = \sin \theta_w$ and $c_w$ denote the sine and cosine of the weak mixing angle $\theta_w$, respectively, which receives $\mathcal{O}(1/\Lambda^2)$ corrections as
\begin{align}
t_w = \frac{g_Y}{g_W} + \frac12\bar{C}_{HWB} \left(1-\frac{g_Y^2}{g_W^2}\right)\,,
\end{align} 
with $t_w = \tan \theta_w$.
The resulting gauge boson masses up to corrections of $\mathcal{O}(1/\Lambda^4)$ are 
\begin{align}
M_W^2 &= \frac{g_W^2 v_T^2}{4} , \nonumber \\
M_Z^2 &= \frac{v_T^2}{4}\Big[(g_{Y}^2+g_W^2)\Big(1+\frac{1}{2} \bar{C}_{HD}\Big)+2 g_{Y}g_{W} \bar{C}_{HWB}\Big].
\end{align}
In the fermion sector, the mass matrices and Higgs-fermion Yukawa couplings are modified by operators of the class $\Psi^2H^3$. The relevant terms in the SMEFT Lagrangian are 
\begin{align}
	{\cal L}_{\text{SMEFT}} &\supset -\Big[ y_{d,i j} \, \overline d_i\,  H^{\dagger k} q_{k j} +  y_{u,i j}\, \overline u_{i}\, \widetilde H^{\dagger k} q_{k j} + y_{e,i j} \, \overline e_{i}\, H^{\dagger k}l_{k j} + \hbox{h.c.}\Big] \nonumber \\
&+\frac{1}{\Lambda^2} \left[ C^{\dagger}_{\substack{dH\\ij}} \left( H^\dagger H \right)\overline{d_i}\ H^{\dagger k} \ q_{k j} + C^{\dagger}_{\substack{uH\\ij}} \left( H^\dagger H \right) \overline{u}_{i}\ \tilde H^{\dagger k} \ q_{k j} + C^{\dagger}_{\substack{eH\\ij}} \left( H^\dagger H \right) \overline{e}_{i}\ H^{\dagger k} \ l_{k j} + \hbox{h.c.} \right]\, .
\end{align}
Here, the Yukawa couplings and Wilson coefficients are treated as general $3\times3$ matrices in flavor space. Their structures can be further simplified by imposing the flavor assumptions discussed in~\cref{sec:SMEFT_flavorAssumptions}. In the general case, the fermion mass matrix generated in the broken theory at $\mathcal{O}(1/\Lambda^2)$ is
\begin{align}\label{eq:SMEFTfermionmass}
 M_{\psi,ij} &= \frac{v_T}{\sqrt 2} \left( y_{\psi,ij}  - \frac{1}{2}  \bar{C}^{\dagger}_{\substack{\psi H\\ ij}} \right)\,, \qquad \psi=u,d,e. 
\end{align}
The corresponding modification in the Yukawa matrices, denoted as $\mathcal{Y}_{\psi,ij}$, is given as
\begin{align}
\mathcal{Y}_{\psi,ij} &= \frac{1}{\sqrt 2}  y_{\psi,ij} \left[ 1+ \bar{C}_{H,\textrm{kin}} \right]  - \frac3{2 \sqrt 2} \bar{C}^{\dagger}_{\substack{\psi H\\ij}}.
\end{align}
Using~\cref{eq:SMEFTfermionmass} to express SM Yukawa matrices in terms of modified fermion mass matrices, the Higgs-fermion Yukawa modifications are expressed as 
\begin{align}\label{eq:SMEFTyuk}
\mathcal{Y}_{\psi,ij}& = \frac{1}{v_T} M_{\psi,ij} \left[ 1+ \bar{C}_{H,\textrm{kin}}  \right] - \frac{1}{\sqrt 2} \bar{C}^{\dagger}_{\substack{\psi H\\ij}}\,,
\qquad \psi=u,d,e.
\end{align}
In contrast to the SM, the SMEFT-corrected fermion mass and Yukawa matrices are not simultaneously diagonalizable by the same flavor rotation. Thus, even after transforming to the fermion mass basis, the Yukawa matrices can still retain off-diagonal entries and lead to flavor-changing Higgs interactions. 
As an illustrative example of how the flavor structure can be simplified, we consider the MFV assumption and take the Wilson coefficients to be real.
Under these assumptions, the operators take the Yukawa insertions as 
\begin{align}
C_{\substack{\psi H\\ij}}= c_{\psi H}\;y_{\psi,ij}
\end{align}
where $c_{\psi H}$ is a real flavor-singlet coefficient. The fermion mass matrix becomes 
\begin{align}
 M_{\psi,ij} &= \frac{v_T}{\sqrt 2} y_{\psi,ij}\left( 1  - \frac{1}{2}  \bar{c}_{\psi H} \right)\,, \qquad \psi=u,d,e. 
\end{align}
Using the above relation in~\cref{eq:SMEFTyuk}, the modified Yukawa coupling matrix is written as
\begin{align}
\mathcal{Y}_{\psi,ij}& = \frac{M_{\psi,ij}}{v_T} \left[ 1+ \bar{C}_{H,\textrm{kin}} -  \bar{c}_{\psi H}\right]\,,
\qquad \psi=u,d,e.
\end{align}
Thus, under the MFV assumption, the modified Higgs-fermion Yukawa coupling matrices remain proportional to the corresponding fermion mass matrices. Consequently, after transforming to the fermion-mass basis, the off-diagonal entries leading to flavor-changing Higgs interactions vanish.

%------------------------------------
\subsection{Electroweak Input Schemes}
\label{sec:SMEFT_inputSchemes}
%------------------------------------
\begin{figure*}[thb]
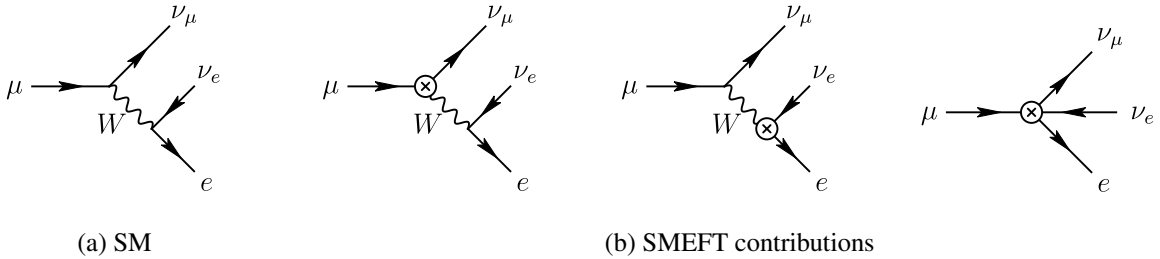

    \centering
    \include{figures/muondecay_2}
    \caption{Tree-level Feynman diagrams generating the decay of the muon (a) at dimension four and (b) dimension six.
    }
    \label{fig:fermi_constant}
\end{figure*}
In phenomenology, we relate observables to observables. The SM has $19$ free parameters which need to be determined by data. These parameters can be written as, for instance, 
$\alpha_s, \, \bar{\theta}$ for QCD, 
the couplings $g_W, g_Y$, the vacuum expectation value (vev) $v$ and mass $m_h$ of the Higgs boson for the EW sector, 
the Yukawa couplings $y_u, y_c, y_t, \, y_d, y_s, y_b , \, y_e, y_\mu, y_\tau$ 
and the CKM angles $\theta_{12}, \theta_{13}, \theta_{23}, \delta$.

There are several choices for determining the values of the EW input parameters experimentally. 
They are typically determined by three observables out of the set $\{ \hat{G}_F , \hat{M}_W, \hat{M}_Z, \hat{\alpha}, \hat{s}_w^\text{eff} \}$, which correspond to the Fermi constant, the masses of the $W$ and $Z$ bosons, the electromagnetic coupling constant and the effective weak mixing angle, respectively.
We use the hat to denote measured values rather than the Lagrangian parameters.
Once three parameters are chosen as inputs, the other two parameters are predicted from those three. For instance, in the SM the $W$-boson mass can be predicted from $\{ G_F , \alpha, M_Z\}$ as 
\begin{align}
    M_{W,\text{SM}}^2 &= \frac{M_Z^2}{2} \left[  1 + \sqrt{1- \frac{4 \pi \alpha }{M_Z^2 \sqrt{2} G_F} } \right]   \, .
    \label{eq:inputs_MW_pred}
\end{align}
SMEFT coefficients can enter the relation between a Lagrangian parameter and its measured value in two ways: through direct SMEFT contributions to the process used to measure the input, or through the field redefinitions discussed in \cref{sec:SMEFT_fieldRedef}.
As a result, the theory parameters entering, for instance, \cref{eq:inputs_MW_pred}, receive SMEFT corrections relative to their measured values.
As an example, SMEFT contributes to the decay rate (or lifetime) of the muon  which is used to define the Fermi constant, see \cref{fig:fermi_constant} 
\begin{align}
    \hat{G}_{F} = G_F \left(1 + \bar{C}_{\substack{H\ell \\ 11}}^{(3)} + \bar{C}_{\substack{H\ell \\ 22}}^{(3)} - \bar{C}_{\substack{\ell\ell \\ 1221}} \right) = G_F (1 + \Delta G_F) \, .
\end{align}
Therefore, the measured value $\hat{G}_F$ does not directly correspond to its Lagrangian parameter $G_F$, but includes a non-zero SMEFT contribution that must be accounted for when computing predictions as a function of the input parameters.
For instance, the prediction for $M_W$ in a scheme using $\{ \hat{G}_F , \hat{\alpha}, \hat{M}_Z\}$ as inputs receives corrections which can be calculated by replacing the Lagrangian parameters in \cref{eq:inputs_MW_pred} as $X \to \hat{X} (1 - \Delta X)$ and expanding to linear order in the $\Delta$s (or, equivalently, to the order $1/\Lambda^2$)
\begin{align}
    M_W^2 &= \frac{\hat{M}_Z^2(1- \Delta M_Z)^2}{2} \left[ 1+ \sqrt{1-\frac{4 \pi \hat{\alpha}(1 - \Delta \alpha) }{\hat{M}_Z^2(1 - \Delta M_Z)^2 \sqrt{2} \hat{G}_F( 1 - \Delta G_F)} } \right] \nonumber \\
    & = M_{W,\text{SM}}^2(\hat{G}_F, \hat{\alpha}, \hat{M}_Z) \left( 1 + \frac{s_w^2}{c_{2w}} (\Delta \alpha - \Delta G_F  ) - 2 \frac{c_w^2}{c_{2w}} \Delta M_Z\right) + \mathcal{O}(\Lambda^{-4}) \, ,
\end{align}
with $c_{2w} = \cos 2 \theta_w$. 
As discussed in \cref{sec:SMEFT_fieldRedef}, $C_{HD}$ enters $\Delta M_Z$, making the measurement of the $W$-boson mass an important constraint on the Wilson coefficient $C_{HD}$. 
The phenomenological consequences of this are discussed in the context of RGE effects in \cref{sec:SMEFT_higherOrder}.

The optimal choice of the EW input scheme in SMEFT is process-dependent. The scheme choice will influence the number of Wilson coefficients that a given prediction will depend on at LO and NLO~\cite{Biekotter:2023xle} and it can be beneficial to keep this number minimal.
Moreover, the accuracy of the measurements of the chosen inputs will influence the precision of the prediction.
For LHC SMEFT analyses,
the high accuracy of the $G_F$ measurement and the fact that the EW gauge boson masses generally appear in the denominator of propagators in LHC processes have made the $\{ \hat{G}_F , \hat{M}_W, \hat{M}_Z\}$ input scheme a very common scheme~\cite{Brivio:2020onw,Brivio:2021yjb}.

%------------------------------------------
\subsection{Higher Orders in SMEFT Predictions}
\label{sec:SMEFT_higherOrder}
\begin{figure*}[thb]
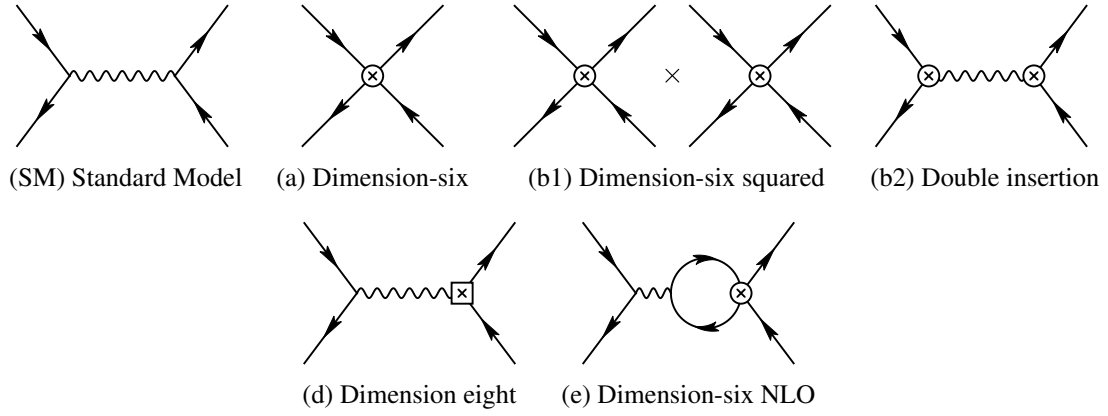

    \centering
    \include{figures/higherorders_2}
    \caption{Example diagrams for leading and higher-order contributions in the SMEFT. Crossed circles and square denote dimension-six and -eight operators, respectively. All diagrams except (b1) are understood to interfere with the leading SM contribution.}
    \label{fig:higher_orders}
\end{figure*}

%------------------------------------------
Higher-order corrections in SMEFT predictions follow two different expansions: the expansion in inverse powers of the new-physics scale $1/\Lambda$ and the expansion in SM loops. 
We can represent these two expansions of an observable $o$ as 
\begin{align}
    o = o_\text{SM} + a_i \frac{C_i^{(6)}}{\Lambda^2} + b_{ij} \frac{C_i^{(6)} C_j^{(6)}}{\Lambda^4} + d_k \frac{C_k^{(8)}}{\Lambda^4} + \frac{1}{16 \pi^2} \left[ e_m \frac{C_m^{(6)}}{\Lambda^2} + f_n \frac{C_n^{(6)}}{\Lambda^2} \log \left( \frac{\mu^2}{\Lambda^2} \right) \right] + \cdots \, ,
    \label{eq:higher_orders}
\end{align}
where a factor $1/(16\pi^2)$ has been factored out to highlight loop corrections. The logarithmic term refers to contributions from the one-loop RGE of the dimension-six operators~\cite{Jenkins:2013zja, Jenkins:2013wua, Alonso:2013hga}. 
A representative diagram for each contribution is shown in \cref{fig:higher_orders} with the subfigure caption identifiers matching the parameters in \cref{eq:higher_orders}.

We can see from the $\Lambda$ counting in \cref{eq:higher_orders} that quadratic SMEFT contributions from the interference of two diagrams with a single insertion of a SMEFT operator as well as those from double insertions in one diagram, are formally of the same order as the interference of a diagram with a single insertion of a dimension-eight operator with a SM diagram.  

Linear dimension-six SMEFT effects are suppressed by $\Lambda^{2}$ only and unsuppressed by loop factors. They will hence dominate for most observables. However, under special circumstances or in specific phase-space regimes higher-order corrections may be relevant.

\textbf{Dimension-six squared} contributions are suppressed by $\Lambda^4$ and thus typically small compared to linear dimension-six effects.  
However, in high-energy tails of kinematic distributions quadratic dimension-six pieces may dominate over the linear contributions. We can take four-fermion operator contributions to Drell-Yan production as an example. 
Due to the fact that the four-fermion contact interaction does not suffer from the same propagator suppression by the virtuality as the SM diagram at high energies, the dimension-six and dimension-six squared contributions, see diagrams (a) and (b1) of \cref{fig:higher_orders}, grow with the center-of-mass energy as $s/\Lambda^2$ and $s^2/\Lambda^4$, respectively. In energy-dependent kinematic distributions, the dimension-six squared contributions may hence dominate and one should carefully check the validity of the EFT approach~\cite{Contino:2016jqw,Keilmann:2019cbp,Allwicher:2024mzw}.
Different approaches have been proposed to ensure EFT validity in the analysis of kinematic distributions~\cite{Brivio:2022pyi}.

\textbf{Double-insertions} of dimension-six operators in a Feynman diagram interfering with a SM contribution are suppressed by $1/\Lambda^4$, just like single insertions of dimension-eight operators or dimension-six squared contributions. 
One-loop diagrams with double insertions of dimension-six operators generally require dimension-eight counterterms. 

\textbf{Dimension-eight} operators are formally suppressed by $1/\Lambda^4$ for a diagram with a single operator insertion interfering with a SM one. 
However, they may lead to the leading SMEFT effects in cases where dimension-six contributions are absent or accidentally small. This is the case, for instance, for neutral triple gauge couplings~\cite{Degrande:2013kka,Ellis:2019zex,Ellis:2022zdw} or quartic gauge interactions contributing to vector boson scattering (VBS)~\cite{Eboli:2016kko}. 
The inclusion of dimension-eight effects may also be relevant when matching UV models onto the SMEFT as certain properties of the full models may first be captured at dimension eight~\cite{Dawson:2022cmu,Banerjee:2022thk,Ellis:2023zim,Banerjee:2023qbg,Dawson:2021xei,Corbett:2021eux,Dawson:2024ozw,Cepedello:2024ogz}. 
At dimension-eight positivity and causality constraints apply~\cite{Adams:2006sv,Zhang:2018shp,Bellazzini:2018paj,Remmen:2019cyz}.

\textbf{SMEFT at NLO:} 
Dimension-six operators appearing in loops are suppressed by factors $16 \pi^2$ and additional powers of SM coupling constants $g_W$, $g_Y$, $g_s$ or the Yukawa couplings. 
While NLO SMEFT contributions are thus suppressed with respect to LO ones, their effect can still be relevant as they increase the number of Wilson coefficients contributing to a single observable. 
These extra degrees of freedom can be important in phenomenological analyses when interpreting bounds on single observables and performing global analyses (see also \cref{sec:SMEFT_globalFits}), as a larger set of Wilson coefficients could explain a potential deviation from the data. 
The calculation of NLO QCD SMEFT corrections has been automated in the tool \texttt{SMEFTatNLO}~\cite{Degrande:2020evl}. 
EW corrections have only been calculated on a case-by-case basis~\cite{Zhang:2013xya,Crivellin:2013hpa,Zhang:2014rja,Pruna:2014asa,Grober:2015cwa,Hartmann:2015oia,Ghezzi:2015vva,Hartmann:2015aia,Aebischer:2015fzz,Zhang:2016omx,BessidskaiaBylund:2016jvp,Maltoni:2016yxb,Degrande:2016dqg,Hartmann:2016pil,Grazzini:2016paz,deFlorian:2017qfk,Deutschmann:2017qum,Baglio:2017bfe,Dawson:2018pyl,Degrande:2018fog,Vryonidou:2018eyv,Dedes:2018seb,Grazzini:2018eyk,Dawson:2018liq,Dawson:2018jlg,Dawson:2018dxp,Neumann:2019kvk,Dedes:2019bew,Boughezal:2019xpp,Dawson:2019clf,Baglio:2019uty,Haisch:2020ahr,Dittmaier:2021fls,Dawson:2021ofa,Boughezal:2021tih,Battaglia:2021nys,Kley:2021yhn,Faham:2021zet,Haisch:2022nwz,Heinrich:2022idm,Asteriadis:2022ras,Bellafronte:2023amz,Kidonakis:2023htm,Gauld:2023gtb,Heinrich:2023rsd,Asteriadis:2024xuk,Asteriadis:2024xts,Dawson:2024pft,ElFaham:2024egs,Bellafronte:2025jbk,Biekotter:2025nln}.

\begin{figure}[t]
    \centering
    \includegraphics[scale=.5]{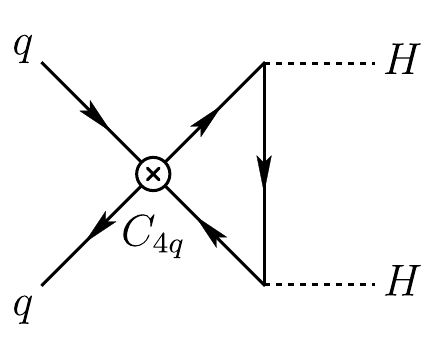}
    \hspace{1.2cm}
    \includegraphics[scale=.5]{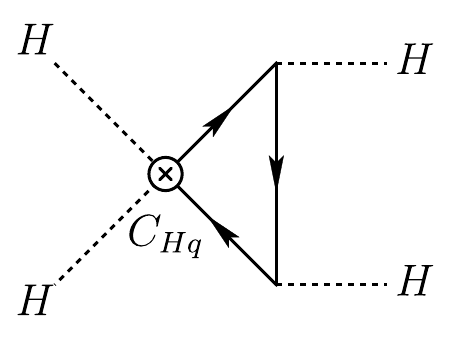}
    \caption{Logarithmically divergent diagrams contributing to the RG mixing of $C_{4q} \to C_{Hq}$ and $C_{Hq} \to C_{HD}$, respectively. }
    \label{fig:four_top_mixing}
\end{figure}

\textbf{RGE:} 
EFTs are generally non-renormalizable as they contain not only an infinite number of operators, but, as a result, also an infinite number of divergences. However, EFTs are renormalizable order-by-order in their power counting, meaning that the renormalization of the dimension-six SMEFT only requires up to dimension-six counterterms. 
The one-loop SMEFT RGE at dimension-six~\cite{Jenkins:2013zja, Jenkins:2013wua, Alonso:2013hga} is available in various computing languages via the tools~\texttt{DsixTools}~\cite{Celis:2017hod,Fuentes-Martin:2020zaz}, \texttt{RGESolver}~\cite{DiNoi:2022ejg} and \texttt{wilson}~\cite{Aebischer:2018bkb}.
We have seen in \cref{sec:resummation} that the Wilson coefficients of an EFT generally mix under the RGE. 
This leads to important effects also for the SMEFT where UV-complete models are typically matched at a high scale $\Lambda$ and we subsequently run down to the appropriate scale(s) for experimental observables. 
Through the RG mixing observables are sensitive to high-scale Wilson coefficients $C_i(\Lambda)$ which do not directly contribute to a corresponding process at some low scale $\mu$.
As an example, it has been shown that the Wilson coefficients of four-top operators receive sizable constraints from EW precision observables through the RG mixing~\cite{Stefanek:2024kds}. 
The four-top Wilson coefficients $C_{\substack{qq\\3333}}^{(1)}, C_{\substack{qq\\3333}}^{(3)}, C_{\substack{qu\\3333}}^{(1)}, C_{\substack{uu\\3333}}$, which we collectively refer to as $C_{4q}$,  mix into $C_{\substack{Hq\\33}}^{(1)}, C_{\substack{Hq\\33}}^{(3)}, C_{\substack{Hu\\33}}$, collectively referred to as $C_{Hq}$, through the logarithmically divergent diagrams shown in the left diagram of \cref{fig:four_top_mixing}. 
The first leading-log (LL) mixing with $C_{Hu}$ and the subsequent mixing of the $C_{Hq}$ coefficients with $C_{HD}$ are given as
\begin{align}
    \left[C_{\substack{Hu\\33}} \right]_{LL} &= \frac{y_t^2}{16 \pi^2} \left[ N_c C_{\substack{qu\\3333}}^{(1)} - 2 (1+N_c) C_{\substack{uu\\3333}} \right] \log \left(  \frac{\mu^2}{\Lambda^2} \right)\, , \nonumber \\
    \left[C_{HD} \right]_{LL} &= \frac{N_c y_t^2}{4 \pi^2} \left[  C_{\substack{Hq\\33}}^{(1)} - C_{\substack{Hu\\33}} \right] \log \left(  \frac{\mu^2}{\Lambda^2} \right)\, ,
\end{align}
where $N_c$ is the number of colors.
The Wilson coefficient $C_{HD}$ is tightly constrained by the measurement of the $W$-boson mass as discussed in \cref{sec:SMEFT_inputSchemes}. As a result, four-top Wilson coefficients receive sizable constraints from this EW precision measurement, even though four-top operators only enter at the two-loop level. 

Recently, the NLO SMEFT RGEs at dimension six have been computed in Ref.~\cite{Born:2026xkr} with partial results published in~\cite{Aebischer:2022anv,Born:2024mgz,DiNoi:2024ajj,Duhr:2025zqw,Haisch:2025lvd,DiNoi:2025arz,DiNoi:2025tka,Haisch:2025vqj,Duhr:2025yor}.
Two-loop RGEs for dimension-five SMEFT have been provided in Ref.~\cite{Ibarra:2024tpt}.

%------------------------------------------
\subsection{Global Analyses}
\label{sec:SMEFT_globalFits}
%------------------------------------------

\begin{figure}[thb]
    \include{figures/globalanalysis_2}
        \caption{Example Feynman diagrams for (a) a process receiving corrections from various dimension-six operators and (b) for an operator entering various processes. } 
        \label{fig:Zqqcoupling}
\end{figure}

The combined analyses of multiple SMEFT Wilson coefficients at the same time, so-called global analyses, have become a standard tool in the bottom-up search for NP at the LHC and beyond. 
Global analyses are typically global both in the Wilson coefficients as well as in the datasets included. 
Being global in the Wilson coefficients is crucial to account for the fact that in a single observable the effects of two Wilson coefficients could (partially) compensate each other. 
For instance, in the top panel of \cref{fig:Zqqcoupling} we display the Feynman diagrams of different modifications of the weak boson fusion~(WBF) process by dimension-six SMEFT operators. The effect of a modified $Vqq^\prime$ coupling could be compensated by another operator affecting the $VV^{(\prime)}h$ couplings or the considered Higgs decay. 

Being global in the dataset and including as many observables as possible is important to respect all constraints on a given Wilson coefficient and to probe different directions in Wilson-coefficient parameter space. For instance, a modified $Zq\bar{q}$ (represented by the blue blob in the bottom panel of \cref{fig:Zqqcoupling}) coupling could be probed in WBF and $Vh$ Higgs production at the LHC as well as via the decay of a $Z$ boson to quarks produced at LEP. 

Global SMEFT analyses allow us to combine experimental constraints on potential directions of NP and they can easily be reinterpreted to assess the status of concrete NP extensions. 
Global SMEFT analyses have been performed for low-energy~\cite{Falkowski:2017pss,Falkowski:2019xoe,Falkowski:2023klj}, flavor~\cite{Aoude:2020dwv,Bruggisser:2021duo,Bruggisser:2022rhb,Grunwald:2023nli}, EW~\cite{Biekotter:2018ohn,Kraml:2019sis,Dawson:2020oco,Almeida:2021asy,Anisha:2021hgc} and top~\cite{Buckley:2015lku,Aguilar-Saavedra:2018ksv,Brivio:2019ius,Bissmann:2019gfc,Durieux:2019rbz}  data as well as combinations thereof~\cite{Ellis:2020unq,Ethier:2021bye,Garosi:2023yxg,Bartocci:2023nvp,Celada:2024mcf,deBlas:2025xhe}. 
The impact of future colliders has been studied in~\cite{deBlas:2022ofj,Celada:2026ubm}. 
Comprehensive SMEFT fits currently include around 50 Wilson coefficients and differ in their EW input schemes, datasets included, higher-order corrections included (e.g.~ dimension-six squared terms) and flavor assumptions. 

%----------------------------
\subsection{Matching and UV Model Connections }
\label{sec:SMEFT_UV_connection}
%----------------------------
One of the key advantages of presenting indirect searches for NP within the SMEFT framework is that the resulting constraints on the Wilson coefficients can be systematically reinterpreted in terms of UV-complete models. A crucial ingredient in establishing this connection is provided by automated matching tools, which determine the relations between the parameters of a UV model and the corresponding SMEFT Wilson coefficients. Several public tools have been developed to automate tree-level and one-loop matching calculations using complementary diagrammatic and functional approaches~\cite{DasBakshi:2018vni,Carmona:2021xtq,Fuentes-Martin:2022jrf}. 

While automated matching was initially demonstrated mostly for relatively simple extensions of the SM, recent developments have considerably expanded the complexity of UV theories that can be treated. In particular, the complete one-loop matching of the general $R$-parity-conserving Minimal Supersymmetric Standard Model (MSSM), involving $\mathcal{O}(100)$ parameters, onto the dimension-six SMEFT using \texttt{Matchete} provides an important proof of principle~\cite{Kraml:2025fpv}. The calculation simultaneously integrates out the full supersymmetric spectrum with non-degenerate masses while retaining a general flavor structure, demonstrating that automated tools can now be applied to realistic multi-scale UV models rather than only simplified toy scenarios. 

A further development is the interface between matching calculations and SMEFT global analyses, providing a more complete interpretation of experimental data. For example, \texttt{match2fit}~\cite{terHoeve:2023pvs} provides an interface between \texttt{MatchMakerEFT}~\cite{Carmona:2021xtq} and \texttt{SMEFiT}~\cite{Ethier:2021bye,Giani:2023gfq}, allowing constraints obtained from global SMEFT fits to be directly translated into bounds on the masses and couplings of UV models. Such developments provide a systematic pipeline connecting UV theories to SMEFT Wilson coefficients and, ultimately, to experimental measurements.

As a simple illustration of matching to a UV scenario, we consider an extension of the SM by a heavy real gauge-singlet scalar and derive its low-energy SMEFT description in the Warsaw basis. The  scalar field~$\phi$ is a singlet under the SM gauge group, and its Lagrangian in the unbroken phase is given by
\begin{align}\label{eq:xSM}
	\mathcal{L} = \mathcal{L}_\text{SM} + \frac{1}{2}\left(\partial_\mu\phi\right)^2 - \frac{1}{2}M_{\phi}^2\phi^2 - A_{\phi}(H^{\dagger}H)\phi -\frac{1}{2}\kappa_{\phi} (H^{\dagger}H) \phi^2 - \frac{1}{3!} \mu_{\phi} \phi^3 - \frac{1}{4!} \lambda_\phi \phi^4\,,
\end{align}
where $M_\phi$ denotes the mass of the heavy scalar, $A_\phi$ and $\kappa_\phi$ denote the Higgs-portal interactions, and $\mu_\phi$ and $\lambda_\phi$ describe the singlet self-interactions.
For $M_\phi \gg v$, the field $\phi$ can be integrated out and its low-energy effects are encoded in SMEFT Wilson coefficients. 
At tree-level, the matching  can be performed straightforwardly by solving the classical EOM for $\phi$,
\begin{align}
	\left(\Box + M_\phi^2 + \kappa_\phi (H^{\dagger}H)\right)\phi + A_{\phi}(H^{\dagger}H)+ \frac{\mu_{\phi}}{2}\phi^2	+ \frac{\lambda_\phi}{3!}\phi^3=0\,.
\end{align}
The classical solution $\phi_c$ can be obtained by solving the EOM perturbatively by expanding $\phi_c=\phi^{(0)}_c+\phi^{(1)}_c+\cdots$, where
\begin{align}
	 \phi^{(0)}_c&=-\frac{1}{\left(\Box + M_\phi^2 + \kappa_\phi(H^{\dagger}H)\right)}
	A_\phi (H^{\dagger}H)\,, \nonumber \\
\phi_c^{(1)}&=
-\frac{1}{\left(\Box + M_\phi^2 + \kappa_\phi(H^{\dagger}H)\right)}
\left[
\frac{\mu_\phi}{2}\left(\phi_c^{(0)}\right)^2
+\frac{\lambda_\phi}{3!}\left(\phi_c^{(0)}\right)^3
\right]\,.
\end{align}
In the low-energy regime, the denominator can be expanded in inverse powers of $M_{\phi}$ as
\begin{align}
\frac{1}{\Box + M_\phi^2 + \kappa_\phi(H^{\dagger}H)}= \frac{1}{M_{\phi}^2}\left[1-\frac{\Box+\kappa_\phi(H^{\dagger}H)}{M_{\phi}^2}+\frac{\left(\Box+\kappa_\phi(H^{\dagger}H)\right)^2}{M_{\phi}^4}+\cdots\right]\,.
\end{align}
The classical solution then becomes
\begin{align}
	\phi_c \simeq -\frac{A_\phi}{M_\phi^2}  (H^{\dagger}H) +\frac{A_\phi}{M_\phi^4}\Box (H^{\dagger}H)+\left(\frac{\kappa_\phi A_\phi}{M_\phi^4}-\frac{\mu_{\phi} A_\phi^2}{2M_\phi^6}	\right)(H^{\dagger}H)^2+\cdots \,,
\end{align}
and substituting $\phi_c$ back into the UV Lagrangian in~\cref{eq:xSM}, we obtain the tree-level effective Lagrangian:
\begin{align}\label{eq:treeEFT_xSM}
	\mathcal{L}^{\textrm{tree}}_\textrm{ EFT}&=	\mathcal{L}_{\rm SM}-\frac{1}{2}\phi_c\left(\Box+M_\phi^2+\kappa_\phi(H^{\dagger}H)\right)\phi_c-A_\phi(H^{\dagger}H)\phi_c-\frac{\mu_\phi}{3!}\phi_c^3-\frac{\lambda_\phi}{4!}\phi_c^4 \, ,\nonumber \\
	&=  \mathcal{L}_{\rm SM}+\frac{A_\phi^2}{2M_\phi^2}(H^{\dagger}H)^2-\frac{A_\phi^2}{2M_\phi^4}(H^{\dagger}H)\Box(H^{\dagger}H) +\left(	-\frac{\kappa_\phi A_\phi^2}{2M_\phi^4}+\frac{\mu_\phi A_\phi^3}{6M_\phi^6}\right)(H^{\dagger}H)^3+\cdots \, .
\end{align}
Identifying the resulting structures with the corresponding Warsaw-basis operators and using the notation of \cref{eq:SMEFTlag}, the tree-level SMEFT Lagrangian is written as 
\begin{align}\label{eq:treeSMEFT_xSM}
	\mathcal{L}^{\textrm{tree}}_\textrm{ SMEFT}&=  \mathcal{L}_{\rm SM}+\frac{A_\phi^2}{2M_\phi^2}(H^{\dagger}H)^2+\frac{C_{H\Box}}{M_\phi^2}O_{H\Box} +\frac{C_{H}}{M_\phi^2}O_{H}.
\end{align}
Here, we identify the SMEFT cutoff scale with the heavy singlet mass, $\Lambda=M_{\phi}$. Thus, two dimension-six SMEFT operators $O_{H}$ and $O_{H\Box}$ are generated at tree-level with Wilson coefficients
\begin{align}
    C_{H\Box}=-\frac{A_\phi^2}{2M_\phi^2}\,, \qquad 
    C_{H}
    =-\frac{\kappa_\phi A_\phi^2}{2M_\phi^2}+\frac{\mu_\phi A_\phi^3}{6M_\phi^4}\,.
	\end{align}
In addition, integrating out $\phi$ generates a shift in the Higgs quartic coupling, $\lambda_{H} \to \lambda_{H}^{\prime}=\lambda_{H}+\delta\lambda_{H}$ through the second term in \cref{eq:treeSMEFT_xSM} as
\begin{align}
	\Delta \lambda_{H}=
	-\frac{A_\phi^2}{2M_\phi^2}\,.
\end{align}
At one-loop level, integrating out the heavy singlet $\phi$ is more involved and generates a considerably larger set of dimension-six SMEFT operators. 
In addition to the one-loop contributions to $C_H$ and $C_{H\Box}$, 23 additional Warsaw basis operators arise for the first time at one-loop. These include purely bosonic, Yukawa-like, Higgs-fermion current and four-fermion operators:
\begin{equation}
	\begin{gathered}
    	O_{HB},\, O_{HW},\, O_{HWB},\, O_{HD},\,
		O_{eH},\, O_{uH},\, O_{dH}, 
		O_{Hl}^{(1)}, \, O_{Hl}^{(3)},\, O_{Hq}^{(1)},\, O_{Hq}^{(3)},\,
		O_{He},\, O_{Hu},\, O_{Hd},\, O_{Hud},\\
        O_{le},\,
        O_{ledq},\, O_{lequ}^{(1)}
        ,\, O_{qd}^{(1)},\,  O_{qu}^{(1)},\, \, O_{qd}^{(8)},\, O_{qu}^{(8)}, \, O_{quqd}^{(1)} \, .
	\end{gathered}
\end{equation}
The one-loop matching of $\phi$ to the dimension-six SMEFT has been studied using both functional and diagrammatic approaches~\cite{Ellis:2017jns,Jiang:2018pbd,Haisch:2020ahr}. The one-loop matching results for the non-zero Wilson coefficients are given in Refs.~\cite{Jiang:2018pbd,Haisch:2020ahr}.
In these calculations, the one-loop matching is performed off-shell, leading initially to a redundant set of local operator structures organized in a Green’s basis. 
These structures are subsequently reduced using IBP identities and appropriate field redefinitions to obtain the Wilson coefficients in the Warsaw basis.
This UV model is also used as a benchmark for automated one-loop matching tools such as \texttt{Matchete} and \texttt{MatchMakerEFT}.
The one-loop Wilson coefficients receive contributions from both purely heavy loops, containing only the heavy $\phi$ and mixed heavy-light loops involving heavy $\phi$ together with SM fields. 
The latter contributions generate operators involving gauge and fermion fields at one loop, even though $\phi$ is a singlet under the SM gauge group. The one-loop matching of $C_H$ and $C_{H\Box}$, both of which are already generated at tree-level, receives additional contributions. 
Since their tree-level Wilson coefficients depend explicitly on the UV parameters $A_\phi,\kappa_\phi,\mu_\phi$, their one-loop matching also requires the consistent renormalization of these parameters, which gives rise to logarithmic dependence on the matching scale. Higgs wave-function renormalization additionally contributes to the one-loop corrections to these coefficients. 

If an unbroken $Z_{2}$ symmetry, $\phi \to -\phi$, is imposed, then $A_{\phi}=\mu_{\phi}=0$, and the tree-level dimension-six contributions vanish while non-zero one-loop contributions are still generated through the remaining portal coupling $\kappa_{\phi}$.

This example demonstrates how rapidly matching becomes cumbersome beyond the tree level. Integrating out a single heavy real singlet scalar generates only $O_H$ and $O_{H\Box}$ at tree-level, whereas the one-loop matching induces a much broader set of SMEFT operators. 
The enlarged operator structure and the presence of mixed heavy-light contributions show the practical advantages of automated tools for systematic one-loop matching calculations.
The SMEFT framework and its automated tools enable us to quickly compare complex BSM theories with experimental data and, in the presence of an anomaly in the data, pinpoint its origin.

%----------------------------
\section{Conclusions}
\label{sec:conclusions}
%----------------------------
EFTs have become a cornerstone of modern QFT and particle physics. They allow us to approximate complex QFTs which exhibit a hierarchy of scales and perform precision calculations which would otherwise be intractable. 
In this review, we have introduced the core concepts of EFTs in general and one of the most popular EFTs for BSM searches, SMEFT. 
We have discussed how integrating out hard modes from the generating functional leaves us with a new QFT including new higher-order interactions or operators of the soft modes. 
When the complete UV model is known and we apply the EFT from the top down we can fix the corresponding coupling constants of the EFT, the Wilson coefficients, by comparing the predictions of both theories in a process termed matching. 
In a bottom-up approach, we can find a basis for the EFT by applying integration-by-parts identities and field redefinitions. 

BSM physics is highly motivated by a number of phenomena which cannot be explained within the SM. In the absence of a discovery of a new TeV-scale particle, the assumed hierarchy of energy scales between the SM particles and potential new particles motivates the application of an EFT for NP searches. 
SMEFT has become the most important framework for indirect NP searches at the LHC. 
We have discussed how global analyses allow interpreting the wide set of experimental data in a model-agnostic framework and how these can straightforwardly be interpreted in terms of complete NP models using automated matching tools. 
As experimental data become ever more precise, precision on the theory side and for the SMEFT double expansion in the NP scale and SM loops becomes more and more relevant. Different pieces of these expansions can be dominant for different processes and phase-space regimes, for which we have given examples in this review. 

In summary, EFTs are a tool for simplifying complex models which exhibit a hierarchy of scales as well as for model-agnostic NP searches. They will hopefully play a crucial role in pinpointing the NP model completing the SM of particle physics. 

\section*{Acknowledgements}
This work was supported by the Deutsche Forschungsgemeinschaft (DFG, German Research Foundation) under grant 396021762 - TRR~257.
The research of MK has received funding from the Cluster of Excellence PRISMA${}^{++}$ (EXC 2118/2, Project ID 390831469) funded by the German Research Foundation (DFG) within the Germany Excellence Strategy, and from the European Research Council (ERC) under the European Union’s Horizon 2022 Research and Innovation Program (ERC Advanced Grant agreement No.~101097780, EFT4jets). Views and opinions expressed in this work are those of the authors only and do not necessarily reflect those of the European Union or the European Research Council Executive Agency. Neither the European Union nor the granting authority can be held responsible for them.

\section*{Declaration of Generative AI and AI-assisted technologies in the writing process}

We acknowledge the use of Claude (Anthropic) and ChatGPT (OpenAI) for language/readability editing, reference suggestions, and structural feedback. All AI suggestions were reviewed and verified by the authors.

%\seealso{article title article title}

\bibliographystyle{JHEP}
\bibliography{references}

\providecommand{\href}[2]{#2}\begingroup\raggedright\begin{thebibliography}{100}

\bibitem{Burgess:2007pt}
C.P.~Burgess, \emph{{Introduction to Effective Field Theory}},
  \href{https://doi.org/10.1146/annurev.nucl.56.080805.140508}{\emph{Ann. Rev.
  Nucl. Part. Sci.} {\bfseries 57} (2007) 329}
  [\href{https://arxiv.org/abs/hep-th/0701053}{{\ttfamily hep-th/0701053}}].

\bibitem{Manohar:2018aog}
A.V.~Manohar, \emph{{Introduction to Effective Field Theories}},
  \href{https://arxiv.org/abs/1804.05863}{{\ttfamily 1804.05863}}.

\bibitem{Cohen:2019wxr}
T.~Cohen, \emph{{As Scales Become Separated: Lectures on Effective Field
  Theory}}, {\emph{PoS} {\bfseries TASI2018} (2019) 011}
  [\href{https://arxiv.org/abs/1903.03622}{{\ttfamily 1903.03622}}].

\bibitem{Neubert:2019mrz}
M.~Neubert, \emph{{Renormalization Theory and Effective Field Theories}},
  \href{https://arxiv.org/abs/1901.06573}{{\ttfamily 1901.06573}}.

\bibitem{Brivio:2017vri}
I.~Brivio and M.~Trott, \emph{{The Standard Model as an Effective Field
  Theory}}, \href{https://doi.org/10.1016/j.physrep.2018.11.002}{\emph{Phys.
  Rept.} {\bfseries 793} (2019) 1}
  [\href{https://arxiv.org/abs/1706.08945}{{\ttfamily 1706.08945}}].

\bibitem{ICTS:2024satpp}
A.V.~Manohar, B.~Ananthanarayan, M.~Neubert, T.~Cohen, R.~Alonso~de Pablo,
  B.~Bellazzini et~al., ``{School for Advanced Topics in Particle Physics
  (SATPP): Selected Topics in Effective Field Theories}.'' ICTS, Bengaluru,
  2024.

\bibitem{Brivio:2023ggi}
I.~Brivio, ``{SMEFT predictions for the LHC, Lectures 1--4}.'' Training Week:
  Theory Challenges in the Precision Era of the Large Hadron Collider, Galileo
  Galilei Institute, Florence, 2023.

\bibitem{Appelquist:1974tg}
T.~Appelquist and J.~Carazzone, \emph{{Infrared Singularities and Massive
  Fields}}, \href{https://doi.org/10.1103/PhysRevD.11.2856}{\emph{Phys. Rev. D}
  {\bfseries 11} (1975) 2856}.

\bibitem{Buchmuller:1985jz}
W.~Buchmuller and D.~Wyler, \emph{{Effective Lagrangian Analysis of New
  Interactions and Flavor Conservation}},
  \href{https://doi.org/10.1016/0550-3213(86)90262-2}{\emph{Nucl. Phys. B}
  {\bfseries 268} (1986) 621}.

\bibitem{Feruglio:1992wf}
F.~Feruglio, \emph{{The Chiral approach to the electroweak interactions}},
  \href{https://doi.org/10.1142/S0217751X93001946}{\emph{Int. J. Mod. Phys. A}
  {\bfseries 8} (1993) 4937}
  [\href{https://arxiv.org/abs/hep-ph/9301281}{{\ttfamily hep-ph/9301281}}].

\bibitem{Grinstein:2007iv}
B.~Grinstein and M.~Trott, \emph{{A Higgs-Higgs bound state due to new physics
  at a TeV}}, \href{https://doi.org/10.1103/PhysRevD.76.073002}{\emph{Phys.
  Rev. D} {\bfseries 76} (2007) 073002}
  [\href{https://arxiv.org/abs/0704.1505}{{\ttfamily 0704.1505}}].

\bibitem{Alonso:2023upf}
R.~Alonso, \emph{{A primer on Higgs Effective Field Theory with Geometry}},
  \href{https://arxiv.org/abs/2307.14301}{{\ttfamily 2307.14301}}.

\bibitem{Jenkins:2017jig}
E.E.~Jenkins, A.V.~Manohar and P.~Stoffer, \emph{{Low-Energy Effective Field
  Theory below the Electroweak Scale: Operators and Matching}},
  \href{https://doi.org/10.1007/JHEP03(2018)016}{\emph{JHEP} {\bfseries 03}
  (2018) 016} [\href{https://arxiv.org/abs/1709.04486}{{\ttfamily
  1709.04486}}].

\bibitem{Jenkins:2017dyc}
E.E.~Jenkins, A.V.~Manohar and P.~Stoffer, \emph{{Low-Energy Effective Field
  Theory below the Electroweak Scale: Anomalous Dimensions}},
  \href{https://doi.org/10.1007/JHEP01(2018)084}{\emph{JHEP} {\bfseries 01}
  (2018) 084} [\href{https://arxiv.org/abs/1711.05270}{{\ttfamily
  1711.05270}}].

\bibitem{Buchalla:1995vs}
G.~Buchalla, A.J.~Buras and M.E.~Lautenbacher, \emph{{Weak Decays beyond
  Leading Logarithms}},
  \href{https://doi.org/10.1103/RevModPhys.68.1125}{\emph{Rev. Mod. Phys.}
  {\bfseries 68} (1996) 1125}
  [\href{https://arxiv.org/abs/hep-ph/9512380}{{\ttfamily hep-ph/9512380}}].

\bibitem{Isgur:1989vq}
N.~Isgur and M.B.~Wise, \emph{{Weak Decays of Heavy Mesons in the Static Quark
  Approximation}},
  \href{https://doi.org/10.1016/0370-2693(89)90566-2}{\emph{Phys. Lett. B}
  {\bfseries 232} (1989) 113}.

\bibitem{Georgi:1990um}
H.~Georgi, \emph{{An Effective Field Theory for Heavy Quarks at Low-energies}},
  \href{https://doi.org/10.1016/0370-2693(90)91128-X}{\emph{Phys. Lett. B}
  {\bfseries 240} (1990) 447}.

\bibitem{Neubert:1993mb}
M.~Neubert, \emph{{Heavy quark symmetry}},
  \href{https://doi.org/10.1016/0370-1573(94)90091-4}{\emph{Phys. Rept.}
  {\bfseries 245} (1994) 259}
  [\href{https://arxiv.org/abs/hep-ph/9306320}{{\ttfamily hep-ph/9306320}}].

\bibitem{Weinberg:1978kz}
S.~Weinberg, \emph{{Phenomenological Lagrangians}},
  \href{https://doi.org/10.1016/0378-4371(79)90223-1}{\emph{Physica A}
  {\bfseries 96} (1979) 327}.

\bibitem{Gasser:1983yg}
J.~Gasser and H.~Leutwyler, \emph{{Chiral Perturbation Theory to One Loop}},
  \href{https://doi.org/10.1016/0003-4916(84)90242-2}{\emph{Annals Phys.}
  {\bfseries 158} (1984) 142}.

\bibitem{Gasser:1984gg}
J.~Gasser and H.~Leutwyler, \emph{{Chiral Perturbation Theory: Expansions in
  the Mass of the Strange Quark}},
  \href{https://doi.org/10.1016/0550-3213(85)90492-4}{\emph{Nucl. Phys. B}
  {\bfseries 250} (1985) 465}.

\bibitem{Bauer:2000ew}
C.W.~Bauer, S.~Fleming and M.E.~Luke, \emph{{Summing Sudakov logarithms in $B
  \to  X_s \gamma $in effective field theory.}},
  \href{https://doi.org/10.1103/PhysRevD.63.014006}{\emph{Phys. Rev. D}
  {\bfseries 63} (2000) 014006}
  [\href{https://arxiv.org/abs/hep-ph/0005275}{{\ttfamily hep-ph/0005275}}].

\bibitem{Bauer:2000yr}
C.W.~Bauer, S.~Fleming, D.~Pirjol and I.W.~Stewart, \emph{{An Effective field
  theory for collinear and soft gluons: Heavy to light decays}},
  \href{https://doi.org/10.1103/PhysRevD.63.114020}{\emph{Phys. Rev. D}
  {\bfseries 63} (2001) 114020}
  [\href{https://arxiv.org/abs/hep-ph/0011336}{{\ttfamily hep-ph/0011336}}].

\bibitem{Bauer:2001ct}
C.W.~Bauer and I.W.~Stewart, \emph{{Invariant operators in collinear effective
  theory}}, \href{https://doi.org/10.1016/S0370-2693(01)00902-9}{\emph{Phys.
  Lett. B} {\bfseries 516} (2001) 134}
  [\href{https://arxiv.org/abs/hep-ph/0107001}{{\ttfamily hep-ph/0107001}}].

\bibitem{Becher:2014oda}
T.~Becher, A.~Broggio and A.~Ferroglia, \emph{{Introduction to Soft-Collinear
  Effective Theory}}, vol.~896, Springer (2015),
  \href{https://doi.org/10.1007/978-3-319-14848-9}{10.1007/978-3-319-14848-9},
  [\href{https://arxiv.org/abs/1410.1892}{{\ttfamily 1410.1892}}].

\bibitem{Gaillard:1985uh}
M.K.~Gaillard, \emph{{The Effective One Loop Lagrangian With Derivative
  Couplings}}, \href{https://doi.org/10.1016/0550-3213(86)90264-6}{\emph{Nucl.
  Phys. B} {\bfseries 268} (1986) 669}.

\bibitem{Henning:2014wua}
B.~Henning, X.~Lu and H.~Murayama, \emph{{How to use the Standard Model
  effective field theory}},
  \href{https://doi.org/10.1007/JHEP01(2016)023}{\emph{JHEP} {\bfseries 01}
  (2016) 023} [\href{https://arxiv.org/abs/1412.1837}{{\ttfamily 1412.1837}}].

\bibitem{Drozd:2015rsp}
A.~Drozd, J.~Ellis, J.~Quevillon and T.~You, \emph{{The Universal One-Loop
  Effective Action}},
  \href{https://doi.org/10.1007/JHEP03(2016)180}{\emph{JHEP} {\bfseries 03}
  (2016) 180} [\href{https://arxiv.org/abs/1512.03003}{{\ttfamily
  1512.03003}}].

\bibitem{Henning:2016lyp}
B.~Henning, X.~Lu and H.~Murayama, \emph{{One-loop Matching and Running with
  Covariant Derivative Expansion}},
  \href{https://doi.org/10.1007/JHEP01(2018)123}{\emph{JHEP} {\bfseries 01}
  (2018) 123} [\href{https://arxiv.org/abs/1604.01019}{{\ttfamily
  1604.01019}}].

\bibitem{Fuentes-Martin:2016uol}
J.~Fuentes-Martin, J.~Portoles and P.~Ruiz-Femenia, \emph{{Integrating out
  heavy particles with functional methods: a simplified framework}},
  \href{https://doi.org/10.1007/JHEP09(2016)156}{\emph{JHEP} {\bfseries 09}
  (2016) 156} [\href{https://arxiv.org/abs/1607.02142}{{\ttfamily
  1607.02142}}].

\bibitem{Zhang:2016pja}
Z.~Zhang, \emph{{Covariant diagrams for one-loop matching}},
  \href{https://doi.org/10.1007/JHEP05(2017)152}{\emph{JHEP} {\bfseries 05}
  (2017) 152} [\href{https://arxiv.org/abs/1610.00710}{{\ttfamily
  1610.00710}}].

\bibitem{Kramer:2019fwz}
M.~Kr{\"a}mer, B.~Summ and A.~Voigt, \emph{{Completing the scalar and fermionic
  Universal One-Loop Effective Action}},
  \href{https://doi.org/10.1007/JHEP01(2020)079}{\emph{JHEP} {\bfseries 01}
  (2020) 079} [\href{https://arxiv.org/abs/1908.04798}{{\ttfamily
  1908.04798}}].

\bibitem{Dittmaier:2021fls}
S.~Dittmaier, S.~Schuhmacher and M.~Stahlhofen, \emph{{Integrating out heavy
  fields in the path integral using the background-field method: general
  formalism}},
  \href{https://doi.org/10.1140/epjc/s10052-021-09587-7}{\emph{Eur. Phys. J. C}
  {\bfseries 81} (2021) 826}
  [\href{https://arxiv.org/abs/2102.12020}{{\ttfamily 2102.12020}}].

\bibitem{Cohen:2020fcu}
T.~Cohen, X.~Lu and Z.~Zhang, \emph{{Functional Prescription for EFT
  Matching}}, \href{https://doi.org/10.1007/JHEP02(2021)228}{\emph{JHEP}
  {\bfseries 02} (2021) 228}
  [\href{https://arxiv.org/abs/2011.02484}{{\ttfamily 2011.02484}}].

\bibitem{Beneke:1997zp}
M.~Beneke and V.A.~Smirnov, \emph{{Asymptotic expansion of Feynman integrals
  near threshold}},
  \href{https://doi.org/10.1016/S0550-3213(98)00138-2}{\emph{Nucl. Phys. B}
  {\bfseries 522} (1998) 321}
  [\href{https://arxiv.org/abs/hep-ph/9711391}{{\ttfamily hep-ph/9711391}}].

\bibitem{Smirnov:2002pj}
V.A.~Smirnov, \emph{{Applied asymptotic expansions in momenta and masses}},
  {\emph{Springer Tracts Mod. Phys.} {\bfseries 177} (2002) 1}.

\bibitem{Nieves:2003in}
J.F.~Nieves and P.B.~Pal, \emph{{Generalized Fierz identities}},
  \href{https://doi.org/10.1119/1.1757445}{\emph{Am. J. Phys.} {\bfseries 72}
  (2004) 1100} [\href{https://arxiv.org/abs/hep-ph/0306087}{{\ttfamily
  hep-ph/0306087}}].

\bibitem{Henning:2015alf}
B.~Henning, X.~Lu, T.~Melia and H.~Murayama, \emph{{2, 84, 30, 993, 560, 15456,
  11962, 261485, ...: Higher dimension operators in the SM EFT}},
  \href{https://doi.org/10.1007/JHEP08(2017)016}{\emph{JHEP} {\bfseries 08}
  (2017) 016} [\href{https://arxiv.org/abs/1512.03433}{{\ttfamily
  1512.03433}}].

\bibitem{Fuentes-Martin:2022jrf}
J.~Fuentes-Mart{\'\i}n, M.~K{\"o}nig, J.~Pag{\`e}s, A.E.~Thomsen and F.~Wilsch,
  \emph{{A proof of concept for matchete: an automated tool for matching
  effective theories}},
  \href{https://doi.org/10.1140/epjc/s10052-023-11726-1}{\emph{Eur. Phys. J. C}
  {\bfseries 83} (2023) 662}
  [\href{https://arxiv.org/abs/2212.04510}{{\ttfamily 2212.04510}}].

\bibitem{Buras:1989xd}
A.J.~Buras and P.H.~Weisz, \emph{{QCD Nonleading Corrections to Weak Decays in
  Dimensional Regularization and 't Hooft-Veltman Schemes}},
  \href{https://doi.org/10.1016/0550-3213(90)90223-Z}{\emph{Nucl. Phys. B}
  {\bfseries 333} (1990) 66}.

\bibitem{Dugan:1990df}
M.J.~Dugan and B.~Grinstein, \emph{{On the vanishing of evanescent operators}},
  \href{https://doi.org/10.1016/0370-2693(91)90680-O}{\emph{Phys. Lett. B}
  {\bfseries 256} (1991) 239}.

\bibitem{Herrlich:1994kh}
S.~Herrlich and U.~Nierste, \emph{{Evanescent operators, scheme dependences and
  double insertions}},
  \href{https://doi.org/10.1016/0550-3213(95)00474-7}{\emph{Nucl. Phys. B}
  {\bfseries 455} (1995) 39}
  [\href{https://arxiv.org/abs/hep-ph/9412375}{{\ttfamily hep-ph/9412375}}].

\bibitem{Fuentes-Martin:2022vvu}
J.~Fuentes-Mart{\'\i}n, M.~K{\"o}nig, J.~Pag{\`e}s, A.E.~Thomsen and F.~Wilsch,
  \emph{{Evanescent operators in one-loop matching computations}},
  \href{https://doi.org/10.1007/JHEP02(2023)031}{\emph{JHEP} {\bfseries 02}
  (2023) 031} [\href{https://arxiv.org/abs/2211.09144}{{\ttfamily
  2211.09144}}].

\bibitem{Weinberg:1979sa}
S.~Weinberg, \emph{{Baryon and Lepton Nonconserving Processes}},
  \href{https://doi.org/10.1103/PhysRevLett.43.1566}{\emph{Phys. Rev. Lett.}
  {\bfseries 43} (1979) 1566}.

\bibitem{Kobach:2016ami}
A.~Kobach, \emph{{Baryon Number, Lepton Number, and Operator Dimension in the
  Standard Model}},
  \href{https://doi.org/10.1016/j.physletb.2016.05.050}{\emph{Phys. Lett. B}
  {\bfseries 758} (2016) 455}
  [\href{https://arxiv.org/abs/1604.05726}{{\ttfamily 1604.05726}}].

\bibitem{Giudice:2007fh}
G.F.~Giudice, C.~Grojean, A.~Pomarol and R.~Rattazzi, \emph{{The
  Strongly-Interacting Light Higgs}},
  \href{https://doi.org/10.1088/1126-6708/2007/06/045}{\emph{JHEP} {\bfseries
  06} (2007) 045} [\href{https://arxiv.org/abs/hep-ph/0703164}{{\ttfamily
  hep-ph/0703164}}].

\bibitem{Elias-Miro:2013mua}
J.~Elias-Miro, J.R.~Espinosa, E.~Masso and A.~Pomarol, \emph{{Higgs windows to
  new physics through d=6 operators: constraints and one-loop anomalous
  dimensions}}, \href{https://doi.org/10.1007/JHEP11(2013)066}{\emph{JHEP}
  {\bfseries 11} (2013) 066} [\href{https://arxiv.org/abs/1308.1879}{{\ttfamily
  1308.1879}}].

\bibitem{Hagiwara:1993ck}
K.~Hagiwara, S.~Ishihara, R.~Szalapski and D.~Zeppenfeld, \emph{{Low-energy
  effects of new interactions in the electroweak boson sector}},
  \href{https://doi.org/10.1103/PhysRevD.48.2182}{\emph{Phys. Rev. D}
  {\bfseries 48} (1993) 2182}.

\bibitem{Grzadkowski:2010es}
B.~Grzadkowski, M.~Iskrzynski, M.~Misiak and J.~Rosiek, \emph{{Dimension-Six
  Terms in the Standard Model Lagrangian}},
  \href{https://doi.org/10.1007/JHEP10(2010)085}{\emph{JHEP} {\bfseries 10}
  (2010) 085} [\href{https://arxiv.org/abs/1008.4884}{{\ttfamily 1008.4884}}].

\bibitem{Eboli:2003nq}
O.J.P.~Eboli, M.C.~Gonzalez-Garcia and S.M.~Lietti, \emph{{Bosonic quartic
  couplings at CERN LHC}},
  \href{https://doi.org/10.1103/PhysRevD.69.095005}{\emph{Phys. Rev. D}
  {\bfseries 69} (2004) 095005}
  [\href{https://arxiv.org/abs/hep-ph/0310141}{{\ttfamily hep-ph/0310141}}].

\bibitem{Eboli:2006wa}
O.J.P.~Eboli, M.C.~Gonzalez-Garcia and J.K.~Mizukoshi, \emph{{p p
  ---{\ensuremath{>}} j j e+- mu+- nu nu and j j e+- mu-+ nu nu at O(
  alpha(em)**6) and O(alpha(em)**4 alpha(s)**2) for the study of the quartic
  electroweak gauge boson vertex at CERN LHC}},
  \href{https://doi.org/10.1103/PhysRevD.74.073005}{\emph{Phys. Rev. D}
  {\bfseries 74} (2006) 073005}
  [\href{https://arxiv.org/abs/hep-ph/0606118}{{\ttfamily hep-ph/0606118}}].

\bibitem{Eboli:2016kko}
O.J.P.~{\'E}boli and M.C.~Gonzalez-Garcia, \emph{{Classifying the bosonic
  quartic couplings}},
  \href{https://doi.org/10.1103/PhysRevD.93.093013}{\emph{Phys. Rev. D}
  {\bfseries 93} (2016) 093013}
  [\href{https://arxiv.org/abs/1604.03555}{{\ttfamily 1604.03555}}].

\bibitem{Durieux:2024zrg}
G.~Durieux, G.N.~Remmen, N.L.~Rodd, O.J.P.~{\'E}boli, M.C.~Gonzalez-Garcia,
  D.~Kondo et~al., \emph{{LHC EFT WG note: Basis for anomalous quartic gauge
  couplings}},  \href{https://arxiv.org/abs/2411.02483}{{\ttfamily
  2411.02483}}.

\bibitem{ATLAS:2026wew}
{\scshape ATLAS} collaboration, \emph{{Combined effective field theory
  interpretation of measurements sensitive to quartic gauge boson couplings in
  pp collisions at s=13 TeV with the ATLAS detector}},
  \href{https://doi.org/10.1016/j.physletb.2026.140683}{\emph{Phys. Lett. B}
  {\bfseries 879} (2026) 140683}
  [\href{https://arxiv.org/abs/2603.18630}{{\ttfamily 2603.18630}}].

\bibitem{CMS:2025dbm}
{\scshape CMS} collaboration, \emph{{Vector boson scattering and anomalous
  quartic couplings in final states with
  {\ensuremath{\ell}}{\ensuremath{\nu}}qq or
  {\ensuremath{\ell}}{\ensuremath{\ell}}qq plus jets using proton-proton
  collisions at $ \sqrt{s}=13 $ TeV}},
  \href{https://doi.org/10.1007/JHEP03(2026)022}{\emph{JHEP} {\bfseries 03}
  (2026) 022} [\href{https://arxiv.org/abs/2510.00118}{{\ttfamily
  2510.00118}}].

\bibitem{Degrande:2013kka}
C.~Degrande, \emph{{A basis of dimension-eight operators for anomalous neutral
  triple gauge boson interactions}},
  \href{https://doi.org/10.1007/JHEP02(2014)101}{\emph{JHEP} {\bfseries 02}
  (2014) 101} [\href{https://arxiv.org/abs/1308.6323}{{\ttfamily 1308.6323}}].

\bibitem{Ellis:2022zdw}
J.~Ellis, H.-J.~He and R.-Q.~Xiao, \emph{{Probing neutral triple gauge
  couplings at the LHC and future hadron colliders}},
  \href{https://doi.org/10.1103/PhysRevD.107.035005}{\emph{Phys. Rev. D}
  {\bfseries 107} (2023) 035005}
  [\href{https://arxiv.org/abs/2206.11676}{{\ttfamily 2206.11676}}].

\bibitem{CMS:2025cxp}
{\scshape CMS} collaboration, \emph{{Measurement of Z$γ$ production in
  proton-proton collisions at $\sqrt{s}$ = 13.6 TeV and constraints on neutral
  triple gauge couplings}},  \href{https://arxiv.org/abs/2512.08582}{{\ttfamily
  2512.08582}}.

\bibitem{ATLAS:2025ply}
{\scshape ATLAS} collaboration, \emph{{Measurements of $Z\gamma$ differential
  cross sections and search for neutral triple gauge couplings in $pp$
  collisions at $\sqrt{s} =$ 13 TeV with the ATLAS detector}}, .

\bibitem{Murphy:2020rsh}
C.W.~Murphy, \emph{{Dimension-8 operators in the Standard Model Effective Field
  Theory}}, \href{https://doi.org/10.1007/JHEP10(2020)174}{\emph{JHEP}
  {\bfseries 10} (2020) 174}
  [\href{https://arxiv.org/abs/2005.00059}{{\ttfamily 2005.00059}}].

\bibitem{Li:2020gnx}
H.-L.~Li, Z.~Ren, J.~Shu, M.-L.~Xiao, J.-H.~Yu and Y.-H.~Zheng, \emph{{Complete
  set of dimension-eight operators in the standard model effective field
  theory}}, \href{https://doi.org/10.1103/PhysRevD.104.015026}{\emph{Phys. Rev.
  D} {\bfseries 104} (2021) 015026}
  [\href{https://arxiv.org/abs/2005.00008}{{\ttfamily 2005.00008}}].

\bibitem{Chala:2021cgt}
M.~Chala, {\'A}.~D{\'\i}az-Carmona and G.~Guedes, \emph{{A
  Green{\textquoteright}s basis for the bosonic SMEFT to dimension 8}},
  \href{https://doi.org/10.1007/JHEP05(2022)138}{\emph{JHEP} {\bfseries 05}
  (2022) 138} [\href{https://arxiv.org/abs/2112.12724}{{\ttfamily
  2112.12724}}].

\bibitem{Ren:2022tvi}
Z.~Ren and J.-H.~Yu, \emph{{A complete set of the dimension-8
  Green{\textquoteright}s basis operators in the Standard Model effective field
  theory}}, \href{https://doi.org/10.1007/JHEP02(2024)134}{\emph{JHEP}
  {\bfseries 02} (2024) 134}
  [\href{https://arxiv.org/abs/2211.01420}{{\ttfamily 2211.01420}}].

\bibitem{Lehman:2014jma}
L.~Lehman, \emph{{Extending the Standard Model Effective Field Theory with the
  Complete Set of Dimension-7 Operators}},
  \href{https://doi.org/10.1103/PhysRevD.90.125023}{\emph{Phys. Rev. D}
  {\bfseries 90} (2014) 125023}
  [\href{https://arxiv.org/abs/1410.4193}{{\ttfamily 1410.4193}}].

\bibitem{Li:2020xlh}
H.-L.~Li, Z.~Ren, M.-L.~Xiao, J.-H.~Yu and Y.-H.~Zheng, \emph{{Complete set of
  dimension-nine operators in the standard model effective field theory}},
  \href{https://doi.org/10.1103/PhysRevD.104.015025}{\emph{Phys. Rev. D}
  {\bfseries 104} (2021) 015025}
  [\href{https://arxiv.org/abs/2007.07899}{{\ttfamily 2007.07899}}].

\bibitem{Liao:2020jmn}
Y.~Liao and X.-D.~Ma, \emph{{An explicit construction of the dimension-9
  operator basis in the standard model effective field theory}},
  \href{https://doi.org/10.1007/JHEP11(2020)152}{\emph{JHEP} {\bfseries 11}
  (2020) 152} [\href{https://arxiv.org/abs/2007.08125}{{\ttfamily
  2007.08125}}].

\bibitem{Harlander:2023psl}
R.V.~Harlander, T.~Kempkens and M.C.~Schaaf, \emph{{Standard model effective
  field theory up to mass dimension 12}},
  \href{https://doi.org/10.1103/PhysRevD.108.055020}{\emph{Phys. Rev. D}
  {\bfseries 108} (2023) 055020}
  [\href{https://arxiv.org/abs/2305.06832}{{\ttfamily 2305.06832}}].

\bibitem{Calibbi:2017uvl}
L.~Calibbi and G.~Signorelli, \emph{{Charged Lepton Flavour Violation: An
  Experimental and Theoretical Introduction}},
  \href{https://doi.org/10.1393/ncr/i2018-10144-0}{\emph{Riv. Nuovo Cim.}
  {\bfseries 41} (2018) 71} [\href{https://arxiv.org/abs/1709.00294}{{\ttfamily
  1709.00294}}].

\bibitem{Silvestrini:2018dos}
L.~Silvestrini and M.~Valli, \emph{{Model-independent Bounds on the Standard
  Model Effective Theory from Flavour Physics}},
  \href{https://doi.org/10.1016/j.physletb.2019.135062}{\emph{Phys. Lett. B}
  {\bfseries 799} (2019) 135062}
  [\href{https://arxiv.org/abs/1812.10913}{{\ttfamily 1812.10913}}].

\bibitem{Gerard:1982mm}
J.M.~Gerard, \emph{{FERMION MASS SPECTRUM IN SU(2)-L x U(1)}},
  \href{https://doi.org/10.1007/BF01572477}{\emph{Z. Phys. C} {\bfseries 18}
  (1983) 145}.

\bibitem{Chivukula:1987py}
R.S.~Chivukula and H.~Georgi, \emph{{Composite Technicolor Standard Model}},
  \href{https://doi.org/10.1016/0370-2693(87)90713-1}{\emph{Phys. Lett. B}
  {\bfseries 188} (1987) 99}.

\bibitem{Hall:1990ac}
L.J.~Hall and L.~Randall, \emph{{Weak scale effective supersymmetry}},
  \href{https://doi.org/10.1103/PhysRevLett.65.2939}{\emph{Phys. Rev. Lett.}
  {\bfseries 65} (1990) 2939}.

\bibitem{DAmbrosio:2002vsn}
G.~D'Ambrosio, G.F.~Giudice, G.~Isidori and A.~Strumia, \emph{{Minimal flavor
  violation: An Effective field theory approach}},
  \href{https://doi.org/10.1016/S0550-3213(02)00836-2}{\emph{Nucl. Phys. B}
  {\bfseries 645} (2002) 155}
  [\href{https://arxiv.org/abs/hep-ph/0207036}{{\ttfamily hep-ph/0207036}}].

\bibitem{Barbieri:2012uh}
R.~Barbieri, D.~Buttazzo, F.~Sala and D.M.~Straub, \emph{{Flavour physics from
  an approximate $U(2)^3$ symmetry}},
  \href{https://doi.org/10.1007/JHEP07(2012)181}{\emph{JHEP} {\bfseries 07}
  (2012) 181} [\href{https://arxiv.org/abs/1203.4218}{{\ttfamily 1203.4218}}].

\bibitem{Alonso:2013hga}
R.~Alonso, E.E.~Jenkins, A.V.~Manohar and M.~Trott, \emph{{Renormalization
  Group Evolution of the Standard Model Dimension Six Operators III: Gauge
  Coupling Dependence and Phenomenology}},
  \href{https://doi.org/10.1007/JHEP04(2014)159}{\emph{JHEP} {\bfseries 04}
  (2014) 159} [\href{https://arxiv.org/abs/1312.2014}{{\ttfamily 1312.2014}}].

\bibitem{Dedes:2017zog}
A.~Dedes, W.~Materkowska, M.~Paraskevas, J.~Rosiek and K.~Suxho, \emph{{Feynman
  rules for the Standard Model Effective Field Theory in R$_{ξ}$ -gauges}},
  \href{https://doi.org/10.1007/JHEP06(2017)143}{\emph{JHEP} {\bfseries 06}
  (2017) 143} [\href{https://arxiv.org/abs/1704.03888}{{\ttfamily
  1704.03888}}].

\bibitem{Brivio:2020onw}
I.~Brivio, \emph{{SMEFTsim 3.0 {\textemdash} a practical guide}},
  \href{https://doi.org/10.1007/JHEP04(2021)073}{\emph{JHEP} {\bfseries 04}
  (2021) 073} [\href{https://arxiv.org/abs/2012.11343}{{\ttfamily
  2012.11343}}].

\bibitem{Biekotter:2023xle}
A.~Biek{\"o}tter, B.D.~Pecjak, D.J.~Scott and T.~Smith, \emph{{Electroweak
  input schemes and universal corrections in SMEFT}},
  \href{https://doi.org/10.1007/JHEP07(2023)115}{\emph{JHEP} {\bfseries 07}
  (2023) 115} [\href{https://arxiv.org/abs/2305.03763}{{\ttfamily
  2305.03763}}].

\bibitem{Brivio:2021yjb}
I.~Brivio, S.~Dawson, J.~de~Blas, G.~Durieux, P.~Savard, A.~Denner et~al.,
  \emph{{Electroweak input parameters}},
  \href{https://arxiv.org/abs/2111.12515}{{\ttfamily 2111.12515}}.

\bibitem{Jenkins:2013zja}
E.E.~Jenkins, A.V.~Manohar and M.~Trott, \emph{{Renormalization Group Evolution
  of the Standard Model Dimension Six Operators I: Formalism and lambda
  Dependence}}, \href{https://doi.org/10.1007/JHEP10(2013)087}{\emph{JHEP}
  {\bfseries 10} (2013) 087} [\href{https://arxiv.org/abs/1308.2627}{{\ttfamily
  1308.2627}}].

\bibitem{Jenkins:2013wua}
E.E.~Jenkins, A.V.~Manohar and M.~Trott, \emph{{Renormalization Group Evolution
  of the Standard Model Dimension Six Operators II: Yukawa Dependence}},
  \href{https://doi.org/10.1007/JHEP01(2014)035}{\emph{JHEP} {\bfseries 01}
  (2014) 035} [\href{https://arxiv.org/abs/1310.4838}{{\ttfamily 1310.4838}}].

\bibitem{Contino:2016jqw}
R.~Contino, A.~Falkowski, F.~Goertz, C.~Grojean and F.~Riva, \emph{{On the
  Validity of the Effective Field Theory Approach to SM Precision Tests}},
  \href{https://doi.org/10.1007/JHEP07(2016)144}{\emph{JHEP} {\bfseries 07}
  (2016) 144} [\href{https://arxiv.org/abs/1604.06444}{{\ttfamily
  1604.06444}}].

\bibitem{Keilmann:2019cbp}
E.~Keilmann and W.~Shepherd, \emph{{Dijets at Tevatron Cannot Constrain SMEFT
  Four-Quark Operators}},
  \href{https://doi.org/10.1007/JHEP09(2019)086}{\emph{JHEP} {\bfseries 09}
  (2019) 086} [\href{https://arxiv.org/abs/1907.13160}{{\ttfamily
  1907.13160}}].

\bibitem{Allwicher:2024mzw}
L.~Allwicher, D.A.~Faroughy, M.~Martines, O.~Sumensari and F.~Wilsch, \emph{{On
  the EFT validity for Drell{\textendash}Yan tails at the LHC}},
  \href{https://doi.org/10.1140/epjc/s10052-025-14171-4}{\emph{Eur. Phys. J. C}
  {\bfseries 85} (2025) 463}
  [\href{https://arxiv.org/abs/2412.14162}{{\ttfamily 2412.14162}}].

\bibitem{Brivio:2022pyi}
I.~Brivio et~al., \emph{{Truncation, validity, uncertainties}},
  \href{https://arxiv.org/abs/2201.04974}{{\ttfamily 2201.04974}}.

\bibitem{Ellis:2019zex}
J.~Ellis, S.-F.~Ge, H.-J.~He and R.-Q.~Xiao, \emph{{Probing the scale of new
  physics in the $ZZ\gamma$ coupling at $e^+e^-$ colliders}},
  \href{https://doi.org/10.1088/1674-1137/44/6/063106}{\emph{Chin. Phys. C}
  {\bfseries 44} (2020) 063106}
  [\href{https://arxiv.org/abs/1902.06631}{{\ttfamily 1902.06631}}].

\bibitem{Dawson:2022cmu}
S.~Dawson, D.~Fontes, S.~Homiller and M.~Sullivan, \emph{{Role of
  dimension-eight operators in an EFT for the 2HDM}},
  \href{https://doi.org/10.1103/PhysRevD.106.055012}{\emph{Phys. Rev. D}
  {\bfseries 106} (2022) 055012}
  [\href{https://arxiv.org/abs/2205.01561}{{\ttfamily 2205.01561}}].

\bibitem{Banerjee:2022thk}
U.~Banerjee, J.~Chakrabortty, C.~Englert, S.U.~Rahaman and M.~Spannowsky,
  \emph{{Integrating out heavy scalars with modified equations of motion:
  Matching computation of dimension-eight SMEFT coefficients}},
  \href{https://doi.org/10.1103/PhysRevD.107.055007}{\emph{Phys. Rev. D}
  {\bfseries 107} (2023) 055007}
  [\href{https://arxiv.org/abs/2210.14761}{{\ttfamily 2210.14761}}].

\bibitem{Ellis:2023zim}
J.~Ellis, K.~Mimasu and F.~Zampedri, \emph{{Dimension-8 SMEFT analysis of
  minimal scalar field extensions of the Standard Model}},
  \href{https://doi.org/10.1007/JHEP10(2023)051}{\emph{JHEP} {\bfseries 10}
  (2023) 051} [\href{https://arxiv.org/abs/2304.06663}{{\ttfamily
  2304.06663}}].

\bibitem{Banerjee:2023qbg}
U.~Banerjee, J.~Chakrabortty, C.~Englert, W.~Naskar, S.U.~Rahaman and
  M.~Spannowsky, \emph{{EFT, decoupling, Higgs boson mixing, and higher
  dimensional operators}},
  \href{https://doi.org/10.1103/PhysRevD.109.055035}{\emph{Phys. Rev. D}
  {\bfseries 109} (2024) 055035}
  [\href{https://arxiv.org/abs/2303.05224}{{\ttfamily 2303.05224}}].

\bibitem{Dawson:2021xei}
S.~Dawson, S.~Homiller and M.~Sullivan, \emph{{Impact of dimension-eight SMEFT
  contributions: A case study}},
  \href{https://doi.org/10.1103/PhysRevD.104.115013}{\emph{Phys. Rev. D}
  {\bfseries 104} (2021) 115013}
  [\href{https://arxiv.org/abs/2110.06929}{{\ttfamily 2110.06929}}].

\bibitem{Corbett:2021eux}
T.~Corbett, A.~Helset, A.~Martin and M.~Trott, \emph{{EWPD in the SMEFT to
  dimension eight}}, \href{https://doi.org/10.1007/JHEP06(2021)076}{\emph{JHEP}
  {\bfseries 06} (2021) 076}
  [\href{https://arxiv.org/abs/2102.02819}{{\ttfamily 2102.02819}}].

\bibitem{Dawson:2024ozw}
S.~Dawson, M.~Forslund and M.~Schnubel, \emph{{SMEFT matching to Z' models at
  dimension eight}},
  \href{https://doi.org/10.1103/PhysRevD.110.015002}{\emph{Phys. Rev. D}
  {\bfseries 110} (2024) 015002}
  [\href{https://arxiv.org/abs/2404.01375}{{\ttfamily 2404.01375}}].

\bibitem{Cepedello:2024ogz}
R.~Cepedello, F.~Esser, M.~Hirsch and V.~Sanz, \emph{{Fermionic UV models for
  neutral triple gauge boson vertices}},
  \href{https://doi.org/10.1007/JHEP07(2024)275}{\emph{JHEP} {\bfseries 07}
  (2024) 275} [\href{https://arxiv.org/abs/2402.04306}{{\ttfamily
  2402.04306}}].

\bibitem{Adams:2006sv}
A.~Adams, N.~Arkani-Hamed, S.~Dubovsky, A.~Nicolis and R.~Rattazzi,
  \emph{{Causality, analyticity and an IR obstruction to UV completion}},
  \href{https://doi.org/10.1088/1126-6708/2006/10/014}{\emph{JHEP} {\bfseries
  10} (2006) 014} [\href{https://arxiv.org/abs/hep-th/0602178}{{\ttfamily
  hep-th/0602178}}].

\bibitem{Zhang:2018shp}
C.~Zhang and S.-Y.~Zhou, \emph{{Positivity bounds on vector boson scattering at
  the LHC}}, \href{https://doi.org/10.1103/PhysRevD.100.095003}{\emph{Phys.
  Rev. D} {\bfseries 100} (2019) 095003}
  [\href{https://arxiv.org/abs/1808.00010}{{\ttfamily 1808.00010}}].

\bibitem{Bellazzini:2018paj}
B.~Bellazzini and F.~Riva, \emph{{New phenomenological and theoretical
  perspective on anomalous ZZ and Z{\ensuremath{\gamma}} processes}},
  \href{https://doi.org/10.1103/PhysRevD.98.095021}{\emph{Phys. Rev. D}
  {\bfseries 98} (2018) 095021}
  [\href{https://arxiv.org/abs/1806.09640}{{\ttfamily 1806.09640}}].

\bibitem{Remmen:2019cyz}
G.N.~Remmen and N.L.~Rodd, \emph{{Consistency of the Standard Model Effective
  Field Theory}}, \href{https://doi.org/10.1007/JHEP12(2019)032}{\emph{JHEP}
  {\bfseries 12} (2019) 032}
  [\href{https://arxiv.org/abs/1908.09845}{{\ttfamily 1908.09845}}].

\bibitem{Degrande:2020evl}
C.~Degrande, G.~Durieux, F.~Maltoni, K.~Mimasu, E.~Vryonidou and C.~Zhang,
  \emph{{Automated one-loop computations in the standard model effective field
  theory}}, \href{https://doi.org/10.1103/PhysRevD.103.096024}{\emph{Phys. Rev.
  D} {\bfseries 103} (2021) 096024}
  [\href{https://arxiv.org/abs/2008.11743}{{\ttfamily 2008.11743}}].

\bibitem{Zhang:2013xya}
C.~Zhang and F.~Maltoni, \emph{{Top-quark decay into Higgs boson and a light
  quark at next-to-leading order in QCD}},
  \href{https://doi.org/10.1103/PhysRevD.88.054005}{\emph{Phys. Rev. D}
  {\bfseries 88} (2013) 054005}
  [\href{https://arxiv.org/abs/1305.7386}{{\ttfamily 1305.7386}}].

\bibitem{Crivellin:2013hpa}
A.~Crivellin, S.~Najjari and J.~Rosiek, \emph{{Lepton Flavor Violation in the
  Standard Model with general Dimension-Six Operators}},
  \href{https://doi.org/10.1007/JHEP04(2014)167}{\emph{JHEP} {\bfseries 04}
  (2014) 167} [\href{https://arxiv.org/abs/1312.0634}{{\ttfamily 1312.0634}}].

\bibitem{Zhang:2014rja}
C.~Zhang, \emph{{Effective field theory approach to top-quark decay at
  next-to-leading order in QCD}},
  \href{https://doi.org/10.1103/PhysRevD.90.014008}{\emph{Phys. Rev. D}
  {\bfseries 90} (2014) 014008}
  [\href{https://arxiv.org/abs/1404.1264}{{\ttfamily 1404.1264}}].

\bibitem{Pruna:2014asa}
G.M.~Pruna and A.~Signer, \emph{{The $\mu\to e\gamma$ decay in a systematic
  effective field theory approach with dimension 6 operators}},
  \href{https://doi.org/10.1007/JHEP10(2014)014}{\emph{JHEP} {\bfseries 10}
  (2014) 014} [\href{https://arxiv.org/abs/1408.3565}{{\ttfamily 1408.3565}}].

\bibitem{Grober:2015cwa}
R.~Grober, M.~Muhlleitner, M.~Spira and J.~Streicher, \emph{{NLO QCD
  Corrections to Higgs Pair Production including Dimension-6 Operators}},
  \href{https://doi.org/10.1007/JHEP09(2015)092}{\emph{JHEP} {\bfseries 09}
  (2015) 092} [\href{https://arxiv.org/abs/1504.06577}{{\ttfamily
  1504.06577}}].

\bibitem{Hartmann:2015oia}
C.~Hartmann and M.~Trott, \emph{{On one-loop corrections in the standard model
  effective field theory; the $\Gamma(h \rightarrow \gamma \, \gamma)$ case}},
  \href{https://doi.org/10.1007/JHEP07(2015)151}{\emph{JHEP} {\bfseries 07}
  (2015) 151} [\href{https://arxiv.org/abs/1505.02646}{{\ttfamily
  1505.02646}}].

\bibitem{Ghezzi:2015vva}
M.~Ghezzi, R.~Gomez-Ambrosio, G.~Passarino and S.~Uccirati, \emph{{NLO Higgs
  effective field theory and {\ensuremath{\kappa}}-framework}},
  \href{https://doi.org/10.1007/JHEP07(2015)175}{\emph{JHEP} {\bfseries 07}
  (2015) 175} [\href{https://arxiv.org/abs/1505.03706}{{\ttfamily
  1505.03706}}].

\bibitem{Hartmann:2015aia}
C.~Hartmann and M.~Trott, \emph{{Higgs Decay to Two Photons at One Loop in the
  Standard Model Effective Field Theory}},
  \href{https://doi.org/10.1103/PhysRevLett.115.191801}{\emph{Phys. Rev. Lett.}
  {\bfseries 115} (2015) 191801}
  [\href{https://arxiv.org/abs/1507.03568}{{\ttfamily 1507.03568}}].

\bibitem{Aebischer:2015fzz}
J.~Aebischer, A.~Crivellin, M.~Fael and C.~Greub, \emph{{Matching of gauge
  invariant dimension-six operators for $b\to s$ and $b\to c$ transitions}},
  \href{https://doi.org/10.1007/JHEP05(2016)037}{\emph{JHEP} {\bfseries 05}
  (2016) 037} [\href{https://arxiv.org/abs/1512.02830}{{\ttfamily
  1512.02830}}].

\bibitem{Zhang:2016omx}
C.~Zhang, \emph{{Single Top Production at Next-to-Leading Order in the Standard
  Model Effective Field Theory}},
  \href{https://doi.org/10.1103/PhysRevLett.116.162002}{\emph{Phys. Rev. Lett.}
  {\bfseries 116} (2016) 162002}
  [\href{https://arxiv.org/abs/1601.06163}{{\ttfamily 1601.06163}}].

\bibitem{BessidskaiaBylund:2016jvp}
O.~Bessidskaia~Bylund, F.~Maltoni, I.~Tsinikos, E.~Vryonidou and C.~Zhang,
  \emph{{Probing top quark neutral couplings in the Standard Model Effective
  Field Theory at NLO in QCD}},
  \href{https://doi.org/10.1007/JHEP05(2016)052}{\emph{JHEP} {\bfseries 05}
  (2016) 052} [\href{https://arxiv.org/abs/1601.08193}{{\ttfamily
  1601.08193}}].

\bibitem{Maltoni:2016yxb}
F.~Maltoni, E.~Vryonidou and C.~Zhang, \emph{{Higgs production in association
  with a top-antitop pair in the Standard Model Effective Field Theory at NLO
  in QCD}}, \href{https://doi.org/10.1007/JHEP10(2016)123}{\emph{JHEP}
  {\bfseries 10} (2016) 123}
  [\href{https://arxiv.org/abs/1607.05330}{{\ttfamily 1607.05330}}].

\bibitem{Degrande:2016dqg}
C.~Degrande, B.~Fuks, K.~Mawatari, K.~Mimasu and V.~Sanz, \emph{{Electroweak
  Higgs boson production in the standard model effective field theory beyond
  leading order in QCD}},
  \href{https://doi.org/10.1140/epjc/s10052-017-4793-x}{\emph{Eur. Phys. J. C}
  {\bfseries 77} (2017) 262}
  [\href{https://arxiv.org/abs/1609.04833}{{\ttfamily 1609.04833}}].

\bibitem{Hartmann:2016pil}
C.~Hartmann, W.~Shepherd and M.~Trott, \emph{{The $Z$ decay width in the SMEFT:
  $y_t$ and $\lambda$ corrections at one loop}},
  \href{https://doi.org/10.1007/JHEP03(2017)060}{\emph{JHEP} {\bfseries 03}
  (2017) 060} [\href{https://arxiv.org/abs/1611.09879}{{\ttfamily
  1611.09879}}].

\bibitem{Grazzini:2016paz}
M.~Grazzini, A.~Ilnicka, M.~Spira and M.~Wiesemann, \emph{{Modeling BSM effects
  on the Higgs transverse-momentum spectrum in an EFT approach}},
  \href{https://doi.org/10.1007/JHEP03(2017)115}{\emph{JHEP} {\bfseries 03}
  (2017) 115} [\href{https://arxiv.org/abs/1612.00283}{{\ttfamily
  1612.00283}}].

\bibitem{deFlorian:2017qfk}
D.~de~Florian, I.~Fabre and J.~Mazzitelli, \emph{{Higgs boson pair production
  at NNLO in QCD including dimension 6 operators}},
  \href{https://doi.org/10.1007/JHEP10(2017)215}{\emph{JHEP} {\bfseries 10}
  (2017) 215} [\href{https://arxiv.org/abs/1704.05700}{{\ttfamily
  1704.05700}}].

\bibitem{Deutschmann:2017qum}
N.~Deutschmann, C.~Duhr, F.~Maltoni and E.~Vryonidou, \emph{{Gluon-fusion Higgs
  production in the Standard Model Effective Field Theory}},
  \href{https://doi.org/10.1007/JHEP12(2017)063}{\emph{JHEP} {\bfseries 12}
  (2017) 063} [\href{https://arxiv.org/abs/1708.00460}{{\ttfamily
  1708.00460}}].

\bibitem{Baglio:2017bfe}
J.~Baglio, S.~Dawson and I.M.~Lewis, \emph{{An NLO QCD effective field theory
  analysis of $W^+W^-$ production at the LHC including fermionic operators}},
  \href{https://doi.org/10.1103/PhysRevD.96.073003}{\emph{Phys. Rev. D}
  {\bfseries 96} (2017) 073003}
  [\href{https://arxiv.org/abs/1708.03332}{{\ttfamily 1708.03332}}].

\bibitem{Dawson:2018pyl}
S.~Dawson and P.P.~Giardino, \emph{{Higgs decays to $ZZ$ and $Z\gamma$ in the
  standard model effective field theory: An NLO analysis}},
  \href{https://doi.org/10.1103/PhysRevD.97.093003}{\emph{Phys. Rev. D}
  {\bfseries 97} (2018) 093003}
  [\href{https://arxiv.org/abs/1801.01136}{{\ttfamily 1801.01136}}].

\bibitem{Degrande:2018fog}
C.~Degrande, F.~Maltoni, K.~Mimasu, E.~Vryonidou and C.~Zhang,
  \emph{{Single-top associated production with a $Z$ or $H$ boson at the LHC:
  the SMEFT interpretation}},
  \href{https://doi.org/10.1007/JHEP10(2018)005}{\emph{JHEP} {\bfseries 10}
  (2018) 005} [\href{https://arxiv.org/abs/1804.07773}{{\ttfamily
  1804.07773}}].

\bibitem{Vryonidou:2018eyv}
E.~Vryonidou and C.~Zhang, \emph{{Dimension-six electroweak top-loop effects in
  Higgs production and decay}},
  \href{https://doi.org/10.1007/JHEP08(2018)036}{\emph{JHEP} {\bfseries 08}
  (2018) 036} [\href{https://arxiv.org/abs/1804.09766}{{\ttfamily
  1804.09766}}].

\bibitem{Dedes:2018seb}
A.~Dedes, M.~Paraskevas, J.~Rosiek, K.~Suxho and L.~Trifyllis, \emph{{The decay
  $h\to \gamma\gamma$ in the Standard-Model Effective Field Theory}},
  \href{https://doi.org/10.1007/JHEP08(2018)103}{\emph{JHEP} {\bfseries 08}
  (2018) 103} [\href{https://arxiv.org/abs/1805.00302}{{\ttfamily
  1805.00302}}].

\bibitem{Grazzini:2018eyk}
M.~Grazzini, A.~Ilnicka and M.~Spira, \emph{{Higgs boson production at large
  transverse momentum within the SMEFT: analytical results}},
  \href{https://doi.org/10.1140/epjc/s10052-018-6261-7}{\emph{Eur. Phys. J. C}
  {\bfseries 78} (2018) 808}
  [\href{https://arxiv.org/abs/1806.08832}{{\ttfamily 1806.08832}}].

\bibitem{Dawson:2018liq}
S.~Dawson and P.P.~Giardino, \emph{{Electroweak corrections to Higgs boson
  decays to $\gamma\gamma$ and $W^+W^-$ in standard model EFT}},
  \href{https://doi.org/10.1103/PhysRevD.98.095005}{\emph{Phys. Rev. D}
  {\bfseries 98} (2018) 095005}
  [\href{https://arxiv.org/abs/1807.11504}{{\ttfamily 1807.11504}}].

\bibitem{Dawson:2018jlg}
S.~Dawson and A.~Ismail, \emph{{Standard model EFT corrections to Z boson
  decays}}, \href{https://doi.org/10.1103/PhysRevD.98.093003}{\emph{Phys. Rev.
  D} {\bfseries 98} (2018) 093003}
  [\href{https://arxiv.org/abs/1808.05948}{{\ttfamily 1808.05948}}].

\bibitem{Dawson:2018dxp}
S.~Dawson, P.P.~Giardino and A.~Ismail, \emph{{Standard model EFT and the
  Drell-Yan process at high energy}},
  \href{https://doi.org/10.1103/PhysRevD.99.035044}{\emph{Phys. Rev. D}
  {\bfseries 99} (2019) 035044}
  [\href{https://arxiv.org/abs/1811.12260}{{\ttfamily 1811.12260}}].

\bibitem{Neumann:2019kvk}
T.~Neumann and Z.E.~Sullivan, \emph{{Off-Shell Single-Top-Quark Production in
  the Standard Model Effective Field Theory}},
  \href{https://doi.org/10.1007/JHEP06(2019)022}{\emph{JHEP} {\bfseries 06}
  (2019) 022} [\href{https://arxiv.org/abs/1903.11023}{{\ttfamily
  1903.11023}}].

\bibitem{Dedes:2019bew}
A.~Dedes, K.~Suxho and L.~Trifyllis, \emph{{The decay $h\to Z \gamma$ in the
  Standard-Model Effective Field Theory}},
  \href{https://doi.org/10.1007/JHEP06(2019)115}{\emph{JHEP} {\bfseries 06}
  (2019) 115} [\href{https://arxiv.org/abs/1903.12046}{{\ttfamily
  1903.12046}}].

\bibitem{Boughezal:2019xpp}
R.~Boughezal, C.-Y.~Chen, F.~Petriello and D.~Wiegand, \emph{{Top quark decay
  at next-to-leading order in the Standard Model Effective Field Theory}},
  \href{https://doi.org/10.1103/PhysRevD.100.056023}{\emph{Phys. Rev. D}
  {\bfseries 100} (2019) 056023}
  [\href{https://arxiv.org/abs/1907.00997}{{\ttfamily 1907.00997}}].

\bibitem{Dawson:2019clf}
S.~Dawson and P.P.~Giardino, \emph{{Electroweak and QCD corrections to $Z$ and
  $W$ pole observables in the standard model EFT}},
  \href{https://doi.org/10.1103/PhysRevD.101.013001}{\emph{Phys. Rev. D}
  {\bfseries 101} (2020) 013001}
  [\href{https://arxiv.org/abs/1909.02000}{{\ttfamily 1909.02000}}].

\bibitem{Baglio:2019uty}
J.~Baglio, S.~Dawson and S.~Homiller, \emph{{QCD corrections in Standard Model
  EFT fits to $WZ$ and $WW$ production}},
  \href{https://doi.org/10.1103/PhysRevD.100.113010}{\emph{Phys. Rev. D}
  {\bfseries 100} (2019) 113010}
  [\href{https://arxiv.org/abs/1909.11576}{{\ttfamily 1909.11576}}].

\bibitem{Haisch:2020ahr}
U.~Haisch, M.~Ruhdorfer, E.~Salvioni, E.~Venturini and A.~Weiler,
  \emph{{Singlet night in Feynman-ville: one-loop matching of a real scalar}},
  \href{https://doi.org/10.1007/JHEP04(2020)164}{\emph{JHEP} {\bfseries 04}
  (2020) 164} [\href{https://arxiv.org/abs/2003.05936}{{\ttfamily
  2003.05936}}].

\bibitem{Dawson:2021ofa}
S.~Dawson and P.P.~Giardino, \emph{{New physics through Drell-Yan standard
  model EFT measurements at NLO}},
  \href{https://doi.org/10.1103/PhysRevD.104.073004}{\emph{Phys. Rev. D}
  {\bfseries 104} (2021) 073004}
  [\href{https://arxiv.org/abs/2105.05852}{{\ttfamily 2105.05852}}].

\bibitem{Boughezal:2021tih}
R.~Boughezal, E.~Mereghetti and F.~Petriello, \emph{{Dilepton production in the
  SMEFT at O(1/{\ensuremath{\Lambda}}4)}},
  \href{https://doi.org/10.1103/PhysRevD.104.095022}{\emph{Phys. Rev. D}
  {\bfseries 104} (2021) 095022}
  [\href{https://arxiv.org/abs/2106.05337}{{\ttfamily 2106.05337}}].

\bibitem{Battaglia:2021nys}
M.~Battaglia, M.~Grazzini, M.~Spira and M.~Wiesemann, \emph{{Sensitivity to BSM
  effects in the Higgs p$_{T}$ spectrum within SMEFT}},
  \href{https://doi.org/10.1007/JHEP11(2021)173}{\emph{JHEP} {\bfseries 11}
  (2021) 173} [\href{https://arxiv.org/abs/2109.02987}{{\ttfamily
  2109.02987}}].

\bibitem{Kley:2021yhn}
J.~Kley, T.~Theil, E.~Venturini and A.~Weiler, \emph{{Electric dipole moments
  at one-loop in the dimension-6 SMEFT}},
  \href{https://doi.org/10.1140/epjc/s10052-022-10861-5}{\emph{Eur. Phys. J. C}
  {\bfseries 82} (2022) 926}
  [\href{https://arxiv.org/abs/2109.15085}{{\ttfamily 2109.15085}}].

\bibitem{Faham:2021zet}
H.E.~Faham, F.~Maltoni, K.~Mimasu and M.~Zaro, \emph{{Single top production in
  association with a WZ pair at the LHC in the SMEFT}},
  \href{https://doi.org/10.1007/JHEP01(2022)100}{\emph{JHEP} {\bfseries 01}
  (2022) 100} [\href{https://arxiv.org/abs/2111.03080}{{\ttfamily
  2111.03080}}].

\bibitem{Haisch:2022nwz}
U.~Haisch, D.J.~Scott, M.~Wiesemann, G.~Zanderighi and S.~Zanoli, \emph{{NNLO
  event generation for $ pp\to Zh\to
  {\mathrm{\ell}}^{+}{\mathrm{\ell}}^{-}b\overline{b} $ production in the SM
  effective field theory}},
  \href{https://doi.org/10.1007/JHEP07(2022)054}{\emph{JHEP} {\bfseries 07}
  (2022) 054} [\href{https://arxiv.org/abs/2204.00663}{{\ttfamily
  2204.00663}}].

\bibitem{Heinrich:2022idm}
G.~Heinrich, J.~Lang and L.~Scyboz, \emph{{SMEFT predictions for gg
  {\textrightarrow} hh at full NLO QCD and truncation uncertainties}},
  \href{https://doi.org/10.1007/JHEP08(2022)079}{\emph{JHEP} {\bfseries 08}
  (2022) 079} [\href{https://arxiv.org/abs/2204.13045}{{\ttfamily
  2204.13045}}].

\bibitem{Asteriadis:2022ras}
K.~Asteriadis, S.~Dawson and D.~Fontes, \emph{{Double insertions of SMEFT
  operators in gluon fusion Higgs boson production}},
  \href{https://doi.org/10.1103/PhysRevD.107.055038}{\emph{Phys. Rev. D}
  {\bfseries 107} (2023) 055038}
  [\href{https://arxiv.org/abs/2212.03258}{{\ttfamily 2212.03258}}].

\bibitem{Bellafronte:2023amz}
L.~Bellafronte, S.~Dawson and P.P.~Giardino, \emph{{The importance of flavor in
  SMEFT Electroweak Precision Fits}},
  \href{https://doi.org/10.1007/JHEP05(2023)208}{\emph{JHEP} {\bfseries 05}
  (2023) 208} [\href{https://arxiv.org/abs/2304.00029}{{\ttfamily
  2304.00029}}].

\bibitem{Kidonakis:2023htm}
N.~Kidonakis and A.~Tonero, \emph{{SMEFT chromomagnetic dipole operator
  contributions to $t{{\bar{t}}}$ production at approximate NNLO in QCD}},
  \href{https://doi.org/10.1140/epjc/s10052-024-12938-9}{\emph{Eur. Phys. J. C}
  {\bfseries 84} (2024) 591}
  [\href{https://arxiv.org/abs/2309.16758}{{\ttfamily 2309.16758}}].

\bibitem{Gauld:2023gtb}
R.~Gauld, U.~Haisch and L.~Schnell, \emph{{SMEFT at NNLO+PS: Vh production}},
  \href{https://doi.org/10.1007/JHEP01(2024)192}{\emph{JHEP} {\bfseries 01}
  (2024) 192} [\href{https://arxiv.org/abs/2311.06107}{{\ttfamily
  2311.06107}}].

\bibitem{Heinrich:2023rsd}
G.~Heinrich and J.~Lang, \emph{{Combining chromomagnetic and four-fermion
  operators with leading SMEFT operators for gg {\textrightarrow} hh at NLO
  QCD}}, \href{https://doi.org/10.1007/JHEP05(2024)121}{\emph{JHEP} {\bfseries
  05} (2024) 121} [\href{https://arxiv.org/abs/2311.15004}{{\ttfamily
  2311.15004}}].

\bibitem{Asteriadis:2024xuk}
K.~Asteriadis, S.~Dawson, P.P.~Giardino and R.~Szafron, \emph{{Impact of
  Next-to-Leading-Order Weak Standard-Model-Effective-Field-Theory Corrections
  in e+e-{\textrightarrow}ZH}},
  \href{https://doi.org/10.1103/PhysRevLett.133.231801}{\emph{Phys. Rev. Lett.}
  {\bfseries 133} (2024) 231801}
  [\href{https://arxiv.org/abs/2406.03557}{{\ttfamily 2406.03557}}].

\bibitem{Asteriadis:2024xts}
K.~Asteriadis, S.~Dawson, P.P.~Giardino and R.~Szafron, \emph{{e$^{+}$e$^{-}$
  {\textrightarrow} ZH process in the SMEFT beyond leading order}},
  \href{https://doi.org/10.1007/JHEP02(2025)162}{\emph{JHEP} {\bfseries 02}
  (2025) 162} [\href{https://arxiv.org/abs/2409.11466}{{\ttfamily
  2409.11466}}].

\bibitem{Dawson:2024pft}
S.~Dawson, M.~Forslund and P.P.~Giardino, \emph{{NLO SMEFT electroweak
  corrections to Higgs boson decays to four leptons in the narrow width
  approximation}},
  \href{https://doi.org/10.1103/PhysRevD.111.015016}{\emph{Phys. Rev. D}
  {\bfseries 111} (2025) 015016}
  [\href{https://arxiv.org/abs/2411.08952}{{\ttfamily 2411.08952}}].

\bibitem{ElFaham:2024egs}
H.~El~Faham, K.~Mimasu, D.~Pagani, C.~Severi, E.~Vryonidou and M.~Zaro,
  \emph{{Electroweak corrections in the SMEFT: four-fermion operators at high
  energies}}, \href{https://doi.org/10.1007/JHEP06(2025)241}{\emph{JHEP}
  {\bfseries 06} (2025) 241}
  [\href{https://arxiv.org/abs/2412.16076}{{\ttfamily 2412.16076}}].

\bibitem{Bellafronte:2025jbk}
L.~Bellafronte, S.~Dawson, C.~Del~Pio, M.~Forslund and P.P.~Giardino,
  \emph{{Complete Next-to-Leading-Order Standard-Model-Effective-Field-Theory
  Electroweak Corrections to Higgs Decays}},
  \href{https://doi.org/10.1103/2wqp-5zfm}{\emph{Phys. Rev. Lett.} {\bfseries
  136} (2026) 051801} [\href{https://arxiv.org/abs/2508.14966}{{\ttfamily
  2508.14966}}].

\bibitem{Biekotter:2025nln}
A.~Biek{\"o}tter and B.D.~Pecjak, \emph{{Analytic results for electroweak
  precision observables at NLO in SMEFT}},
  \href{https://doi.org/10.1007/JHEP07(2025)134}{\emph{JHEP} {\bfseries 07}
  (2025) 134} [\href{https://arxiv.org/abs/2503.07724}{{\ttfamily
  2503.07724}}].

\bibitem{Celis:2017hod}
A.~Celis, J.~Fuentes-Martin, A.~Vicente and J.~Virto, \emph{{DsixTools: The
  Standard Model Effective Field Theory Toolkit}},
  \href{https://doi.org/10.1140/epjc/s10052-017-4967-6}{\emph{Eur. Phys. J. C}
  {\bfseries 77} (2017) 405}
  [\href{https://arxiv.org/abs/1704.04504}{{\ttfamily 1704.04504}}].

\bibitem{Fuentes-Martin:2020zaz}
J.~Fuentes-Martin, P.~Ruiz-Femenia, A.~Vicente and J.~Virto, \emph{{DsixTools
  2.0: The Effective Field Theory Toolkit}},
  \href{https://doi.org/10.1140/epjc/s10052-020-08778-y}{\emph{Eur. Phys. J. C}
  {\bfseries 81} (2021) 167}
  [\href{https://arxiv.org/abs/2010.16341}{{\ttfamily 2010.16341}}].

\bibitem{DiNoi:2022ejg}
S.~Di~Noi and L.~Silvestrini, \emph{{RGESolver: a C++ library to perform
  renormalization group evolution in the Standard Model Effective Theory}},
  \href{https://doi.org/10.1140/epjc/s10052-023-11189-4}{\emph{Eur. Phys. J. C}
  {\bfseries 83} (2023) 200}
  [\href{https://arxiv.org/abs/2210.06838}{{\ttfamily 2210.06838}}].

\bibitem{Aebischer:2018bkb}
J.~Aebischer, J.~Kumar and D.M.~Straub, \emph{{Wilson: a Python package for the
  running and matching of Wilson coefficients above and below the electroweak
  scale}}, \href{https://doi.org/10.1140/epjc/s10052-018-6492-7}{\emph{Eur.
  Phys. J. C} {\bfseries 78} (2018) 1026}
  [\href{https://arxiv.org/abs/1804.05033}{{\ttfamily 1804.05033}}].

\bibitem{Stefanek:2024kds}
B.A.~Stefanek, \emph{{Non-universal probes of composite Higgs models: new
  bounds and prospects for FCC-ee}},
  \href{https://doi.org/10.1007/JHEP09(2024)103}{\emph{JHEP} {\bfseries 09}
  (2024) 103} [\href{https://arxiv.org/abs/2407.09593}{{\ttfamily
  2407.09593}}].

\bibitem{Born:2026xkr}
L.~Born, J.~Fuentes-Mart{\'\i}n and A.E.~Thomsen, \emph{{Next-to-leading order
  running in the SMEFT}},
  \href{https://doi.org/10.1007/JHEP09(2026)106}{\emph{JHEP} {\bfseries 09}
  (2026) 106} [\href{https://arxiv.org/abs/2601.19974}{{\ttfamily
  2601.19974}}].

\bibitem{Aebischer:2022anv}
J.~Aebischer, A.J.~Buras and J.~Kumar, \emph{{NLO QCD renormalization group
  evolution for nonleptonic {\ensuremath{\Delta}}F=2 transitions in the
  SMEFT}}, \href{https://doi.org/10.1103/PhysRevD.106.035003}{\emph{Phys. Rev.
  D} {\bfseries 106} (2022) 035003}
  [\href{https://arxiv.org/abs/2203.11224}{{\ttfamily 2203.11224}}].

\bibitem{Born:2024mgz}
L.~Born, J.~Fuentes-Mart{\'\i}n, S.~Kvedarait{\.{e}} and A.E.~Thomsen,
  \emph{{Two-loop running in the bosonic SMEFT using functional methods}},
  \href{https://doi.org/10.1007/JHEP05(2025)121}{\emph{JHEP} {\bfseries 05}
  (2025) 121} [\href{https://arxiv.org/abs/2410.07320}{{\ttfamily
  2410.07320}}].

\bibitem{DiNoi:2024ajj}
S.~Di~Noi, R.~Gr{\"o}ber and M.K.~Mandal, \emph{{Two-loop running effects in
  Higgs physics in Standard Model Effective Field Theory}},
  \href{https://doi.org/10.1007/JHEP12(2024)220}{\emph{JHEP} {\bfseries 12}
  (2024) 220} [\href{https://arxiv.org/abs/2408.03252}{{\ttfamily
  2408.03252}}].

\bibitem{Duhr:2025zqw}
C.~Duhr, A.~Vasquez, G.~Ventura and E.~Vryonidou, \emph{{Two-loop
  renormalisation of quark and gluon fields in the SMEFT}},
  \href{https://doi.org/10.1007/JHEP07(2025)160}{\emph{JHEP} {\bfseries 07}
  (2025) 160} [\href{https://arxiv.org/abs/2503.01954}{{\ttfamily
  2503.01954}}].

\bibitem{Haisch:2025lvd}
U.~Haisch, \emph{{Higgs production from anomalous gluon dynamics}},
  \href{https://doi.org/10.1007/JHEP06(2025)004}{\emph{JHEP} {\bfseries 06}
  (2025) 004} [\href{https://arxiv.org/abs/2503.06249}{{\ttfamily
  2503.06249}}].

\bibitem{DiNoi:2025arz}
S.~Di~Noi and R.~Gr{\"o}ber, \emph{{Two loops, four tops and two
  {\ensuremath{\gamma}}5 schemes: A renormalization story}},
  \href{https://doi.org/10.1016/j.physletb.2025.139878}{\emph{Phys. Lett. B}
  {\bfseries 869} (2025) 139878}
  [\href{https://arxiv.org/abs/2507.10295}{{\ttfamily 2507.10295}}].

\bibitem{DiNoi:2025tka}
S.~Di~Noi, B.A.~Erdelyi and R.~Gr{\"o}ber, \emph{{Complete two-loop
  Yukawa-induced running of the Higgs-gluon coupling in SMEFT}},
  \href{https://doi.org/10.1007/JHEP06(2026)177}{\emph{JHEP} {\bfseries 06}
  (2026) 177} [\href{https://arxiv.org/abs/2510.14680}{{\ttfamily
  2510.14680}}].

\bibitem{Haisch:2025vqj}
U.~Haisch and M.~Niggetiedt, \emph{{Precision tests of third-generation
  four-quark operators: $gg \to h$ and $h \to γγ$}},
  \href{https://arxiv.org/abs/2507.20803}{{\ttfamily 2507.20803}}.

\bibitem{Duhr:2025yor}
C.~Duhr, G.~Ventura and E.~Vryonidou, \emph{{Two-loop renormalisation of quark
  and gluon fields in the SMEFT in the on-shell scheme}},
  \href{https://doi.org/10.1007/JHEP11(2025)046}{\emph{JHEP} {\bfseries 11}
  (2025) 046} [\href{https://arxiv.org/abs/2508.04500}{{\ttfamily
  2508.04500}}].

\bibitem{Ibarra:2024tpt}
A.~Ibarra, N.~Leister and D.~Zhang, \emph{{Complete two-loop renormalization
  group equation of the Weinberg operator}},
  \href{https://doi.org/10.1007/JHEP03(2025)214}{\emph{JHEP} {\bfseries 03}
  (2025) 214} [\href{https://arxiv.org/abs/2411.08011}{{\ttfamily
  2411.08011}}].

\bibitem{Falkowski:2017pss}
A.~Falkowski, M.~Gonz{\'a}lez-Alonso and K.~Mimouni, \emph{{Compilation of
  low-energy constraints on 4-fermion operators in the SMEFT}},
  \href{https://doi.org/10.1007/JHEP08(2017)123}{\emph{JHEP} {\bfseries 08}
  (2017) 123} [\href{https://arxiv.org/abs/1706.03783}{{\ttfamily
  1706.03783}}].

\bibitem{Falkowski:2019xoe}
A.~Falkowski, M.~Gonz{\'a}lez-Alonso and Z.~Tabrizi, \emph{{Reactor neutrino
  oscillations as constraints on Effective Field Theory}},
  \href{https://doi.org/10.1007/JHEP05(2019)173}{\emph{JHEP} {\bfseries 05}
  (2019) 173} [\href{https://arxiv.org/abs/1901.04553}{{\ttfamily
  1901.04553}}].

\bibitem{Falkowski:2023klj}
A.~Falkowski, M.~Gonz{\'a}lez-Alonso, O.~Naviliat-Cuncic and N.~Severijns,
  \emph{{Superallowed decays within and beyond the standard model}},
  \href{https://doi.org/10.1140/epja/s10050-023-01030-7}{\emph{Eur. Phys. J. A}
  {\bfseries 59} (2023) 113}.

\bibitem{Aoude:2020dwv}
R.~Aoude, T.~Hurth, S.~Renner and W.~Shepherd, \emph{{The impact of flavour
  data on global fits of the MFV SMEFT}},
  \href{https://doi.org/10.1007/JHEP12(2020)113}{\emph{JHEP} {\bfseries 12}
  (2020) 113} [\href{https://arxiv.org/abs/2003.05432}{{\ttfamily
  2003.05432}}].

\bibitem{Bruggisser:2021duo}
S.~Bruggisser, R.~Sch{\"a}fer, D.~van Dyk and S.~Westhoff, \emph{{The Flavor of
  UV Physics}}, \href{https://doi.org/10.1007/JHEP05(2021)257}{\emph{JHEP}
  {\bfseries 05} (2021) 257}
  [\href{https://arxiv.org/abs/2101.07273}{{\ttfamily 2101.07273}}].

\bibitem{Bruggisser:2022rhb}
S.~Bruggisser, D.~van Dyk and S.~Westhoff, \emph{{Resolving the flavor
  structure in the MFV-SMEFT}},
  \href{https://doi.org/10.1007/JHEP02(2023)225}{\emph{JHEP} {\bfseries 02}
  (2023) 225} [\href{https://arxiv.org/abs/2212.02532}{{\ttfamily
  2212.02532}}].

\bibitem{Grunwald:2023nli}
C.~Grunwald, G.~Hiller, K.~Kr{\"o}ninger and L.~Nollen, \emph{{More synergies
  from beauty, top, Z and Drell-Yan measurements in SMEFT}},
  \href{https://doi.org/10.1007/JHEP11(2023)110}{\emph{JHEP} {\bfseries 11}
  (2023) 110} [\href{https://arxiv.org/abs/2304.12837}{{\ttfamily
  2304.12837}}].

\bibitem{Biekotter:2018ohn}
A.~Biek{\"o}tter, T.~Corbett and T.~Plehn, \emph{{The Gauge-Higgs Legacy of the
  LHC Run II}},
  \href{https://doi.org/10.21468/SciPostPhys.6.6.064}{\emph{SciPost Phys.}
  {\bfseries 6} (2019) 064} [\href{https://arxiv.org/abs/1812.07587}{{\ttfamily
  1812.07587}}].

\bibitem{Kraml:2019sis}
S.~Kraml, T.Q.~Loc, D.T.~Nhung and L.D.~Ninh, \emph{{Constraining new physics
  from Higgs measurements with Lilith: update to LHC Run 2 results}},
  \href{https://doi.org/10.21468/SciPostPhys.7.4.052}{\emph{SciPost Phys.}
  {\bfseries 7} (2019) 052} [\href{https://arxiv.org/abs/1908.03952}{{\ttfamily
  1908.03952}}].

\bibitem{Dawson:2020oco}
S.~Dawson, S.~Homiller and S.D.~Lane, \emph{{Putting standard model EFT fits to
  work}}, \href{https://doi.org/10.1103/PhysRevD.102.055012}{\emph{Phys. Rev.
  D} {\bfseries 102} (2020) 055012}
  [\href{https://arxiv.org/abs/2007.01296}{{\ttfamily 2007.01296}}].

\bibitem{Almeida:2021asy}
E.d.S.~Almeida, A.~Alves, O.J.P.~{\'E}boli and M.C.~Gonzalez-Garcia,
  \emph{{Electroweak legacy of the LHC run II}},
  \href{https://doi.org/10.1103/PhysRevD.105.013006}{\emph{Phys. Rev. D}
  {\bfseries 105} (2022) 013006}
  [\href{https://arxiv.org/abs/2108.04828}{{\ttfamily 2108.04828}}].

\bibitem{Anisha:2021hgc}
Anisha, S.~Das~Bakshi, S.~Banerjee, A.~Biek{\"o}tter, J.~Chakrabortty,
  S.~Kumar~Patra et~al., \emph{{Effective limits on single scalar extensions in
  the light of recent LHC data}},
  \href{https://doi.org/10.1103/PhysRevD.107.055028}{\emph{Phys. Rev. D}
  {\bfseries 107} (2023) 055028}
  [\href{https://arxiv.org/abs/2111.05876}{{\ttfamily 2111.05876}}].

\bibitem{Buckley:2015lku}
A.~Buckley, C.~Englert, J.~Ferrando, D.J.~Miller, L.~Moore, M.~Russell et~al.,
  \emph{{Constraining top quark effective theory in the LHC Run II era}},
  \href{https://doi.org/10.1007/JHEP04(2016)015}{\emph{JHEP} {\bfseries 04}
  (2016) 015} [\href{https://arxiv.org/abs/1512.03360}{{\ttfamily
  1512.03360}}].

\bibitem{Aguilar-Saavedra:2018ksv}
D.~Barducci et~al., \emph{{Interpreting top-quark LHC measurements in the
  standard-model effective field theory}},
  \href{https://arxiv.org/abs/1802.07237}{{\ttfamily 1802.07237}}.

\bibitem{Brivio:2019ius}
I.~Brivio, S.~Bruggisser, F.~Maltoni, R.~Moutafis, T.~Plehn, E.~Vryonidou
  et~al., \emph{{O new physics, where art thou? A global search in the top
  sector}}, \href{https://doi.org/10.1007/JHEP02(2020)131}{\emph{JHEP}
  {\bfseries 02} (2020) 131}
  [\href{https://arxiv.org/abs/1910.03606}{{\ttfamily 1910.03606}}].

\bibitem{Bissmann:2019gfc}
S.~Bi{\ss}mann, J.~Erdmann, C.~Grunwald, G.~Hiller and K.~Kr{\"o}ninger,
  \emph{{Constraining top-quark couplings combining top-quark and
  $\boldsymbol{B}$ decay observables}},
  \href{https://doi.org/10.1140/epjc/s10052-020-7680-9}{\emph{Eur. Phys. J. C}
  {\bfseries 80} (2020) 136}
  [\href{https://arxiv.org/abs/1909.13632}{{\ttfamily 1909.13632}}].

\bibitem{Durieux:2019rbz}
G.~Durieux, A.~Irles, V.~Miralles, A.~Pe{\~n}uelas, R.~P{\"o}schl,
  M.~Perell{\'o} et~al., \emph{{The electro-weak couplings of the top and
  bottom quarks {\textemdash} Global fit and future prospects}},
  \href{https://doi.org/10.1007/JHEP12(2019)098}{\emph{JHEP} {\bfseries 12}
  (2019) 98} [\href{https://arxiv.org/abs/1907.10619}{{\ttfamily 1907.10619}}].

\bibitem{Ellis:2020unq}
J.~Ellis, M.~Madigan, K.~Mimasu, V.~Sanz and T.~You, \emph{{Top, Higgs, Diboson
  and Electroweak Fit to the Standard Model Effective Field Theory}},
  \href{https://doi.org/10.1007/JHEP04(2021)279}{\emph{JHEP} {\bfseries 04}
  (2021) 279} [\href{https://arxiv.org/abs/2012.02779}{{\ttfamily
  2012.02779}}].

\bibitem{Ethier:2021bye}
{\scshape SMEFiT} collaboration, \emph{{Combined SMEFT interpretation of Higgs,
  diboson, and top quark data from the LHC}},
  \href{https://doi.org/10.1007/JHEP11(2021)089}{\emph{JHEP} {\bfseries 11}
  (2021) 089} [\href{https://arxiv.org/abs/2105.00006}{{\ttfamily
  2105.00006}}].

\bibitem{Garosi:2023yxg}
F.~Garosi, D.~Marzocca, A.R.~S{\'a}nchez and A.~Stanzione, \emph{{Indirect
  constraints on top quark operators from a global SMEFT analysis}},
  \href{https://doi.org/10.1007/JHEP12(2023)129}{\emph{JHEP} {\bfseries 12}
  (2023) 129} [\href{https://arxiv.org/abs/2310.00047}{{\ttfamily
  2310.00047}}].

\bibitem{Bartocci:2023nvp}
R.~Bartocci, A.~Biek{\"o}tter and T.~Hurth, \emph{{A global analysis of the
  SMEFT under the minimal MFV assumption}},
  \href{https://doi.org/10.1007/JHEP05(2024)074}{\emph{JHEP} {\bfseries 05}
  (2024) 074} [\href{https://arxiv.org/abs/2311.04963}{{\ttfamily
  2311.04963}}].

\bibitem{Celada:2024mcf}
E.~Celada, T.~Giani, J.~ter Hoeve, L.~Mantani, J.~Rojo, A.N.~Rossia et~al.,
  \emph{{Mapping the SMEFT at high-energy colliders: from LEP and the (HL-)LHC
  to the FCC-ee}}, \href{https://doi.org/10.1007/JHEP09(2024)091}{\emph{JHEP}
  {\bfseries 09} (2024) 091}
  [\href{https://arxiv.org/abs/2404.12809}{{\ttfamily 2404.12809}}].

\bibitem{deBlas:2025xhe}
J.~de~Blas, A.~Goncalves, V.~Miralles, L.~Reina, L.~Silvestrini and M.~Valli,
  \emph{{Constraining new physics effective interactions via a global fit of
  electroweak, Drell-Yan, Higgs, top, and flavour observables}},
  \href{https://doi.org/10.1007/JHEP03(2026)013}{\emph{JHEP} {\bfseries 03}
  (2026) 013} [\href{https://arxiv.org/abs/2507.06191}{{\ttfamily
  2507.06191}}].

\bibitem{deBlas:2022ofj}
J.~de~Blas, Y.~Du, C.~Grojean, J.~Gu, V.~Miralles, M.E.~Peskin et~al.,
  \emph{{Global SMEFT Fits at Future Colliders}},  in \emph{{Snowmass 2021}},
  6, 2022 [\href{https://arxiv.org/abs/2206.08326}{{\ttfamily 2206.08326}}].

\bibitem{Celada:2026ubm}
E.~Celada, \emph{{Higgs, electroweak, and top quark global fits at future
  colliders}},
  \href{https://doi.org/10.1140/epjp/s13360-026-07789-0}{\emph{Eur. Phys. J.
  Plus} {\bfseries 141} (2026) 572}.

\bibitem{DasBakshi:2018vni}
S.~Das~Bakshi, J.~Chakrabortty and S.K.~Patra, \emph{{CoDEx: Wilson coefficient
  calculator connecting SMEFT to UV theory}},
  \href{https://doi.org/10.1140/epjc/s10052-018-6444-2}{\emph{Eur. Phys. J. C}
  {\bfseries 79} (2019) 21} [\href{https://arxiv.org/abs/1808.04403}{{\ttfamily
  1808.04403}}].

\bibitem{Carmona:2021xtq}
A.~Carmona, A.~Lazopoulos, P.~Olgoso and J.~Santiago, \emph{{Matchmakereft:
  automated tree-level and one-loop matching}},
  \href{https://doi.org/10.21468/SciPostPhys.12.6.198}{\emph{SciPost Phys.}
  {\bfseries 12} (2022) 198}
  [\href{https://arxiv.org/abs/2112.10787}{{\ttfamily 2112.10787}}].

\bibitem{Kraml:2025fpv}
S.~Kraml, A.~Lessa, S.~Prakash and F.~Wilsch, \emph{{SUSY meets SMEFT: complete
  one-loop matching of the general MSSM}},
  \href{https://doi.org/10.1007/JHEP04(2026)028}{\emph{JHEP} {\bfseries 04}
  (2026) 028} [\href{https://arxiv.org/abs/2506.05201}{{\ttfamily
  2506.05201}}].

\bibitem{terHoeve:2023pvs}
J.~ter Hoeve, G.~Magni, J.~Rojo, A.N.~Rossia and E.~Vryonidou, \emph{{The
  automation of SMEFT-assisted constraints on UV-complete models}},
  \href{https://doi.org/10.1007/JHEP01(2024)179}{\emph{JHEP} {\bfseries 01}
  (2024) 179} [\href{https://arxiv.org/abs/2309.04523}{{\ttfamily
  2309.04523}}].

\bibitem{Giani:2023gfq}
T.~Giani, G.~Magni and J.~Rojo, \emph{{SMEFiT: a flexible toolbox for global
  interpretations of particle physics data with effective field theories}},
  \href{https://doi.org/10.1140/epjc/s10052-023-11534-7}{\emph{Eur. Phys. J. C}
  {\bfseries 83} (2023) 393}
  [\href{https://arxiv.org/abs/2302.06660}{{\ttfamily 2302.06660}}].

\bibitem{Ellis:2017jns}
S.A.R.~Ellis, J.~Quevillon, T.~You and Z.~Zhang, \emph{{Extending the Universal
  One-Loop Effective Action: Heavy-Light Coefficients}},
  \href{https://doi.org/10.1007/JHEP08(2017)054}{\emph{JHEP} {\bfseries 08}
  (2017) 054} [\href{https://arxiv.org/abs/1706.07765}{{\ttfamily
  1706.07765}}].

\bibitem{Jiang:2018pbd}
M.~Jiang, N.~Craig, Y.-Y.~Li and D.~Sutherland, \emph{{Complete one-loop
  matching for a singlet scalar in the Standard Model EFT}},
  \href{https://doi.org/10.1007/JHEP02(2019)031}{\emph{JHEP} {\bfseries 02}
  (2019) 031} [\href{https://arxiv.org/abs/1811.08878}{{\ttfamily
  1811.08878}}].

\end{thebibliography}\endgroup

\bibstyle{JHEP}
\bibdata{references}

\end{document}